\documentclass[prd, reprint, raggedbottom, preprintnumbers]{revtex4-2}
\usepackage{graphicx}
\usepackage{amssymb}
\usepackage{amsfonts}
\usepackage{dcolumn}
\usepackage[caption=false]{subfig}

\begin{document}

\preprint{FERMILAB-PUB-19-482-T}

\title{\Large{Effects of matter density profiles on neutrino\\ oscillations for T2HK and T2HKK}}

\author{Stephen F. King}
 \email{s.f.king@soton.ac.uk, 0000-0002-4351-7507}
\affiliation{%
Department of Physics and Astronomy,\\
University of Southampton, SO17~1BJ Southampton, UK
}%

\author{Susana Molina Sedgwick}
 \email{Now at IFIC (CSIC - U. Valencia), Spain\\
 s.molina.sedgwick@ific.uv.es, 0000-0003-1872-3787}
\affiliation{
 Department of Physics and Astronomy,\\
 University of Southampton, SO17~1BJ Southampton, UK\\
 and\\
 Particle Physics Research Centre,\\
 Queen Mary University of London\\
 E1~4NS London, UK
}

\author{Stephen J. Parke}
 \email{parke@fnal.gov, 0000-0003-2028-6782}
\affiliation{%
Theoretical Physics Department,\\
Fermi National Accelerator Laboratory, Batavia, IL~60510, USA
}%

\author{Nick W. Prouse}
\email{nprouse@triumf.ca, 0000-0003-1037-3081}
\affiliation{%
TRIUMF, Vancouver, \\
BC V6T 2A3, Canada
}%

\date{March 17, 2020}

\begin{abstract}
This paper explores the effects of changes in matter density profiles on neutrino oscillation probabilities, and whether these could potentially be seen by the future Hyper-Kamiokande long-baseline oscillation experiment (T2HK). The analysis is extended to include the possibility of having an additional detector in Korea (T2HKK). 
In both cases, we find that these effects will be immeasurable, as the magnitudes of the changes in the oscillation probabilities induced in all density profile scenarios considered here remain smaller than the estimated experimental sensitivity to the oscillation probabilities of each experiment, for both appearance and disappearance channels. Therefore, we conclude that using a constant density profile is sufficient for both the T2HK and T2HKK experiments.
\end{abstract}

\maketitle

\section{Introduction}

Following the success of the Kamiokande and Super-Kamiokande (Super-K) experiments~\cite{Fukuda:2002uc}, Hyper-Kamiokande~\cite{Abe:2018uyc} (Hyper-K) will be the next-generation water Cherenkov detector in Japan. With a fiducial volume of 187~kton, 8.3 times greater than that of Super-K, Hyper-K will have a huge multipurpose research potential. It will be capable of studying everything from solar and atmospheric neutrinos to supernovae, as well as having applications to dark matter searches, neutrino tomography and proton decay.

\begin{figure}[!htbp]
	\includegraphics[width=\columnwidth]{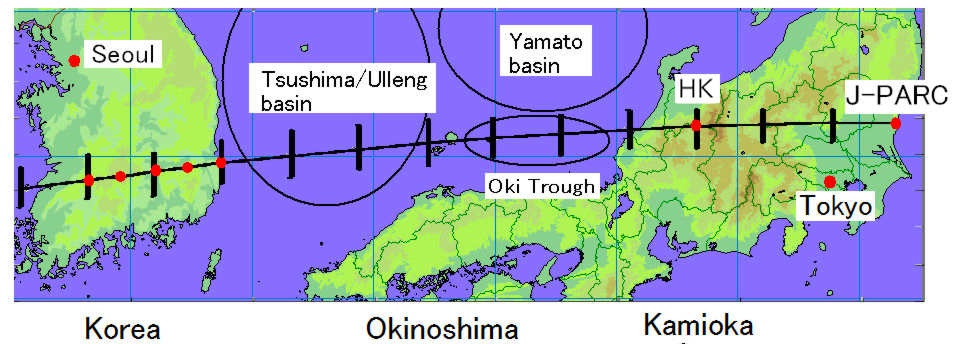}
    \caption{
        \label{fig:map}
        Simplified map showing the geographical features present between Japan and Korea, the positions of J-PARC and Kamioka (the approximate location of the Super-K and Hyper-K detectors) and five potential placements for a Korean detector, from~\cite{Hagiwara:2011kw}.
    }
\end{figure}

The long-baseline aspect of the experiment, T2HK, will involve a neutrino beam originating at J-PARC, Tokai, with an intermediate water Cherenkov detector and upgraded near detectors. While this will have a baseline of approximately 300~km, where oscillations are at the first maximum, it is a natural extension to consider an additional detector further from the beam source, at a Korean Neutrino Observatory~\cite{Abe:2016ero}. At the second oscillation maximum, this baseline would be comparable to that of other current and future long-baseline experiments, such as NO$\nu$A~\cite{Ayres:2007tu} and DUNE~\cite{Acciarri:2015uup}. Fig.~\ref{fig:map} shows these locations on a geographical map of Japan and South Korea.

At these baselines matter effects can become significant; as neutrinos propagate through the Earth, their oscillation probability is affected by interactions with the protons, neutrons and electrons in matter.

While all neutrino flavours experience neutral-current interactions with nucleons, only electron neutrinos undergo charged-current interactions with electrons, producing an asymmetry dependent on the density of electrons $(n_e)$ that directly translates into matter density.

In matter of constant density, the probability for an muon neutrino to oscillate to electron flavour is given by:
\begin{eqnarray}
    &P&(\nu_\mu \rightarrow \nu_e) \simeq \sin^2\theta_{23}\sin^2 2\theta_{13}\frac{\sin^2(\Delta_{31}-aL)}{(\Delta_{31}-aL)^2}\Delta^2_{31}\nonumber\\
    & & +\sin 2\theta_{23}\cos\theta_{13}\sin 2\theta_{13}\sin 2\theta_{12}\frac{\sin(\Delta_{31}-aL)}{(\Delta_{31}-aL)}\Delta_{31}\nonumber\\
    & & \qquad \times\frac{\sin(aL)}{aL}\Delta_{21}\cos{(\Delta_{31}+\delta)}\nonumber\\
    & & +\cos^2\theta_{13}\cos^2\theta_{23}\sin^2 2\theta_{12}\frac{\sin^2(aL)}{(aL)^2}\Delta^2_{21} 
\end{eqnarray}
where $\Delta{jk}=\Delta m^2_{jk}L/4E$, $L$ is the distance travelled, $E$ is the neutrino energy and $\Delta m_{jk}^{2} \equiv m_{j}^{2}-m_{k}^{2}$ is the neutrino mass squared difference, see \cite{Nunokawa:2007qh}. The effects of propagation through matter are given by the Wolfenstein matter potential \cite{Wolfenstein:1977ue} divided by 2:
\begin{equation}
a=\frac{G_{F} n_{e} }{ \sqrt{2}} \approx \frac{1}{3500 ~\text{km}} \, \left( \frac{\rho}{3.0~\text{g/cm}^{3}} \right)
\end{equation} 
Barger \emph{et~al.} were the first to explore, in detail, the effects of matter on three flavour neutrino oscillations, see \cite{Barger:1980tf}.

Since we are interested in varying matter density, the Schr{\"o}dinger-like equation for neutrino propagation in matter has to be solved numerically. For this reason, the GLoBES  software package~\cite{Huber:2007ji} is used for our analysis.

For DUNE it has been shown that varying the density profile would have no measurable effect on oscillations~\cite{Kelly:2018kmb}, and studies of previous designs for long-baseline experiments with beams originating in Japan found similarly small effects~\cite{Huber:2002mx,Coloma:2012ji,Barger:2007jq,Barger:2006kp,Jacobsson:2001zk}. This study aims to determine whether the geographical features present in the baselines for the approved T2HK and proposed T2HKK projects could produce a different conclusion.

\begin{figure}[!htbp]
    \includegraphics[width=\columnwidth]{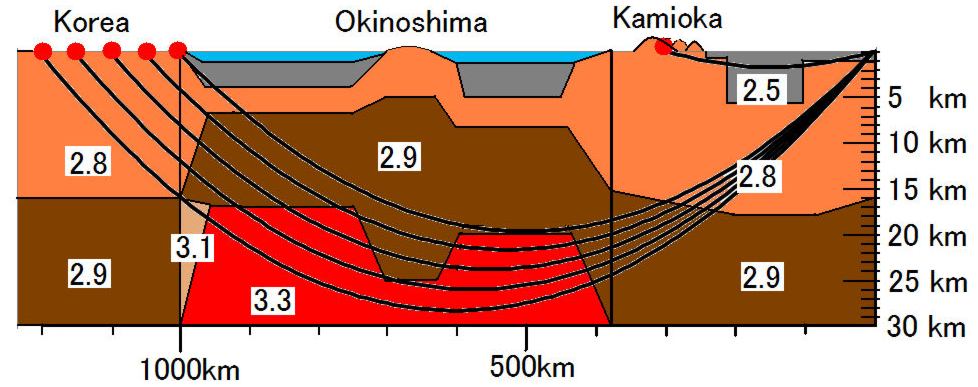}
    \caption{
        \label{fig:Baselines}
        Cross-section view of the matter density profile (in g$/$cm$^3$) from Tokai, with approximate neutrino beamlines for T2HK and 5 possible sites for a Korean detector, from~\cite{Hagiwara:2011kw}.
    }
\end{figure}

To do this, we use GLoBES implementations of T2HK and T2HKK based on those in \cite{Ballett:2016daj}, modified to match the matter density profiles shown in Fig.~\ref{fig:Baselines}. We assume that the Hyper-K detector and the Korean detector each consist of single tanks of 187~kton fiducial volume.

Section~\ref{Sites} gives an overview of the two most likely sites for the Korean detector, Mount Bisul and Mount Bohyun, on which much of this paper will be focused. Section~\ref{Sens} shows our estimations of the measurement precision of T2HK and T2HKK. In Section~\ref{Param} we present the sensitivity to variations in oscillation parameters. Section~\ref{Dens} explores the effects of changes in a matter density profile with respect to its average value, while Section~\ref{Low-dens} investigates the effects when compared to a representative low-density matter profile. Section~\ref{ap:MatterSensitivity} shows the combined experimental sensitivity to the scale of the matter density with the oscillation parameters being measured. In Section~\ref{Conc} we provide a summary of the results we have obtained, and discuss their implications.

\section{Mount Bisul and Mount Bohyun}
\label{Sites}

The most likely candidate site for the Korean detector is Mount Bisul, which is shown to have the highest sensitivity enhancements to non-standard interactions of neutrinos and mass ordering determination, as well as both solar and supernova relic neutrino searches~\cite{Abe:2016ero}.

The location which would provide the second-highest event rate is Mount Bohyun, which also currently hosts the Bohyunsan Optical Astronomy Observatory \cite{Abe:2016ero}. Both sites have been shown to be suitable for a large-cavern excavation using geological surveys. Table~\ref{tab:sites_approx} shows a comparison of the approximate values of the baseline and off-axis angle for these two sites, which are the most relevant characteristics used in our simulations.

\begin{table}[htbp]
    \begin{ruledtabular}\begin{tabular}{lcr}
	    Candidate site & Approx. baseline & Approx. off-axis angle \\
    	\colrule
    	Mount Bisul & 1100 km & 1.5\,$^\circ$ \\
    	Mount Bohyun & 1050 km & 2.5\,$^\circ$ \\
    \end{tabular}\end{ruledtabular}
    \caption{
        \label{tab:sites_approx}
        The baseline and off-axis angle values used in the simulations for the two most likely candidate sites for the Korean detector, Mount Bisul and Mount Bohyun. The actual values differ slightly and are (1088~km,  $1.3^\circ$) and (1043~km, $2.3^\circ$)~\cite{Abe:2016ero}.  This difference has no effect on our conclusions.
    }
\end{table}

\begin{figure*}[!htbp]
    \subfloat{
        \includegraphics[width=0.45\textwidth]{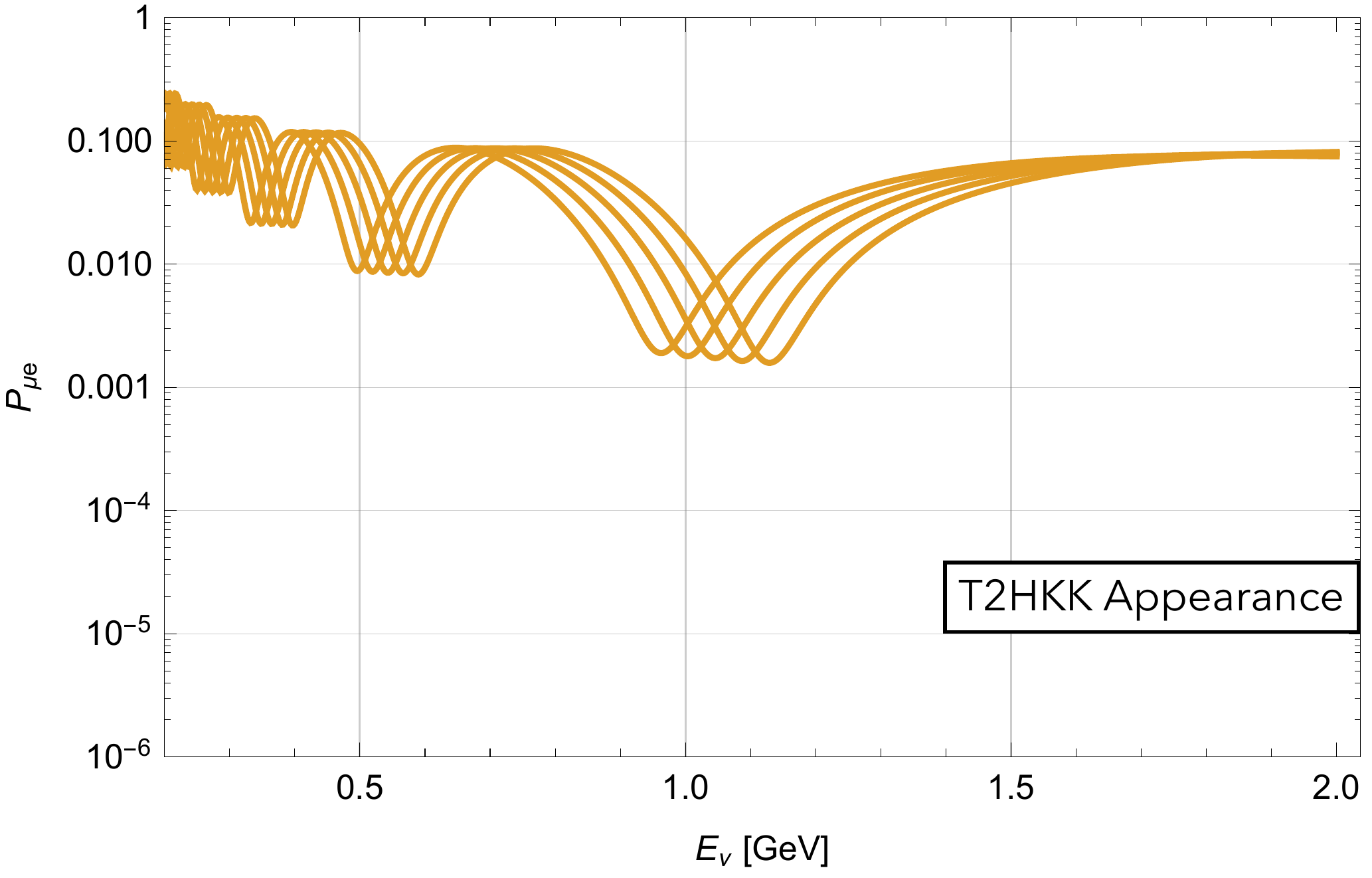}\qquad
        \includegraphics[width=0.45\textwidth]{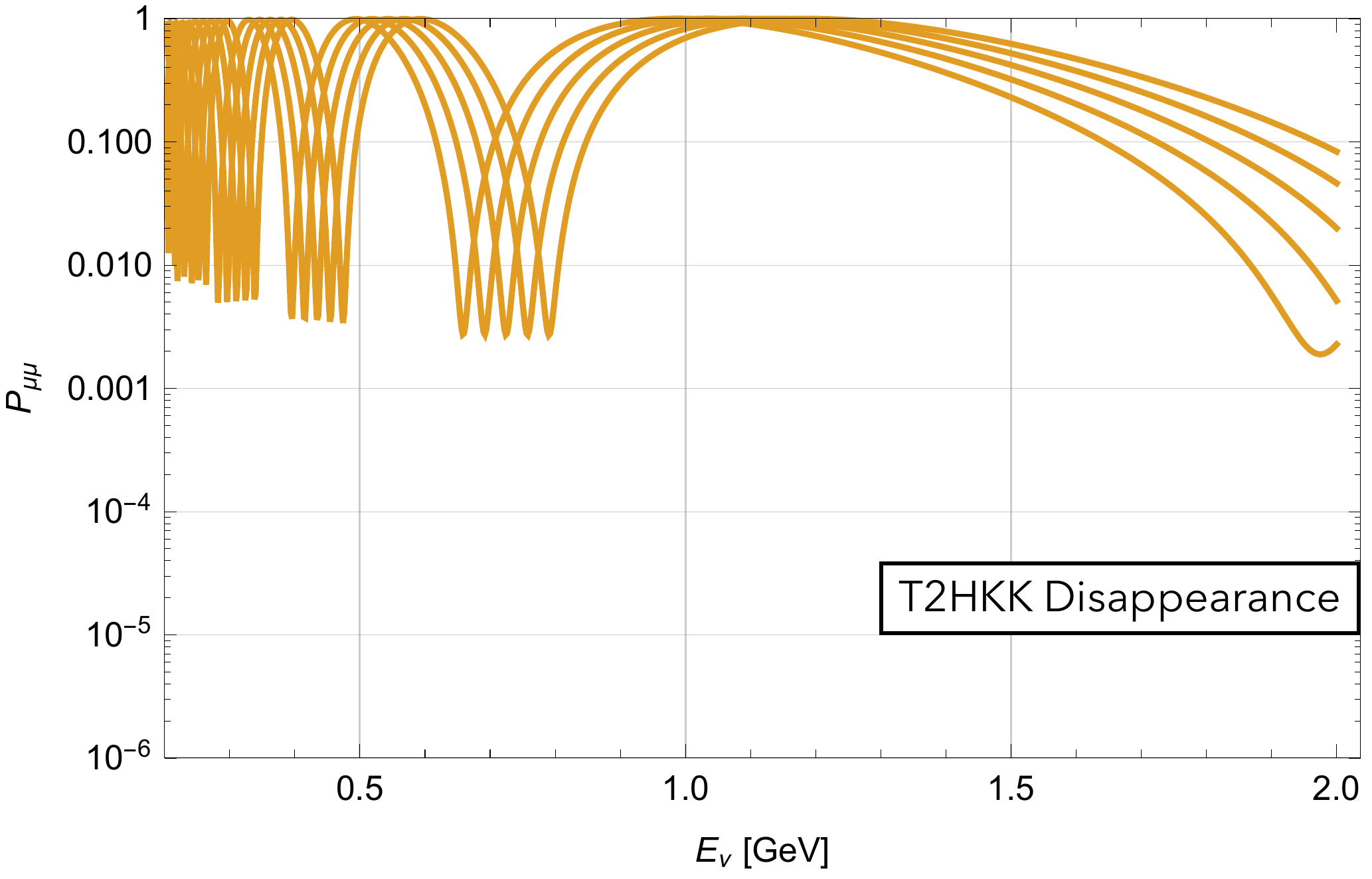}
	}\\
	\subfloat{
        \includegraphics[width=0.45\textwidth]{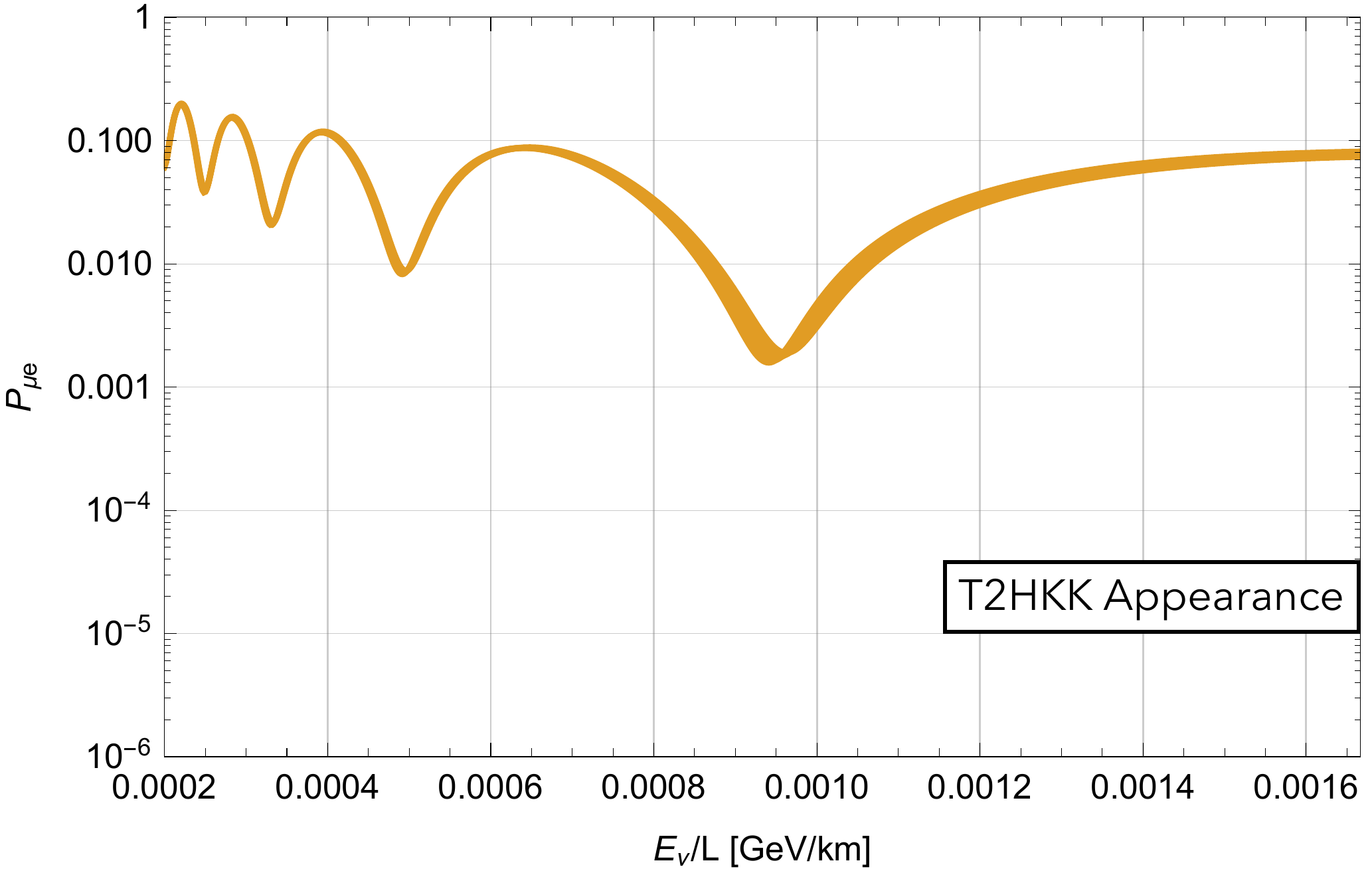}\qquad
        \includegraphics[width=0.45\textwidth]{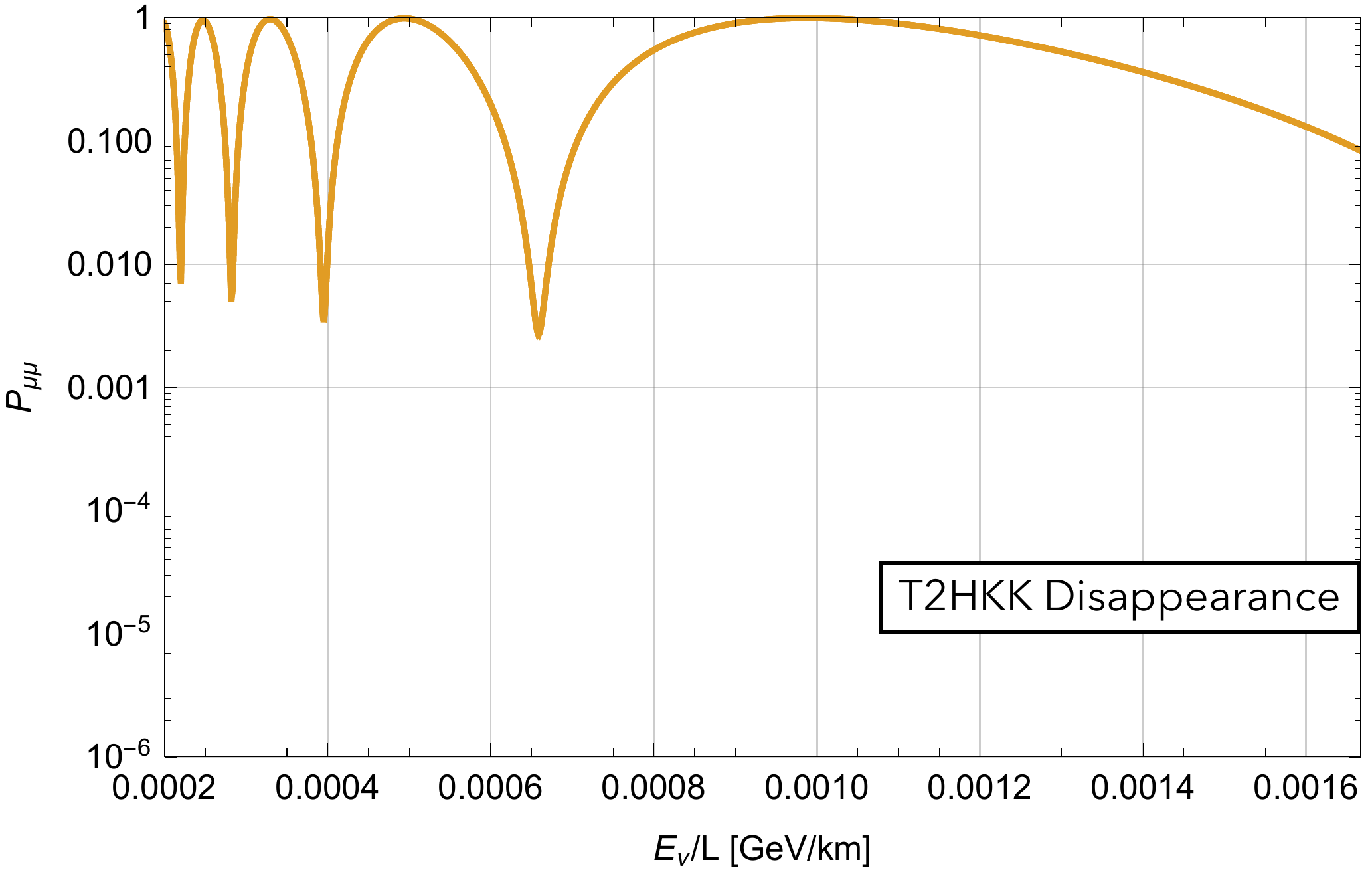}
    }
	\caption{
	    \label{fig:VarProb_All_HKK}
	    Oscillation probability as a function of energy (top) or $E/L$ (bottom) for all five baseline profiles of T2HKK, taken from Fig.~\ref{fig:Baselines} and increasing in length from left to right (1000 km, 1050 km, 1100 km, 1150 km, 1200 km). Left and right panels correspond to appearance and disappearance channels, respectively.
	}
\end{figure*}

Fig.~\ref{fig:VarProb_All_HKK} shows the oscillation probability for different baselines of T2HKK as a function of energy and of $E/L$. It can be seen that in the latter case they follow an almost identical path, with slight variations a direct result of differences in their matter density profiles.

For the following studies, we have chosen to focus on the baseline corresponding most closely to the primary candidate site of Mount Bisul, in addition to the secondary site of Mount Bohyun in Sections \ref{Dens} and \ref{Low-dens}.

\section{Sensitivity Estimates}
\label{Sens}

We explore two methods of determining experimental sensitivity to the oscillation probability:
\begin{itemize}
\item Find a $1\sigma$ sensitivity region in oscillation parameter space (using the $\chi^2$ calculation of GLoBES) then convert this into a corresponding $1\sigma$ region of the oscillation probability.
\item A ``naive'' estimate based on the method described in \cite{Kelly:2018kmb}. This is used as a cross-check, as comparisons between both methods allow us to be confident about the shapes produced.
\end{itemize}

\begin{figure*}[!htbp]
	\subfloat[T2HK]{
    	\includegraphics[width=0.45\textwidth]{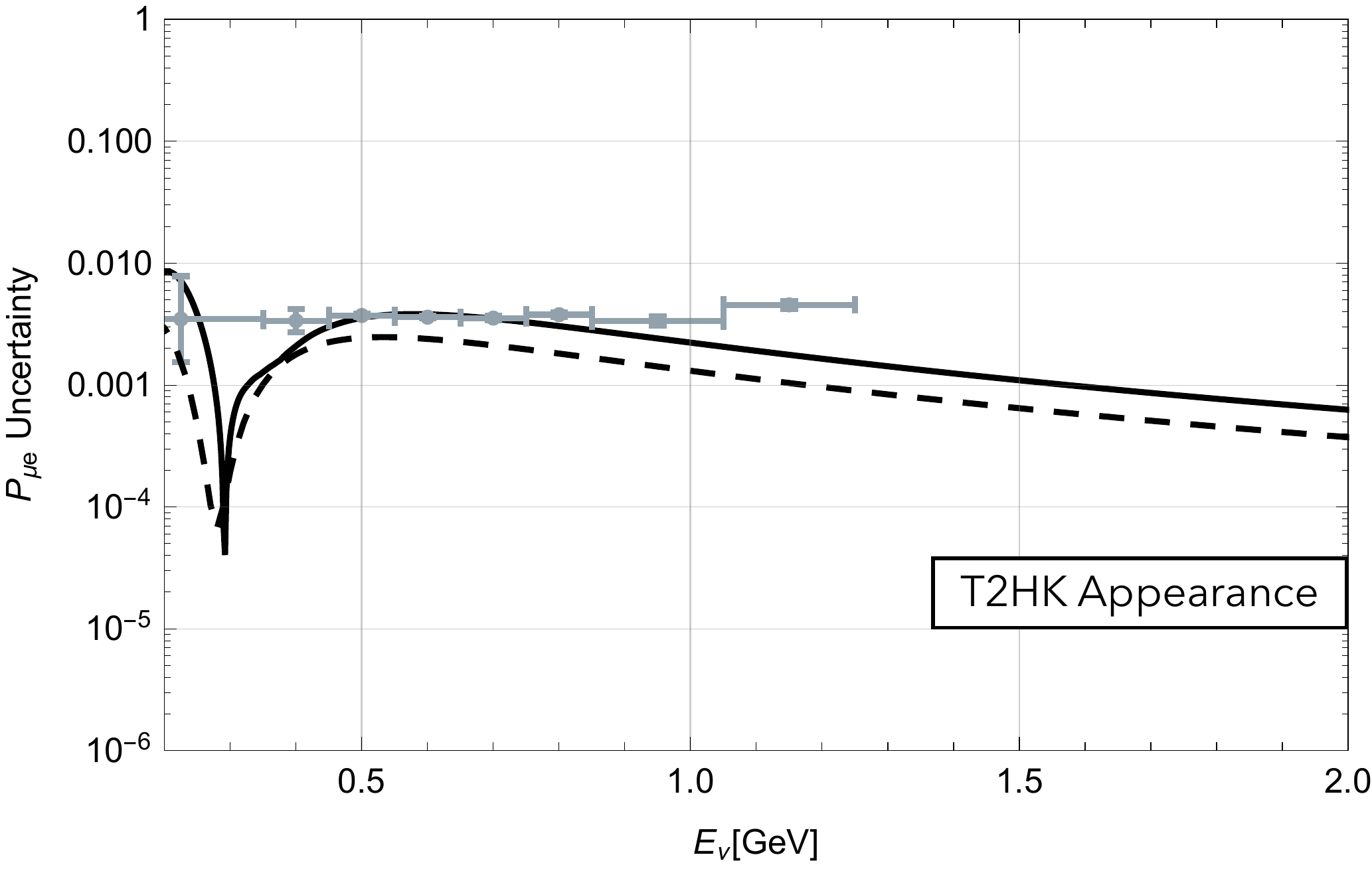}\qquad
    	\includegraphics[width=0.45\textwidth]{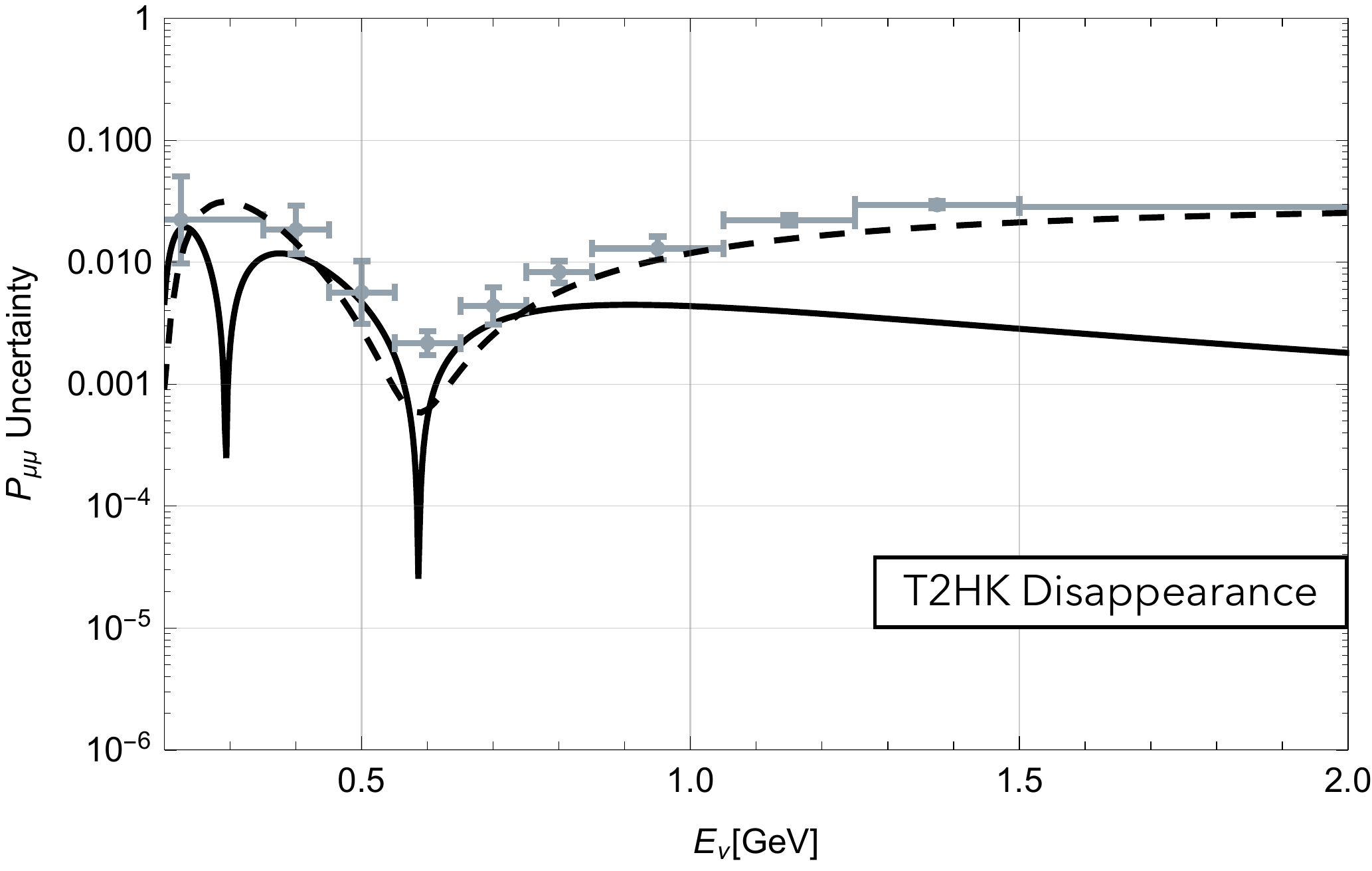}
	}\\
	\subfloat[T2HKK]{
        \includegraphics[width=0.45\textwidth]{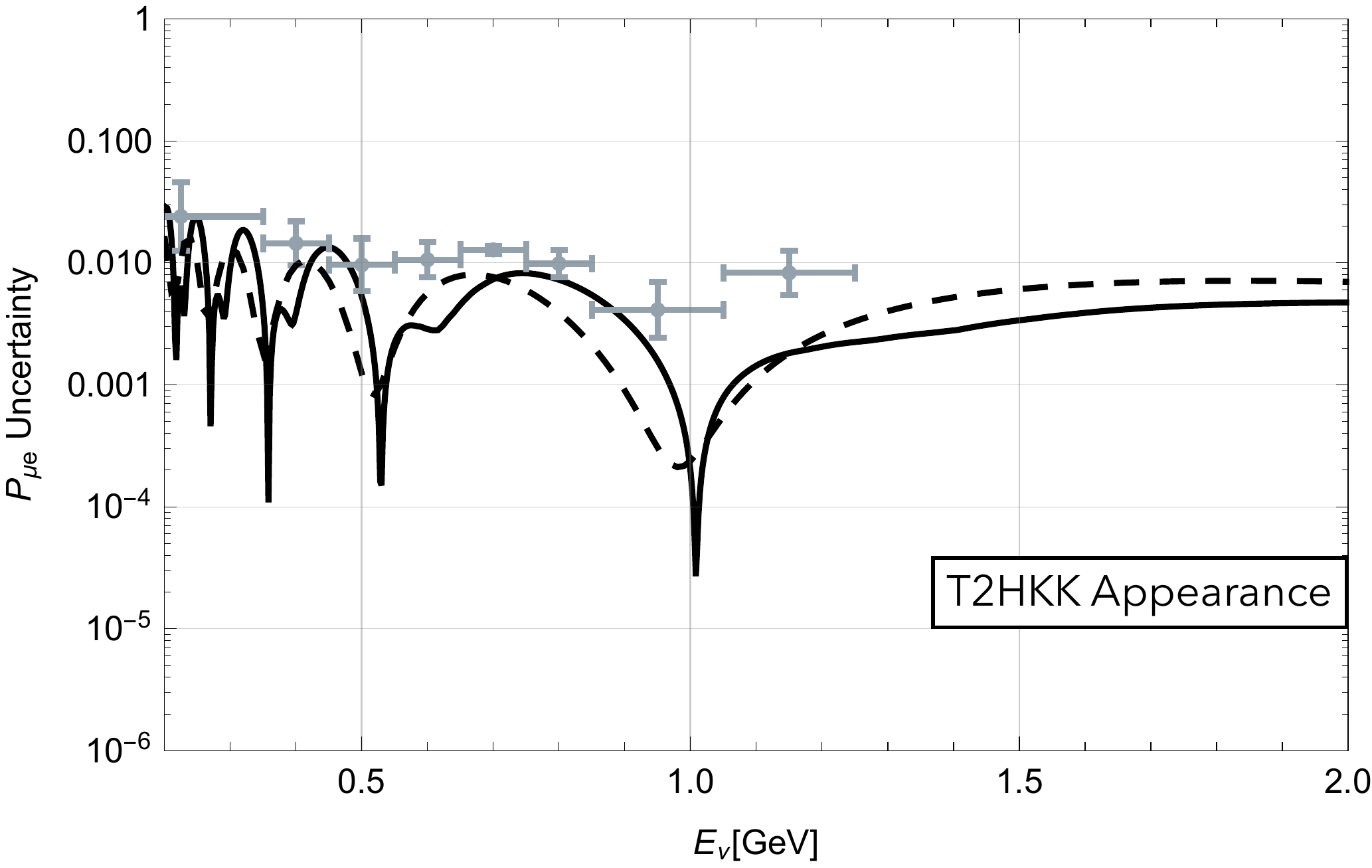}\qquad
    	\includegraphics[width=0.45\textwidth]{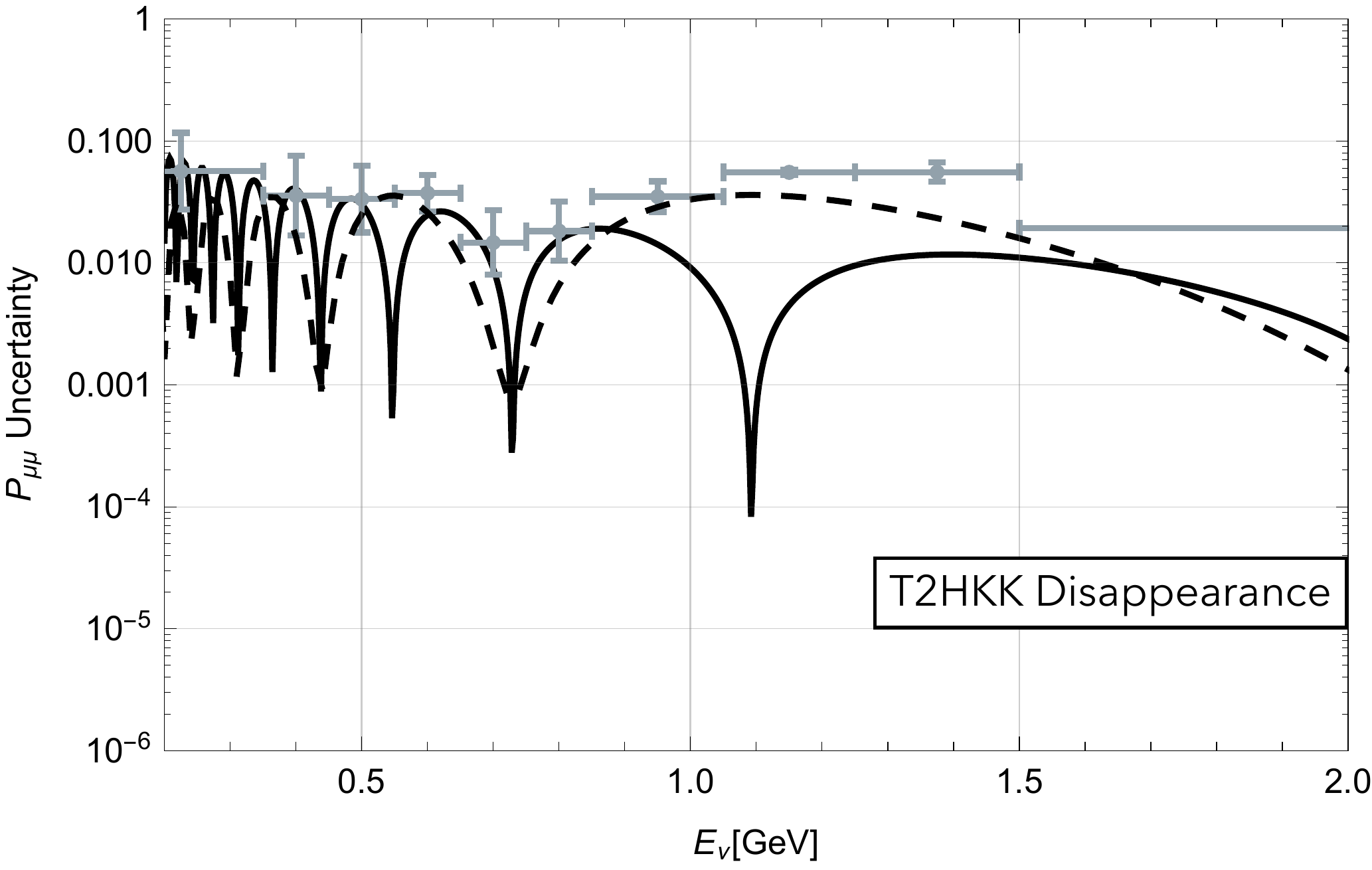}
	}
    \caption{
        \label{fig:FullSens}
        Sensitivity estimates for T2HK (above) and T2HKK (below). The continuous black curves depict the results using the GLoBES method, which uses information from both detectors to determine an uncertainty on oscillation probability. The naive method, shown in the grey energy bins, assumes that -- at a given energy -- only the events in the same energy bin contribute to the sensitivity. The vertical error bars then correspond to the range of values for the estimated uncertainty as the energy is varied across the bin. The dashed line is obtained by using the total number of events across all energies, rather than from each specific bin. The left and right panels show appearance and disappearance channels, respectively.
        }
\end{figure*}

The first method is shown as the continuous black line in Fig.~\ref{fig:FullSens}. We show the curves calculated using combined data from both detectors. We assume that the uncertainty in the (dis)appearance probability is only significantly dependant on the uncertainties of $\delta_{CP}$ and $\theta_{13}$ ($\theta_{23}$ and $\Delta m^2_{32}$).

The values in these curves correspond to the uncertainty on the oscillation probability of neutrinos at each given energy point, obtained from experimental measurements of neutrinos from a beam spanning a range of energies. Since the uncertainty on oscillation parameters is determined using data from all neutrino energies present in the beam, this information is used to constrain the allowed values for the oscillation probability at all other energies, under the assumption of standard three flavour neutrino oscillations.

Two additional ways of estimating the uncertainty on oscillation probability are overlaid in Fig.~\ref{fig:FullSens}; these are primarily used as a cross-check to the GLoBES method, which was developed for this analysis.

The so-called ``naive'' method, based on the method used in \cite{Kelly:2018kmb} and shown here in discrete energy bins, is dependent on the event rates that would be observed in each energy bin if the oscillation probability were 1. In the case of the appearance channel, there are only 8 bins that have non-zero appearance rates at HK, which accounts for the smaller energy range than in our other studies.

The vertical error bars correspond to the range of values for the estimated uncertainty as the energy is varied across the bin. This method of estimating the uncertainty makes the assumption that, at a given energy, only the events in the same energy bin contribute to the sensitivity. This gives a lower estimated sensitivity away from the HK flux peak of 600\,MeV, due to the decreased event rates in these bins. As such, this assumption leads to a more conservative sensitivity estimate.

The dashed line is obtained by using the total number of events across all energies, rather than from each specific bin. In this case, the assumption is that every event contributes information that constrains the oscillation probability across all energies. This provides a more similar estimate to the GLoBES method, where all events are used to constrain the allowed ranges of the oscillation parameters, which then determine the allowed range of the oscillation probability at any given energy.

In other words, while the naive method shows how sensitive each energy bin of each channel is to the probability in that bin / channel, the GLoBES method shows the sensitivity taking into account all data and the correlations between probabilities in all bins / channels.

\section{Variation of Oscillation Parameters}
\label{Param}

\begin{figure*}[!htbp]
    \subfloat[T2HK]{
    	\includegraphics[width=0.45\textwidth]{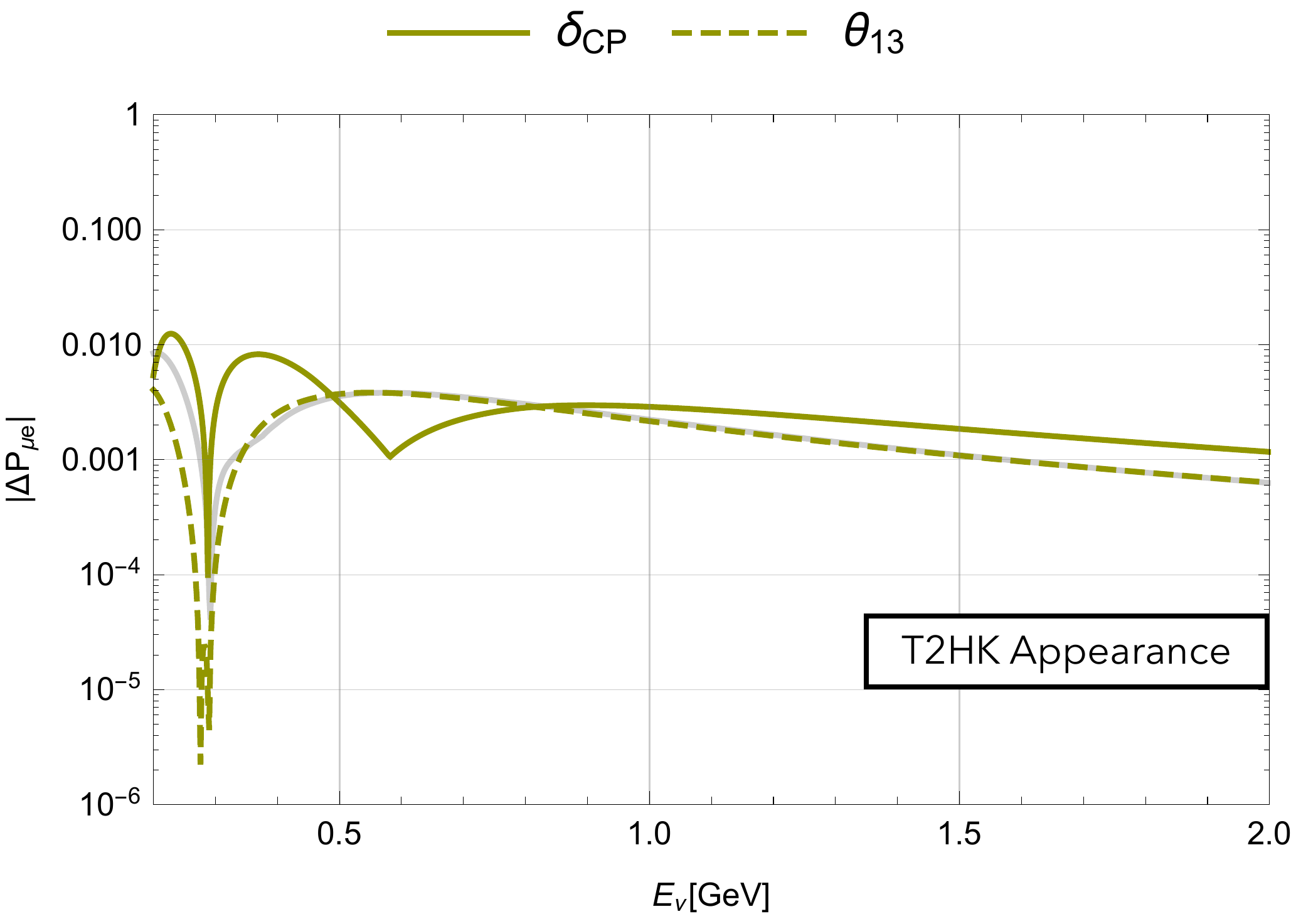}\qquad
    	\includegraphics[width=0.45\textwidth]{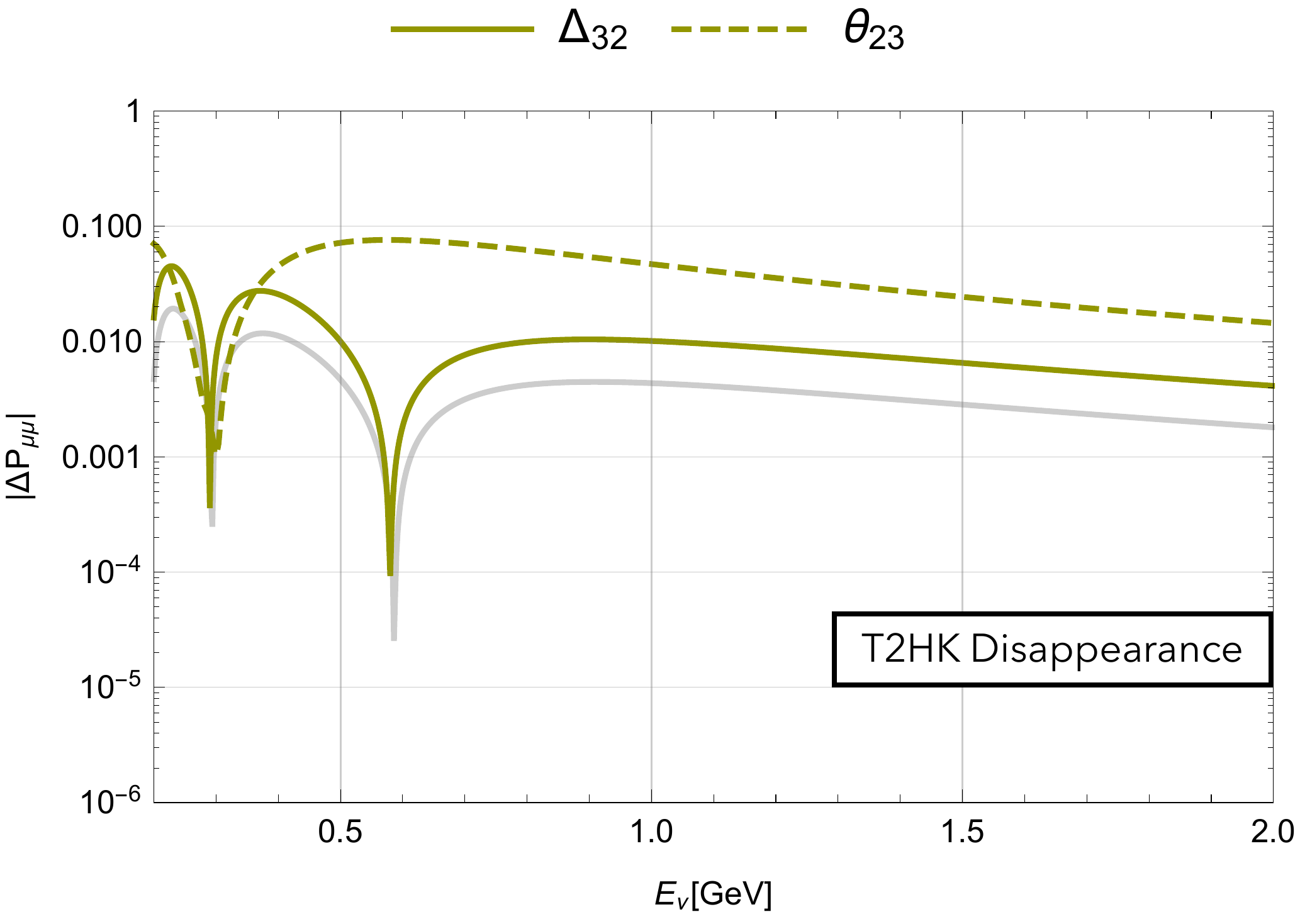}
	}\\
	\subfloat[T2HKK]{
    	\includegraphics[width=0.45\textwidth]{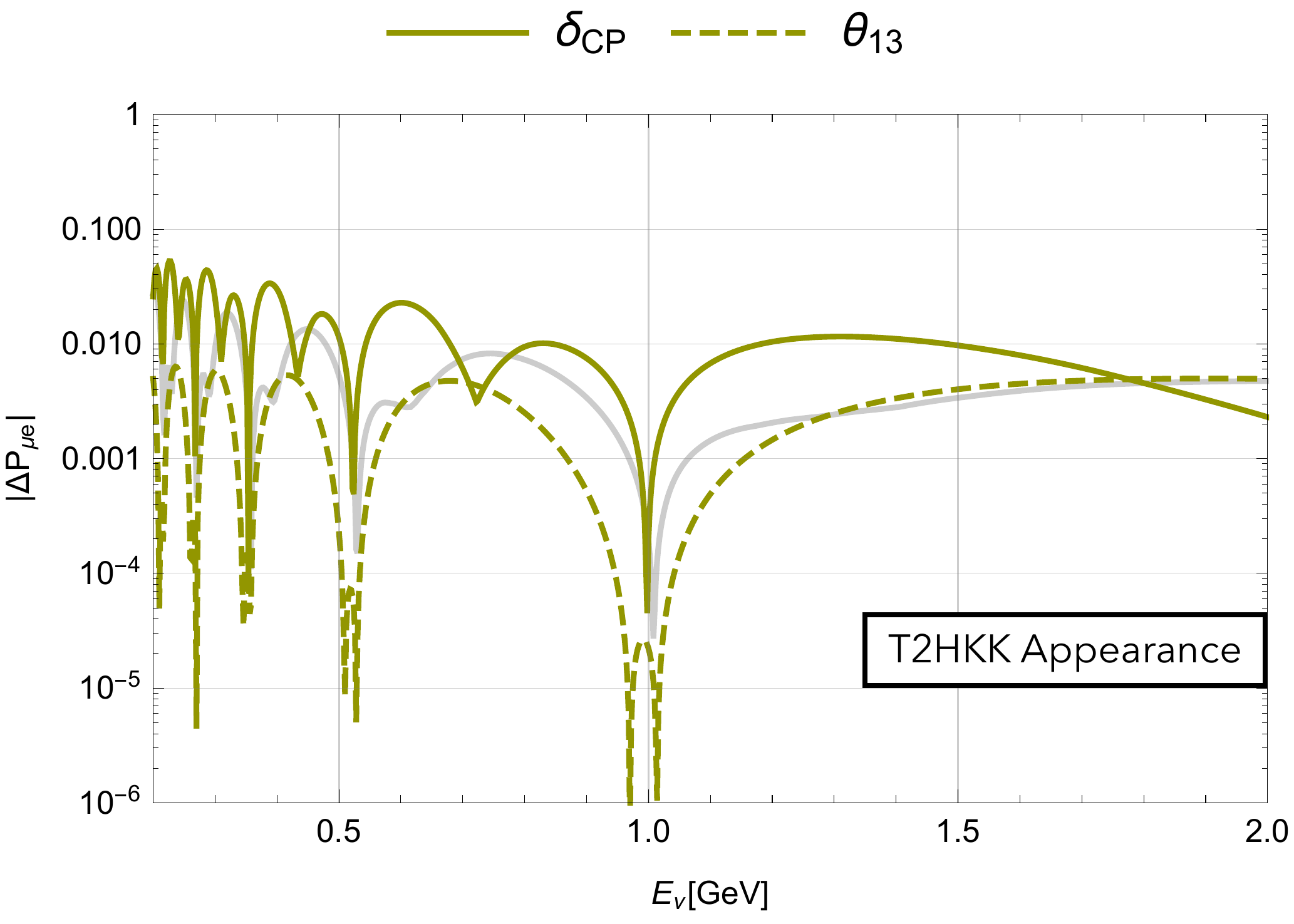}\qquad
    	\includegraphics[width=0.45\textwidth]{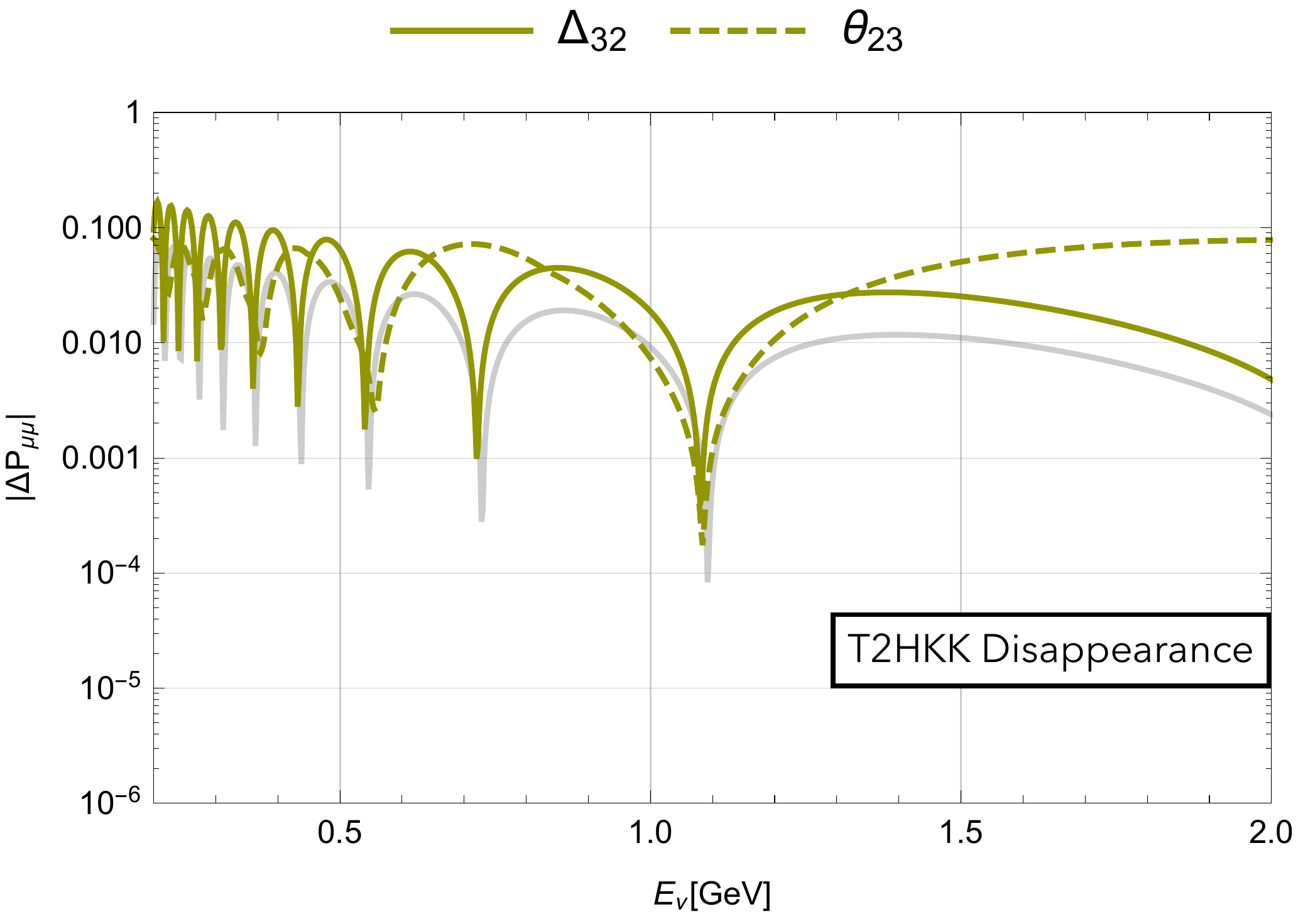}
	}
    \caption{
        \label{fig:ParamScan_Var}
        Changes in oscillation probability due to parameter variation at T2HK (above) and T2HKK (below). Left panel shows the parameters measured by the appearance channel ($\delta_{CP}$ and $\theta_{13}$), right panel the parameters associated with the disappearance channel ($\Delta m_{32}$ and $\theta_{23}$). The corresponding GLoBES sensitivity estimates from Fig.~\ref{fig:FullSens} are overlaid in grey.
    }
\end{figure*}

In this section, we show the changes in neutrino oscillation probability arising from a variation of the oscillation parameters within their $1\sigma$ ranges. The change in the oscillation probabilities while oscillation parameters vary across these ranges provides a verification of the sensitivity estimates of the previous section. This calculation of the change in oscillation probability is performed in the same way as our results shown in Fig.~\ref{fig:FullSens}, but by varying oscillation parameters across the ranges from Hyper-K's official expected sensitivities, rather than using our own sensitivity calculations using GLoBES. The ranges used for the calculation are shown in Table~\ref{tab:param_ranges}.
\begin{table}[!htbp]
    \begin{ruledtabular}
    \begin{tabular}{lrr}
	    Observable & Physical Value & $1\sigma$ Range \\
	    \colrule
	    $\delta/^\circ$ & ${-90}$ & $[-113,-67]$ \\
	    $\theta_{13}/^\circ$ & $8.54$ & $[8.39,8.69]$ \\
	    $\theta_{23}/^\circ$ & $45$ & $[37.51,52.49]$ \\
	    $\Delta m_{32}^2/\textrm{eV}^2$ & $2.494 \times 10^{-3}$ & $[2.48,2.508] \times 10^{-3}$ \\
        \end{tabular}
    \end{ruledtabular}
    \caption{
        \label{tab:param_ranges}
        Expected measurement precision of oscillation parameters with T2HK. The $\pm 1\sigma$ ranges have been taken from the Hyper-K Design Report \cite{Abe:2018uyc}, with the exception of $\theta_{13}$, which is given by the current NuFit range \cite{Esteban:2018azc}.
    }
\end{table}

The results are shown in Fig.~\ref{fig:ParamScan_Var}. At each energy point, we scan over the parameter range and plot the difference between the maximum and minimum oscillation probability found within the scanned range.

The shapes of these curves are seen to correspond approximately to the sensitivities seen in the previous section. This is as expected, since the changes in oscillation probability that would result from changing the oscillation parameters correspond precisely to the changes that would be measured by T2HK and T2HKK.

\section{Matter Density Profile vs. Average Density}
\label{Dens}

We show here the changes in oscillation probability as a result of changes in the matter density with respect to its average value. Throughout this section, left and right panels correspond to appearance and disappearance channels, respectively.
\begin{figure*}[!htbp]
    \includegraphics[width=0.9\textwidth]{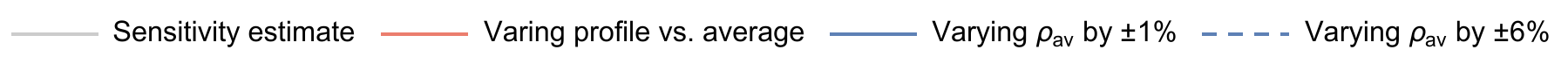}\\
	\includegraphics[width=0.45\textwidth]{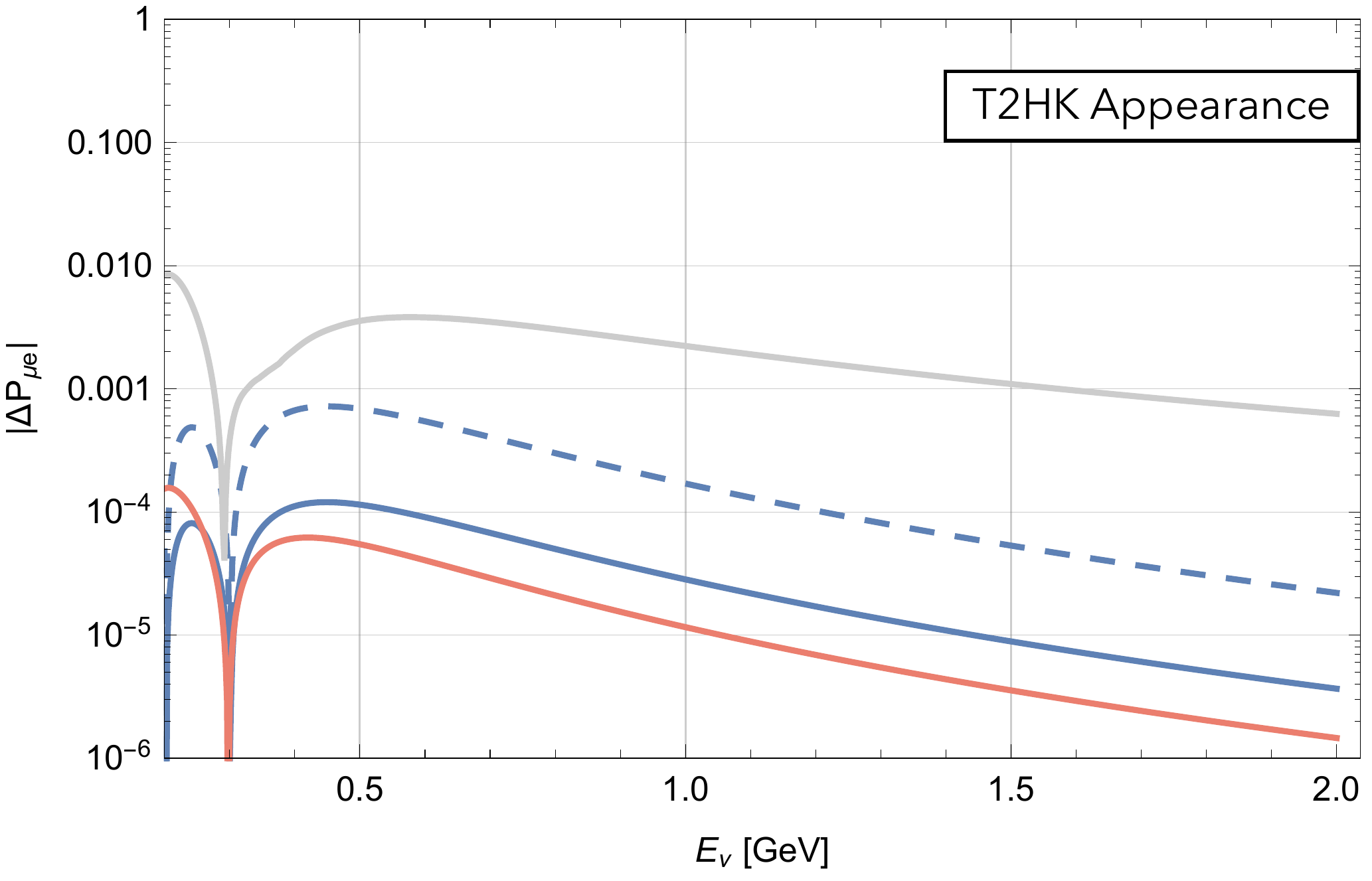}\qquad
	\includegraphics[width=0.45\textwidth]{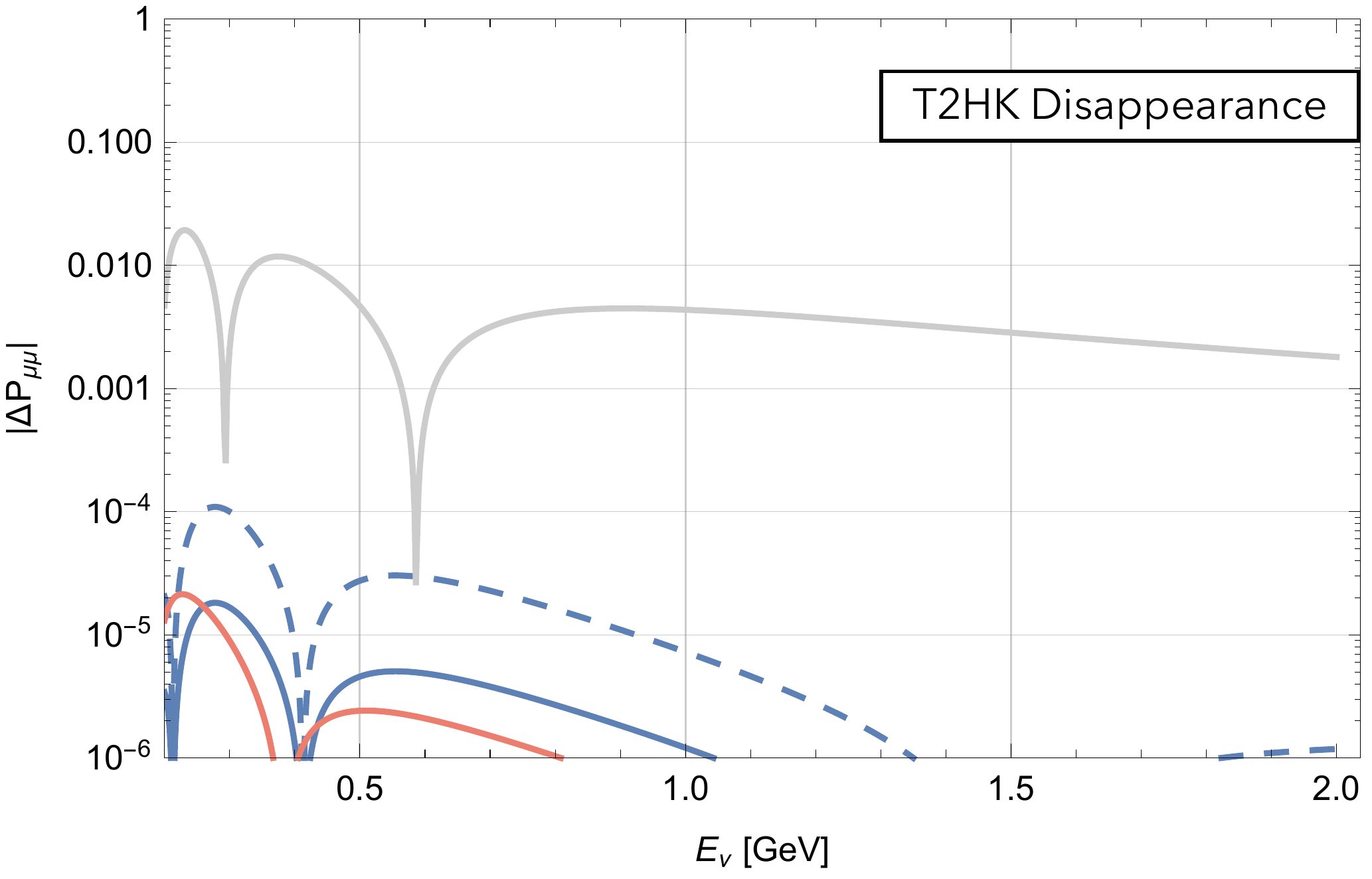}
	\caption{
        \label{fig:Dens_HK}
        Changes in oscillation probability from variations in the density profile with respect to average density of T2HK. To obtain the solid (dashed) blue curve, the average density is varied by $\pm 1\%$ ($\pm 6\%$) and the difference between the two extremes is taken at each point. The red curve is produced by taking the difference at each point between the probability calculated using the changing (real) matter density profile and the probability computed using its average density value as a constant. For comparison purposes, the sensitivity estimate obtained using GLoBES for T2HK in Fig.~\ref{fig:FullSens} is overlaid in grey.
	}
\end{figure*}

\begin{figure*}[!htbp]
    \subfloat{
        \includegraphics[width=0.9\textwidth]{dpv_Legend.pdf}
    }\addtocounter{subfigure}{-1}\\
    \subfloat[Mount Bisul]{
        \includegraphics[width=0.45\textwidth]{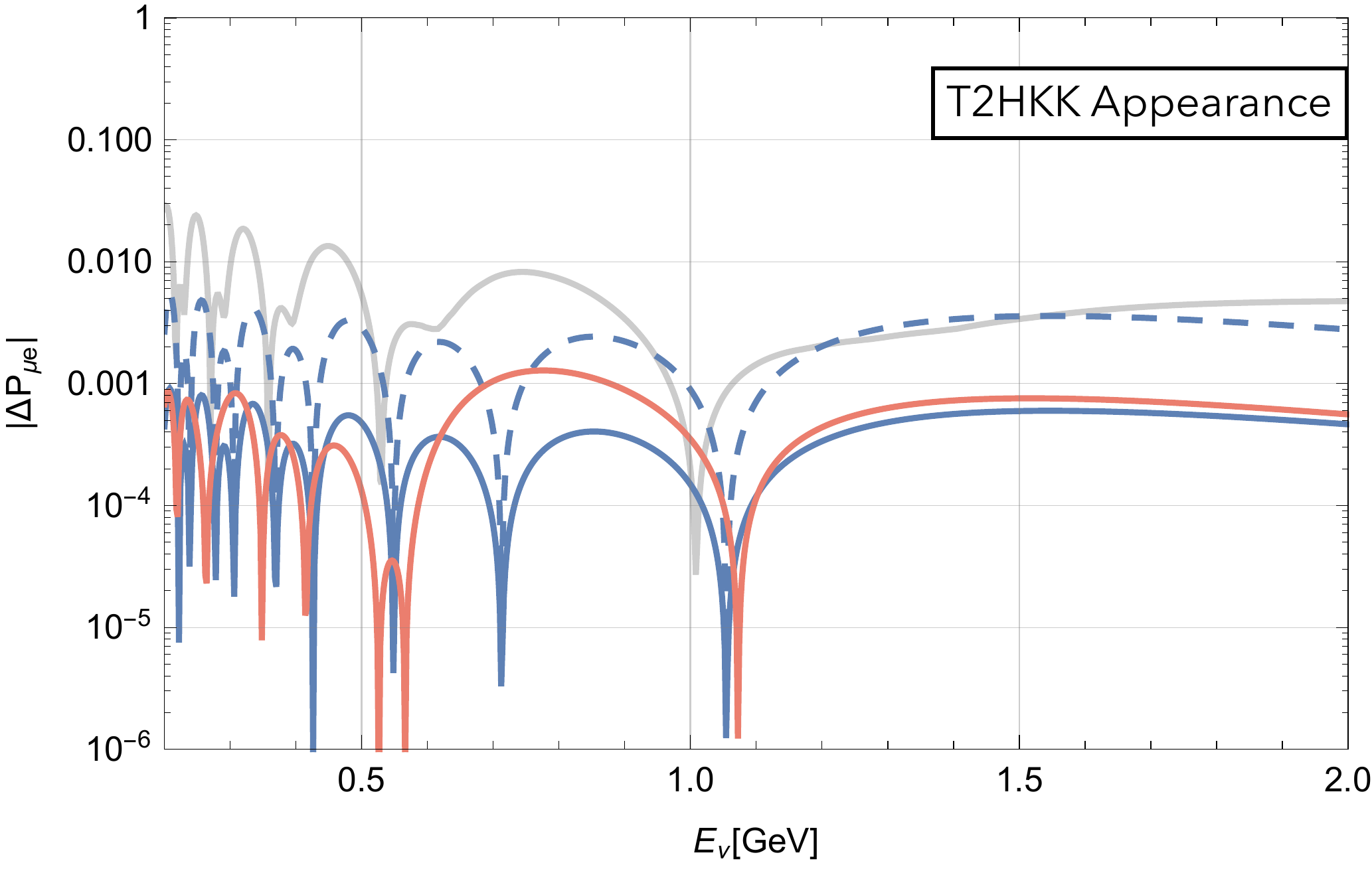}\qquad
        \includegraphics[width=0.45\textwidth]{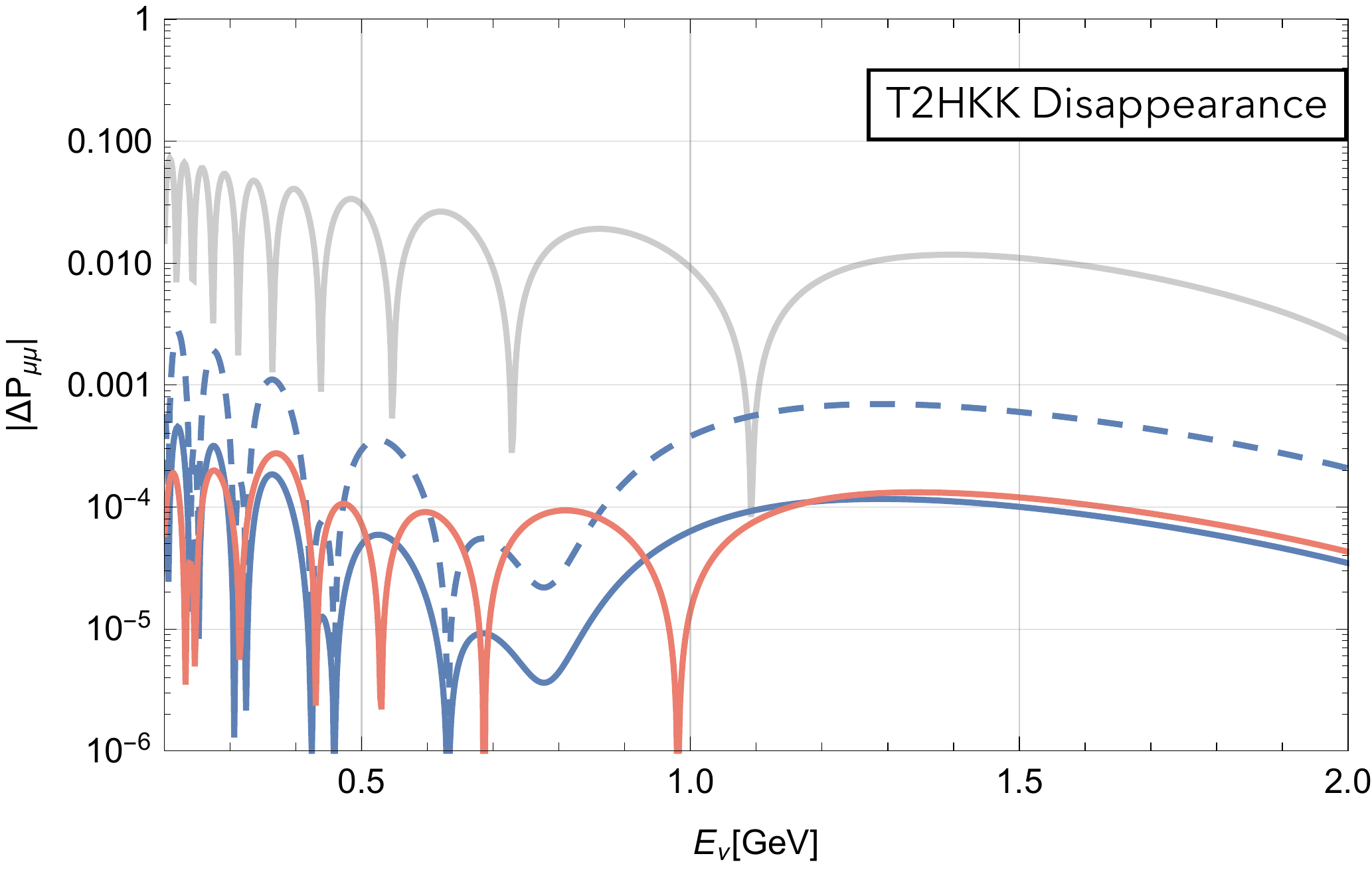}
	}\\
	\subfloat[Mount Bohyun]{
        \includegraphics[width=0.45\textwidth]{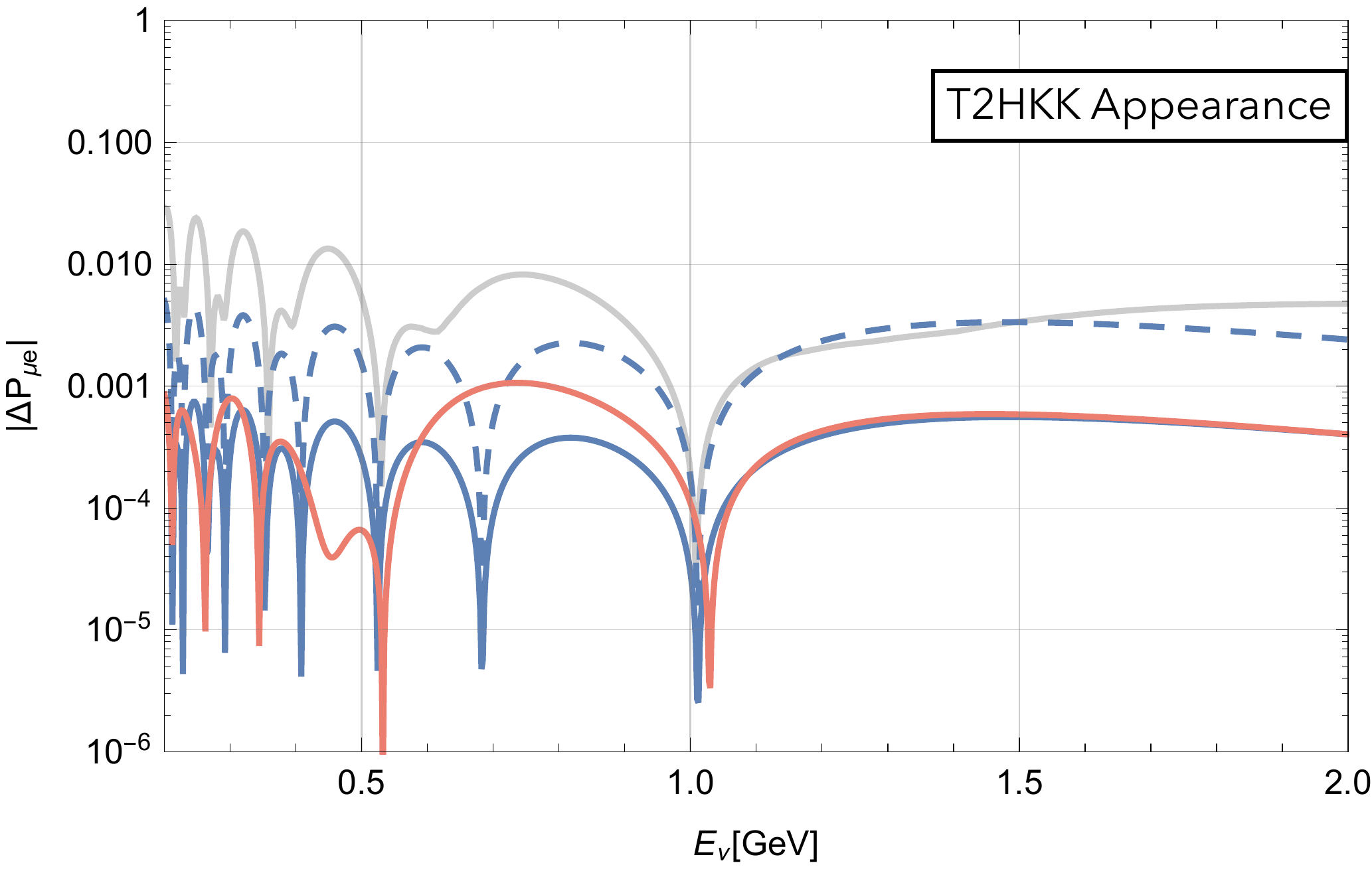}\qquad
        \includegraphics[width=0.45\textwidth]{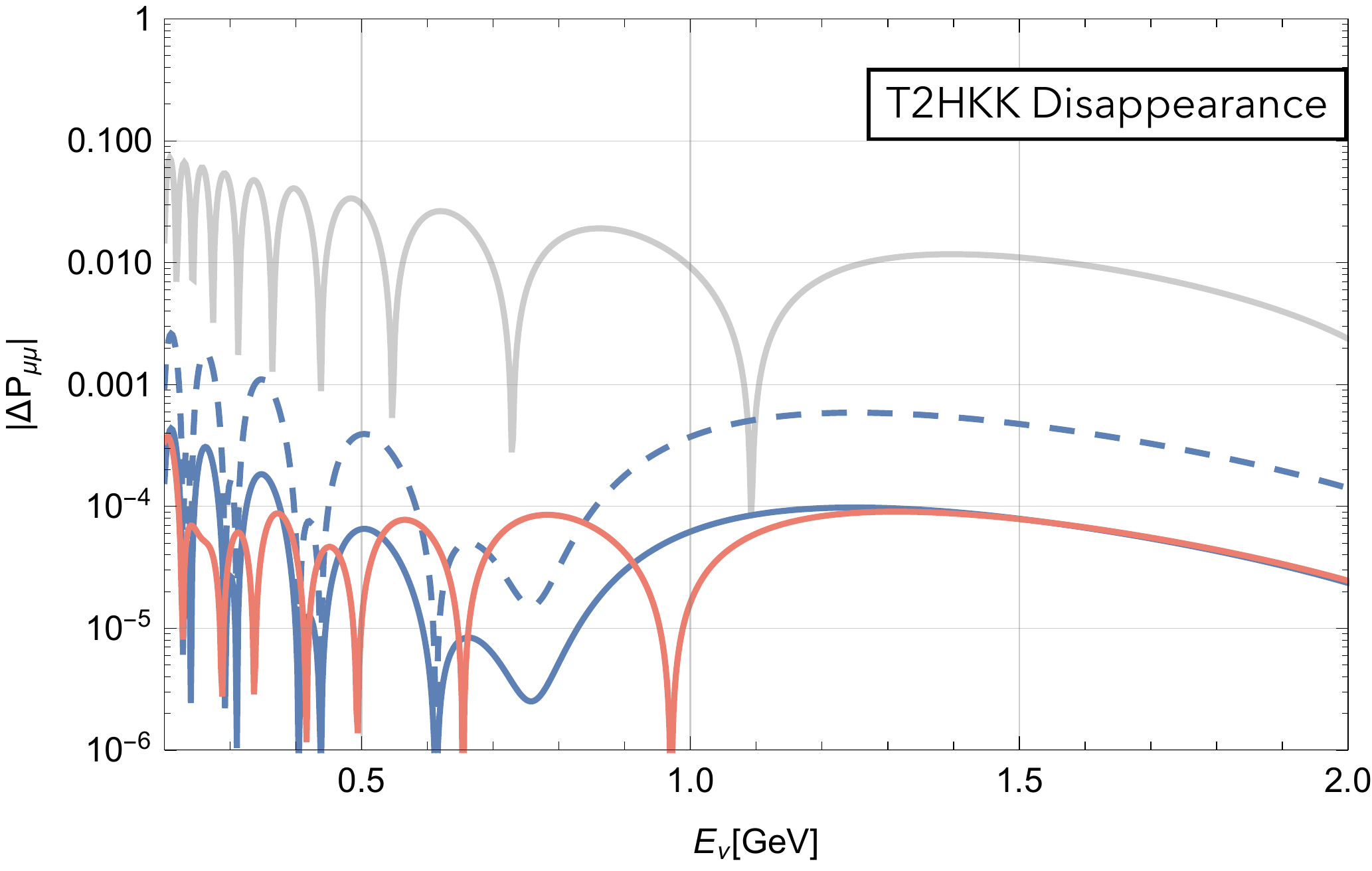}
    }
    \caption{
        \label{fig:Dens_Sites}
        Density profile changes with respect to average density corresponding to the two most likely candidate sites for the Korean detector, Mount Bisul (above) and Mount Bohyun (below).
        The blue (solid + dashed) and red curves are the same as in Fig.~\ref{fig:Dens_HK} and the sensitivity estimate shown here in grey corresponds to that of T2HKK in Fig.~\ref{fig:FullSens} (GLoBES method).
    }
\end{figure*}

For T2HK, the result is shown in Fig.~\ref{fig:Dens_HK}. For T2HKK, these same results for all the individual baselines are presented in Appendix~\ref{ap:Baselines}, and additional figures showing these results in combined plots can be found in Appendix~\ref{ap:Combined}. However, here we include only those corresponding to Mount Bisul and Mount Bohyun; these are shown in Fig.~\ref{fig:Dens_Sites}. It can be seen that while the results for each site seem at first glance to follow roughly the same path, there are noticeable differences in their shapes.

It is also interesting to note that for both T2HKK baselines the effect of comparing a varying profile to its average value is consistently higher than that of varying the average by $\pm1\%$, while in the case of T2HK the latter effect is dominant over the former at all but very low energies. However, if we increase the amount the average value is varied to $\pm6\%$, which is the uncertainty on this value given in \cite{Hagiwara:2011kw}, this effect is almost always about an order of magnitude more significant than the other two cases considered.

\begin{figure*}[!htbp]
    \includegraphics[width=0.9\textwidth]{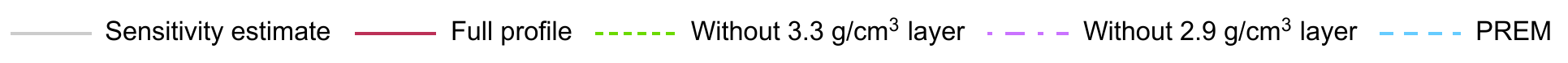}\\
	\subfloat[Mount Bisul]{
        \includegraphics[width=0.45\textwidth]{{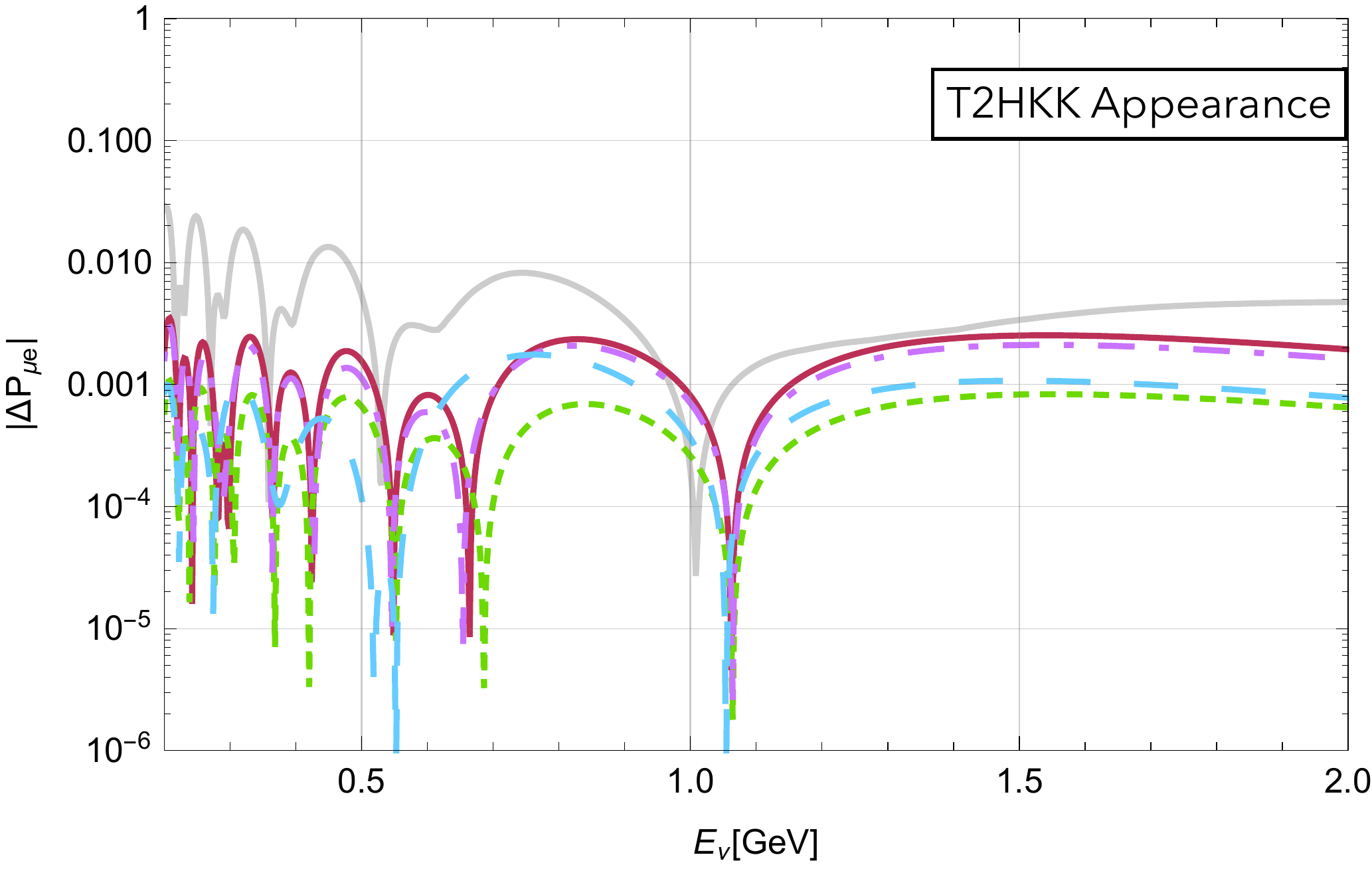}}\qquad
        \includegraphics[width=0.45\textwidth]{{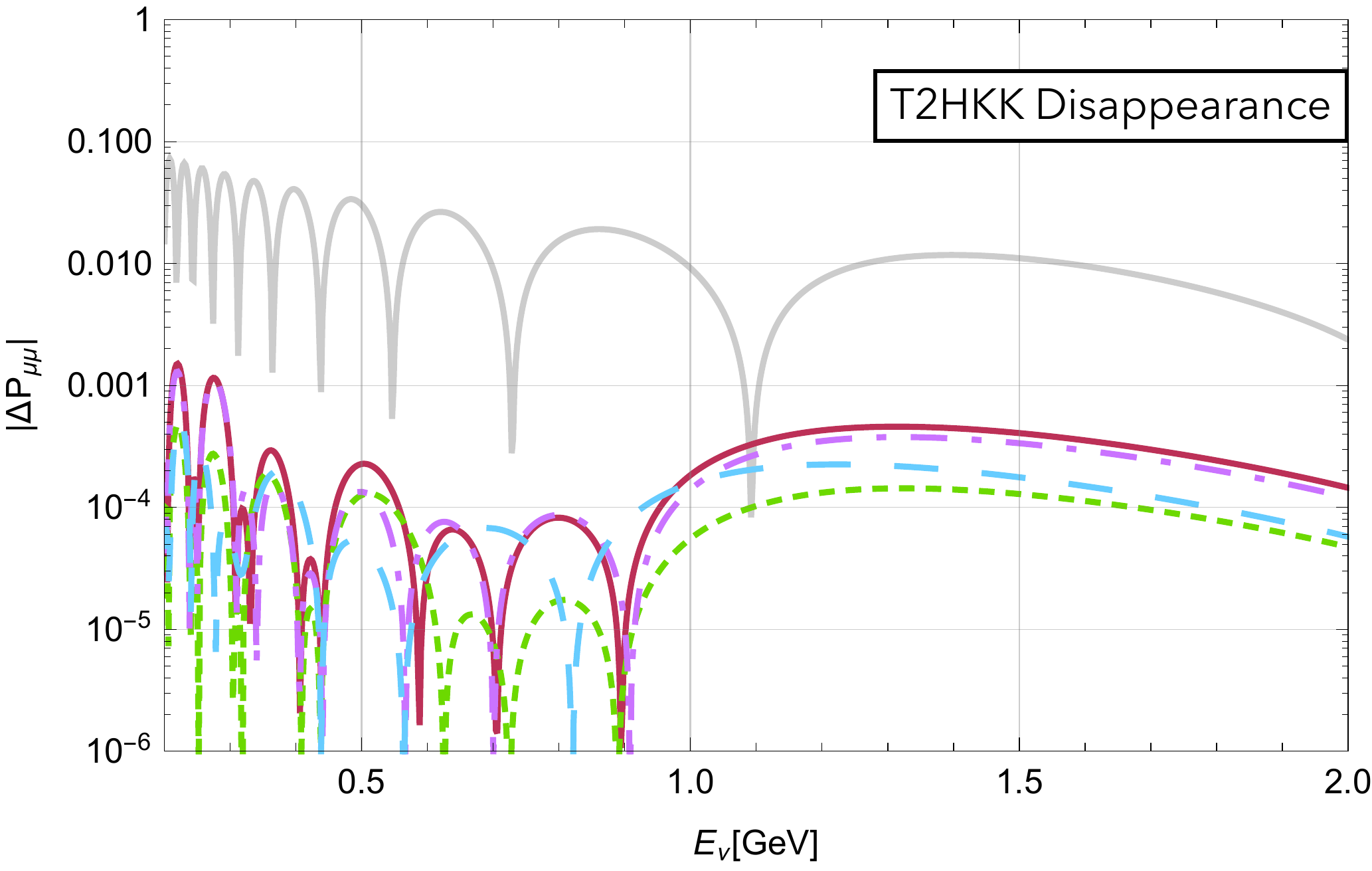}}
    }\\
    \subfloat[Mount Bohyun]{
    	\includegraphics[width=0.45\textwidth]{{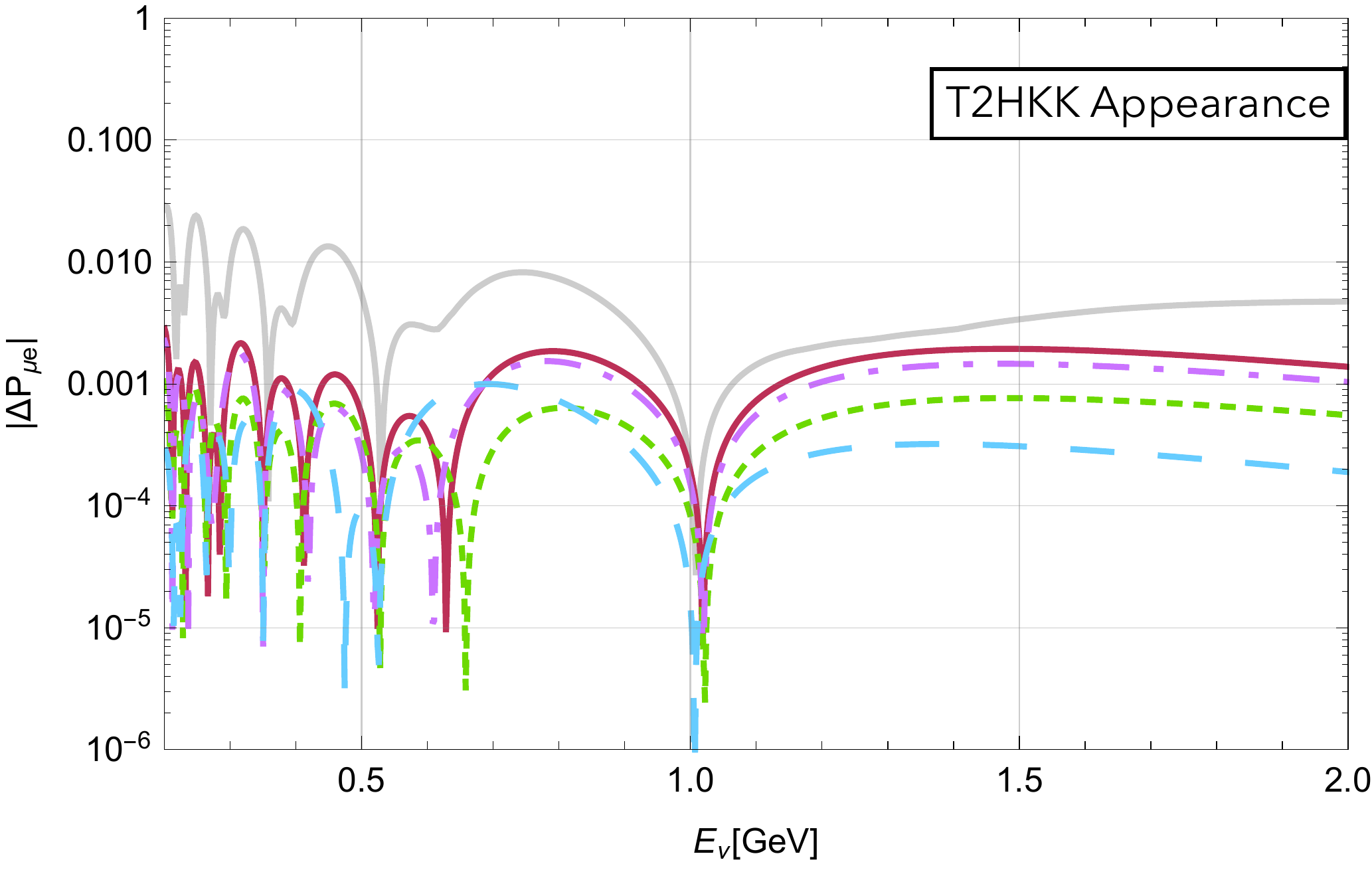}}\qquad
    	\includegraphics[width=0.45\textwidth]{{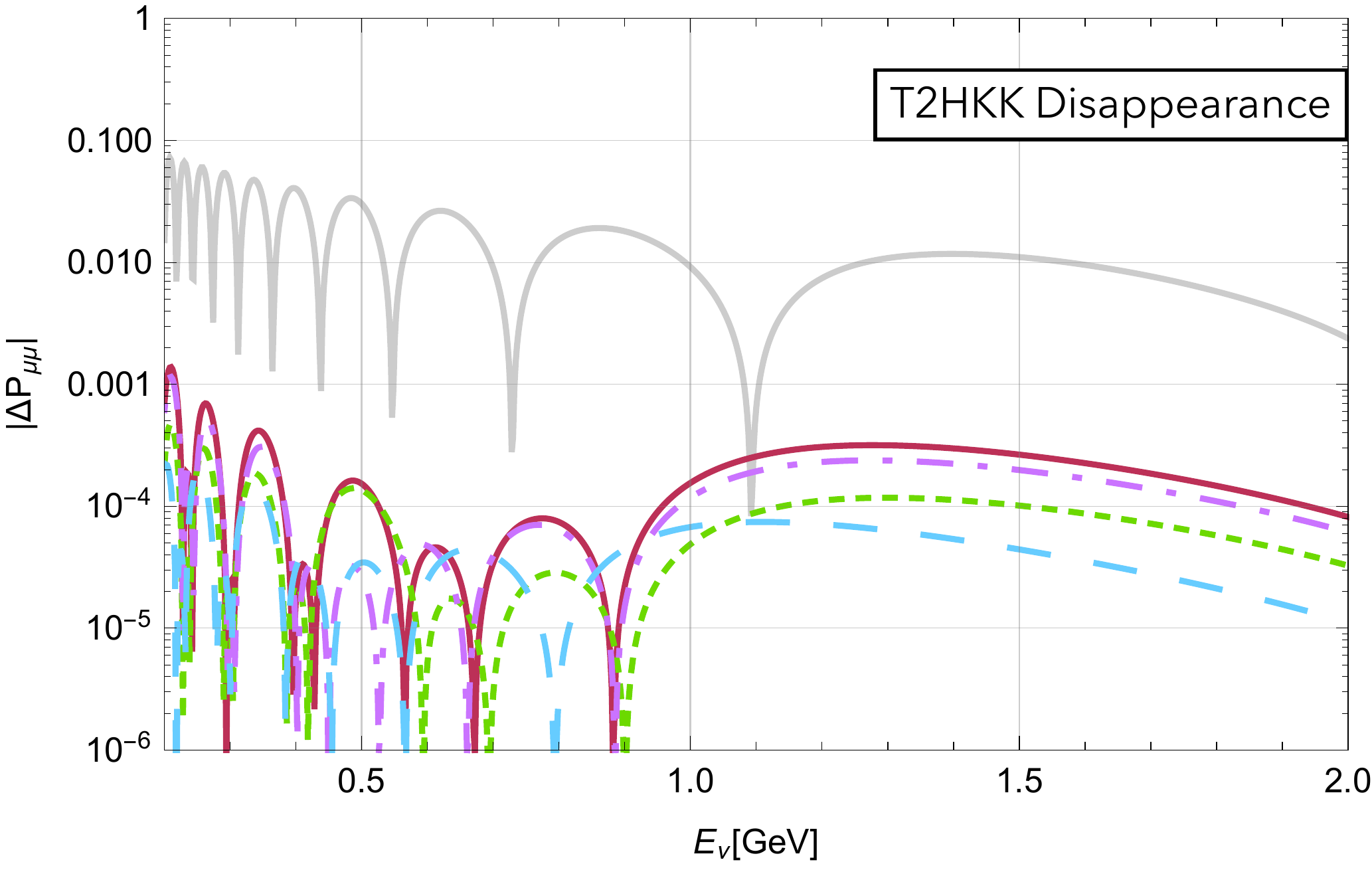}}
    }\\
    \caption{
        \label{fig:Sites}
        Changes in oscillation probability from variations in the density profile corresponding to the two most likely candidate sites for the Korean detector, Mount Bisul (above) and Mount Bohyun (below). The four curves correspond to changes in probability when comparing the four scenarios described in Section~\ref{Low-dens} to a representative basic profile with no high-density region at all, taking the difference between the two at each energy point. As before, the sensitivity estimate obtained for T2HKK in Fig.~\ref{fig:FullSens} is shown in grey for comparison. Appearance channel is shown on the left and disappearance channel on the right.}
\end{figure*}

\section{Comparisons with low-density matter profile}
\label{Low-dens}

By focusing again on the most likely candidate sites for the Korean detector, Mount Bisul and Mount Bohyun (see Table~\ref{tab:sites_approx}), we can redo the calculations after removing each of the density ``chunks'' in turn. This can give us a better idea for how each of the different density areas under the Earth's surface will affect the final result.

Fig.~\ref{fig:Sites} shows the changes in oscillation probability when comparing four different matter density profiles to a representative basic profile with no high-density region at all.

The four scenarios that are considered are:
\begin{itemize}
\item The full profile, as depicted in Fig.~\ref{fig:Baselines} (continuous pink line).
\item The full profile without the highest density region of $3.3$ g/cm$^3$ (dashed green line).
\item The full profile without the second-highest density region of $2.9$ g/cm$^3$ (dot-dashed purple line).
\item The matter density profile according to the Preliminary Reference Earth Model (PREM)~\cite{Dziewonski:1981xy}, where the earth is a spherically symmetric ball and its layers are concentric with equal distance between their boundaries at every point (dashed blue line).
\end{itemize}

Since these different profiles differ in both shape and average density, but with the average density changing between the profiles by less than the $\pm 6\%$ studied in the previous section, it is not surprising that the changes seen here are found to be more significant than changes due to changing the matter profile shape alone, but less significant than changing the average density by $\pm 6\%$.

\begin{figure*}[!htbp]
	\subfloat[$\delta=0$]{
    	\includegraphics[width=0.4\textwidth]{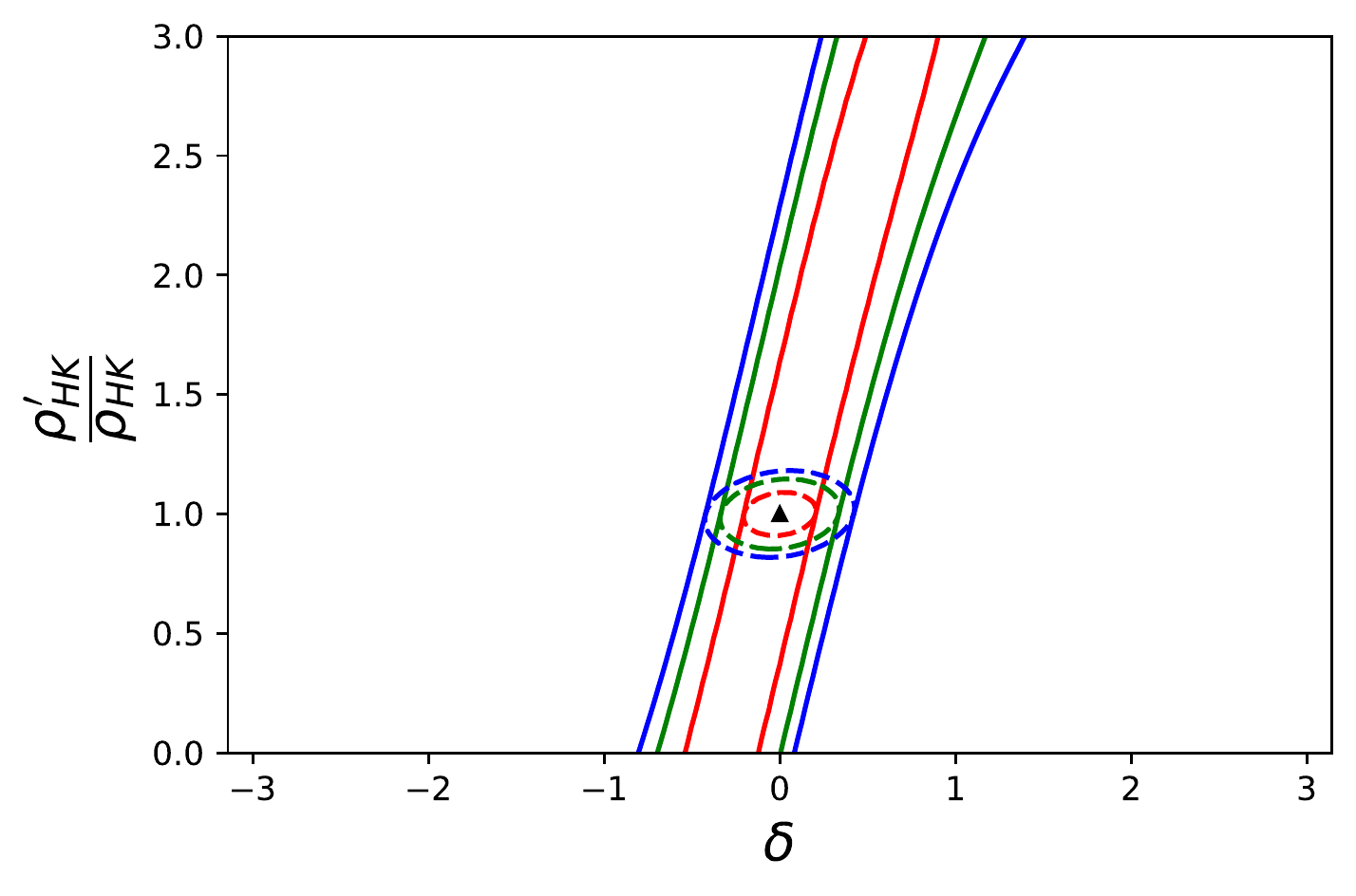}\qquad
	    \includegraphics[width=0.4\textwidth]{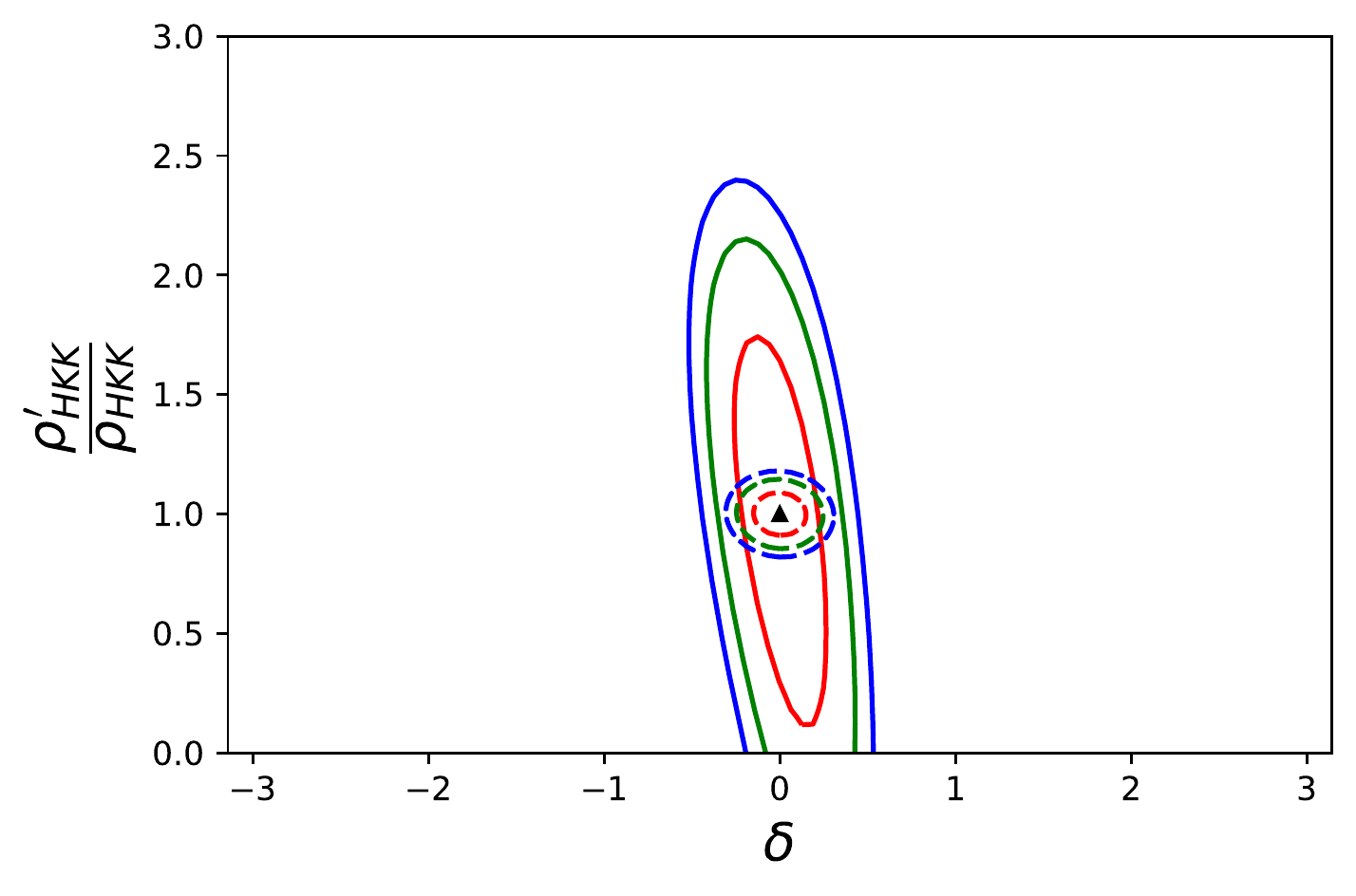}
	}\\
	\subfloat[$\delta=\pi/2$]{
    	\includegraphics[width=0.4\textwidth]{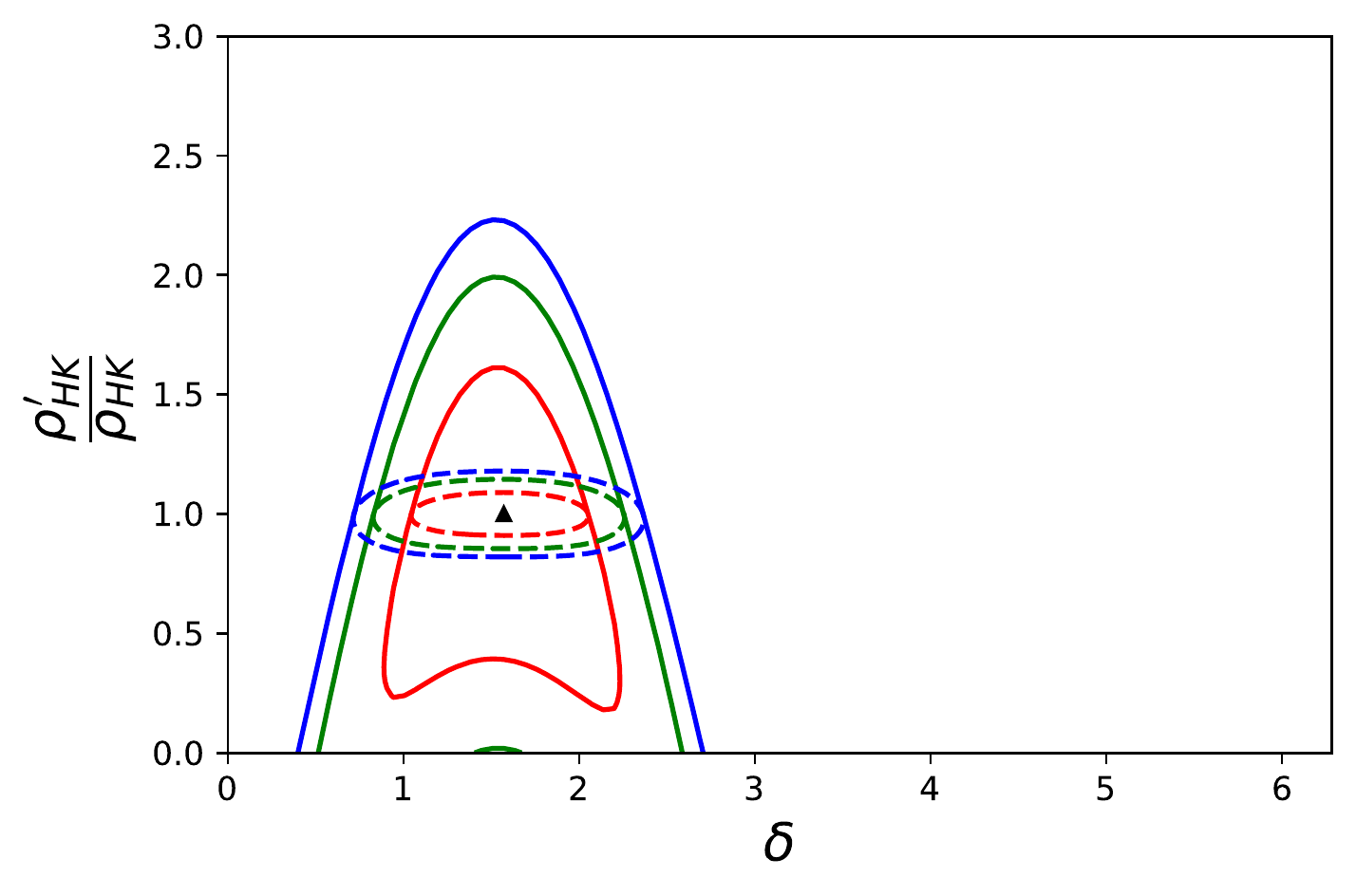}\qquad
	    \includegraphics[width=0.4\textwidth]{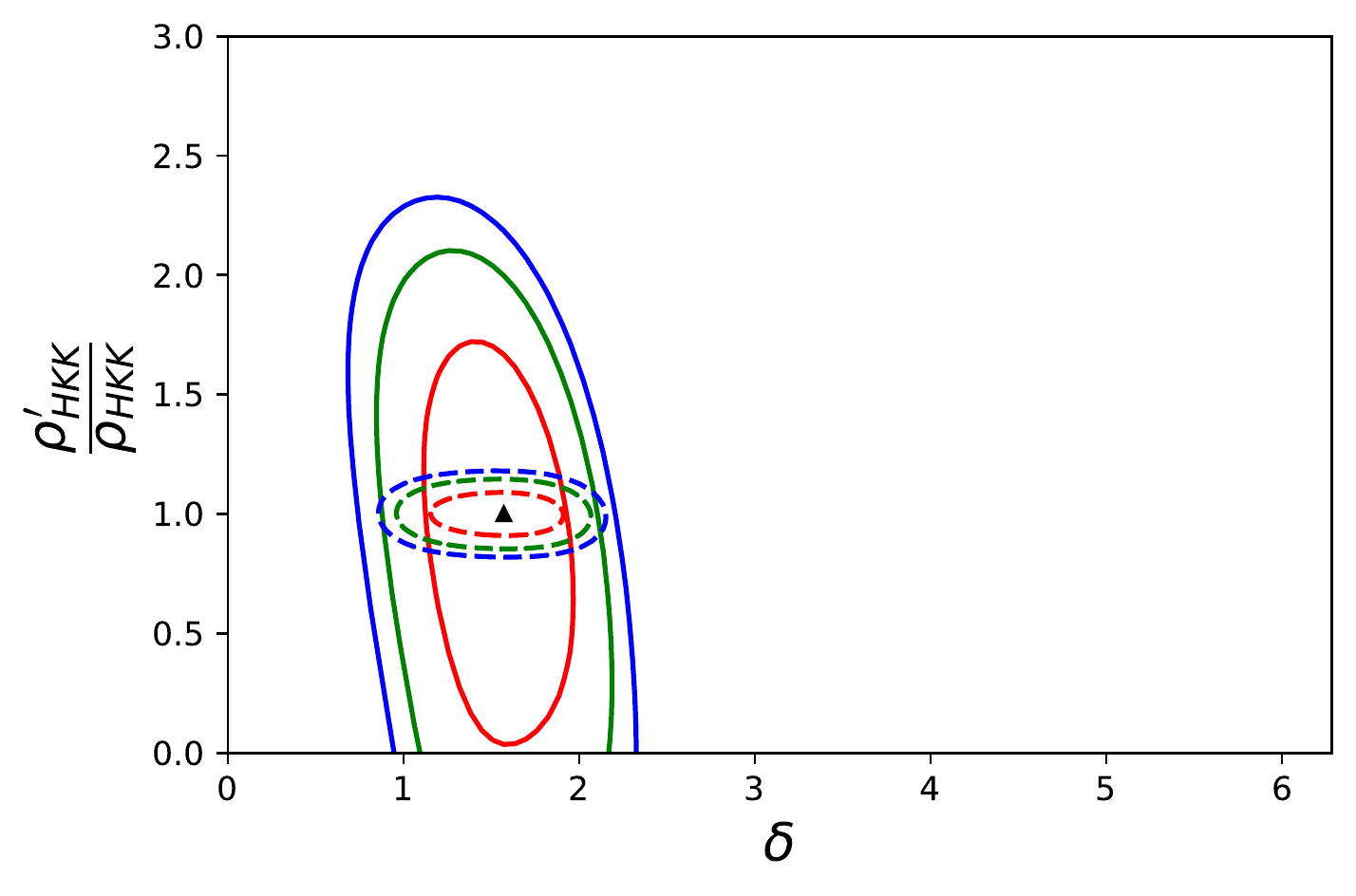}
	}\\
	\subfloat[$\delta=\pi$]{
	    \includegraphics[width=0.4\textwidth]{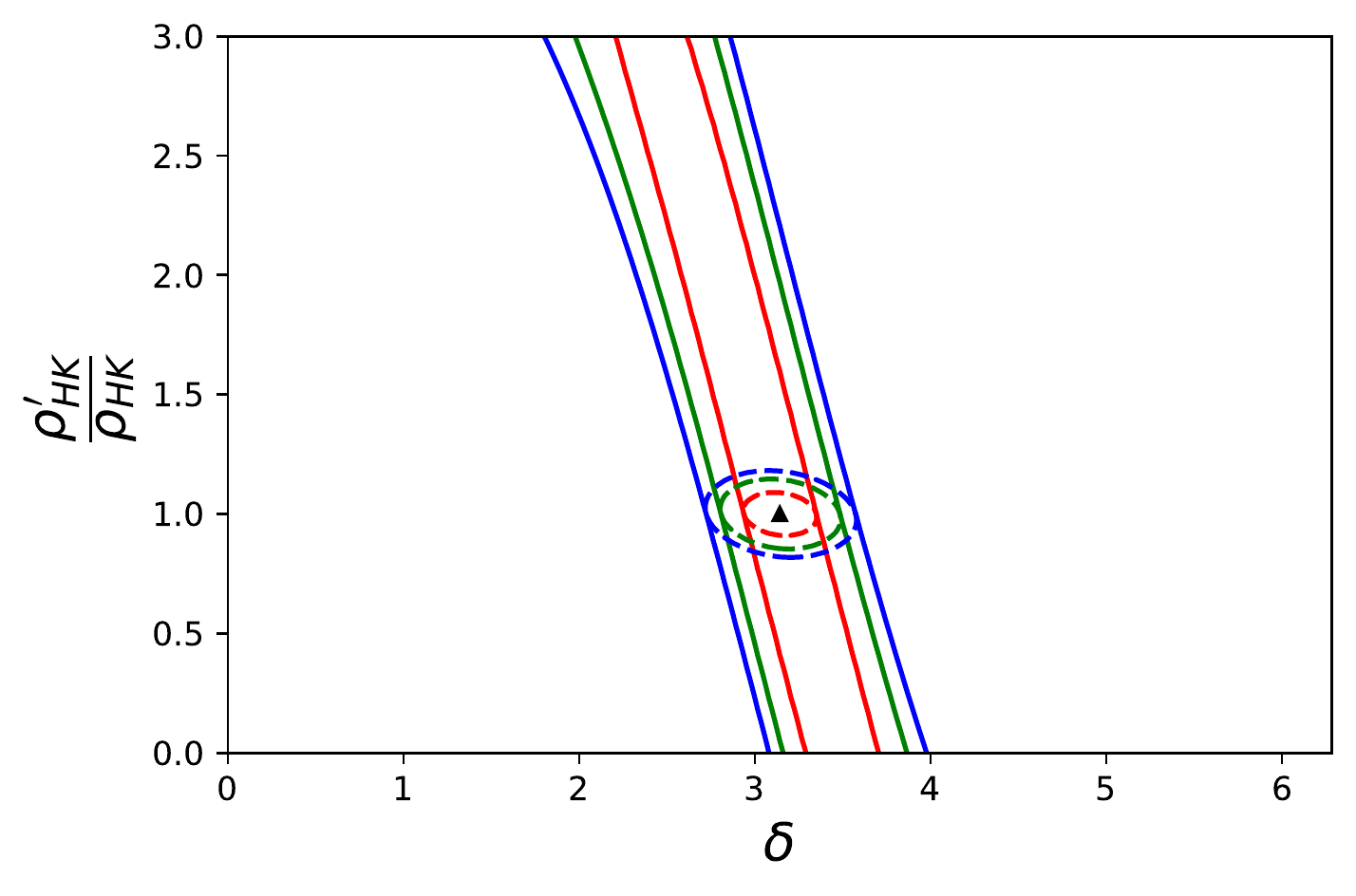}\qquad
	    \includegraphics[width=0.4\textwidth]{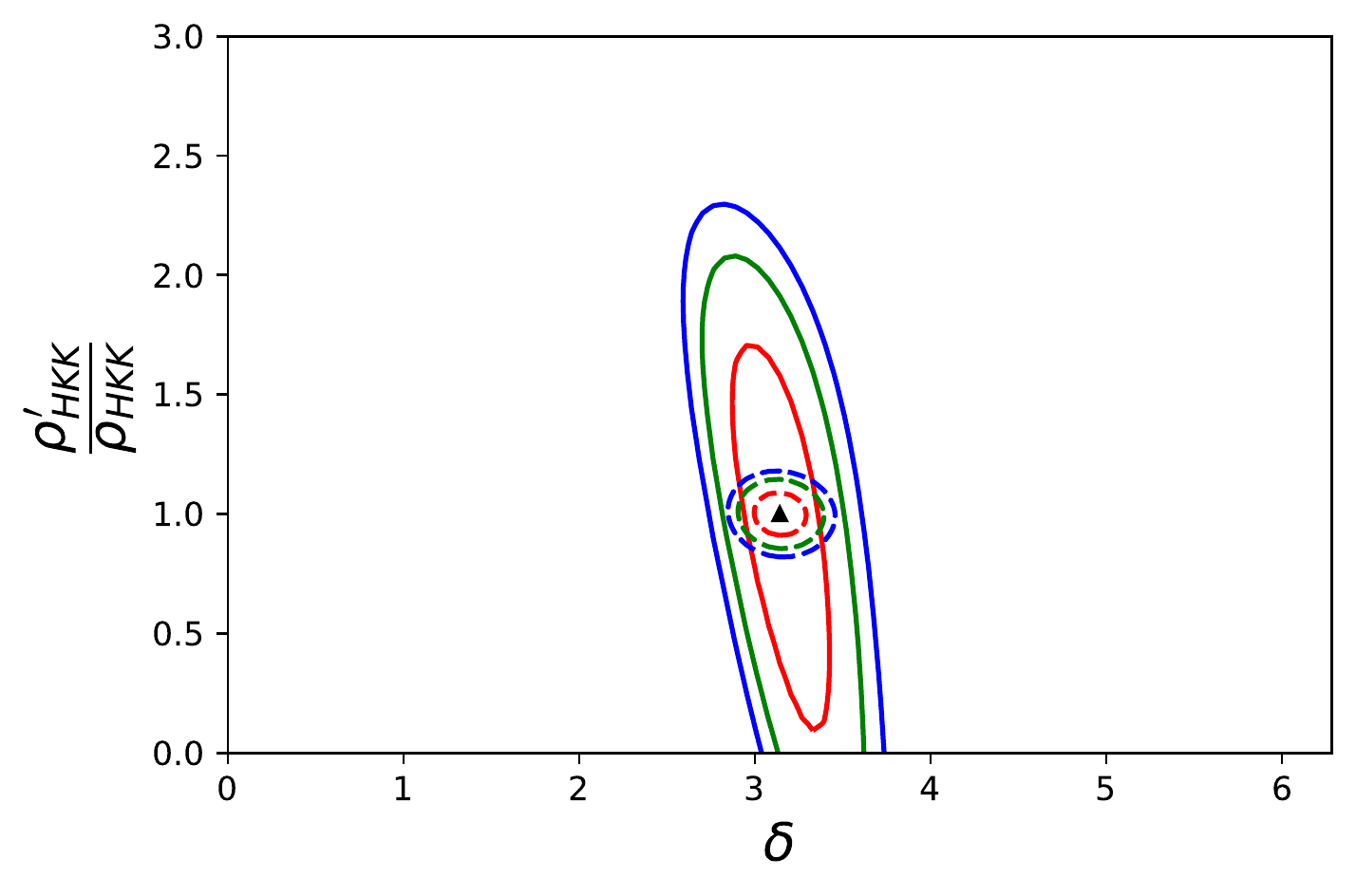}
	}\\
	\subfloat[$\delta=3\pi/2$]{
	    \includegraphics[width=0.4\textwidth]{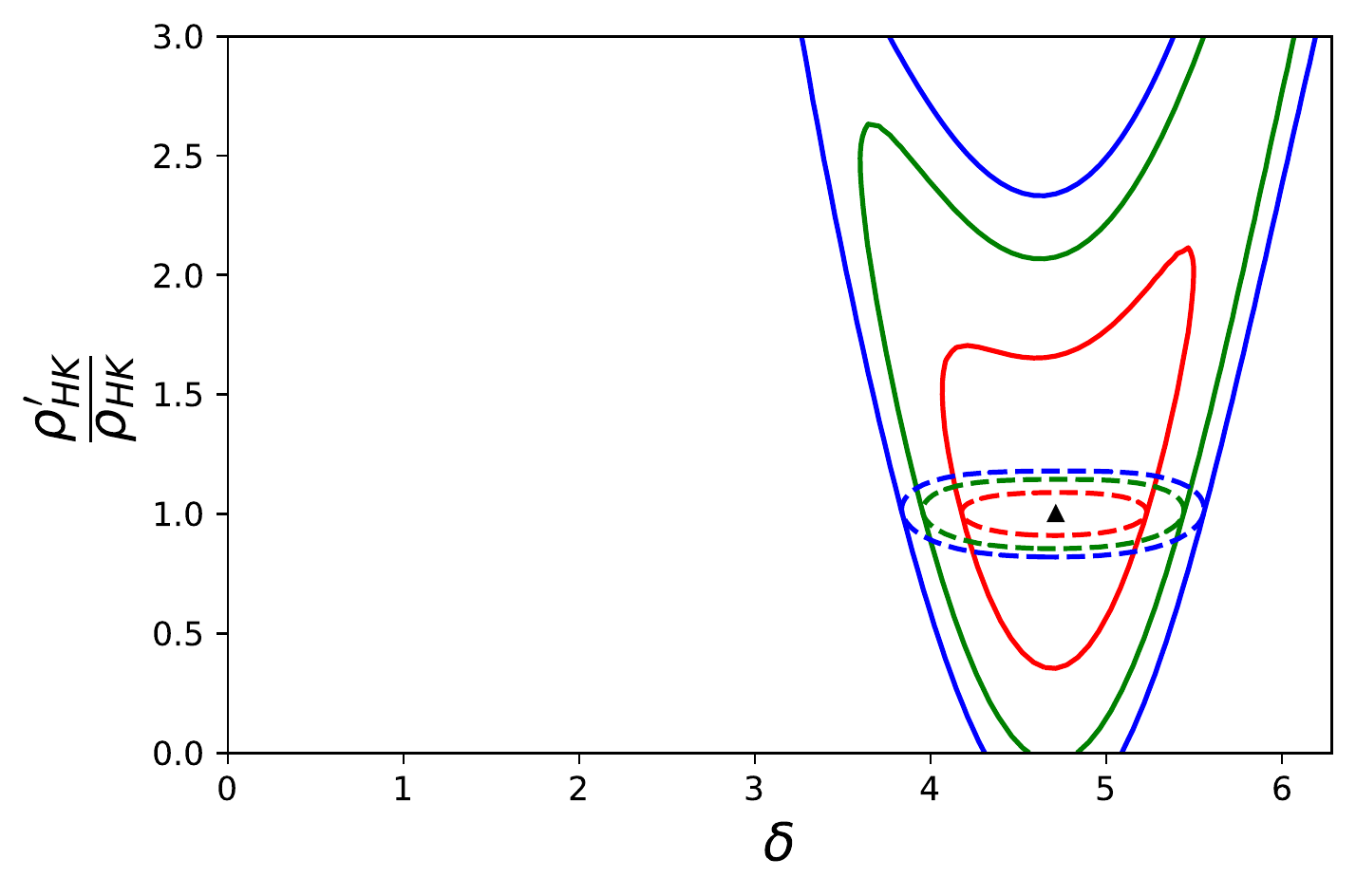}\qquad
	    \includegraphics[width=0.4\textwidth]{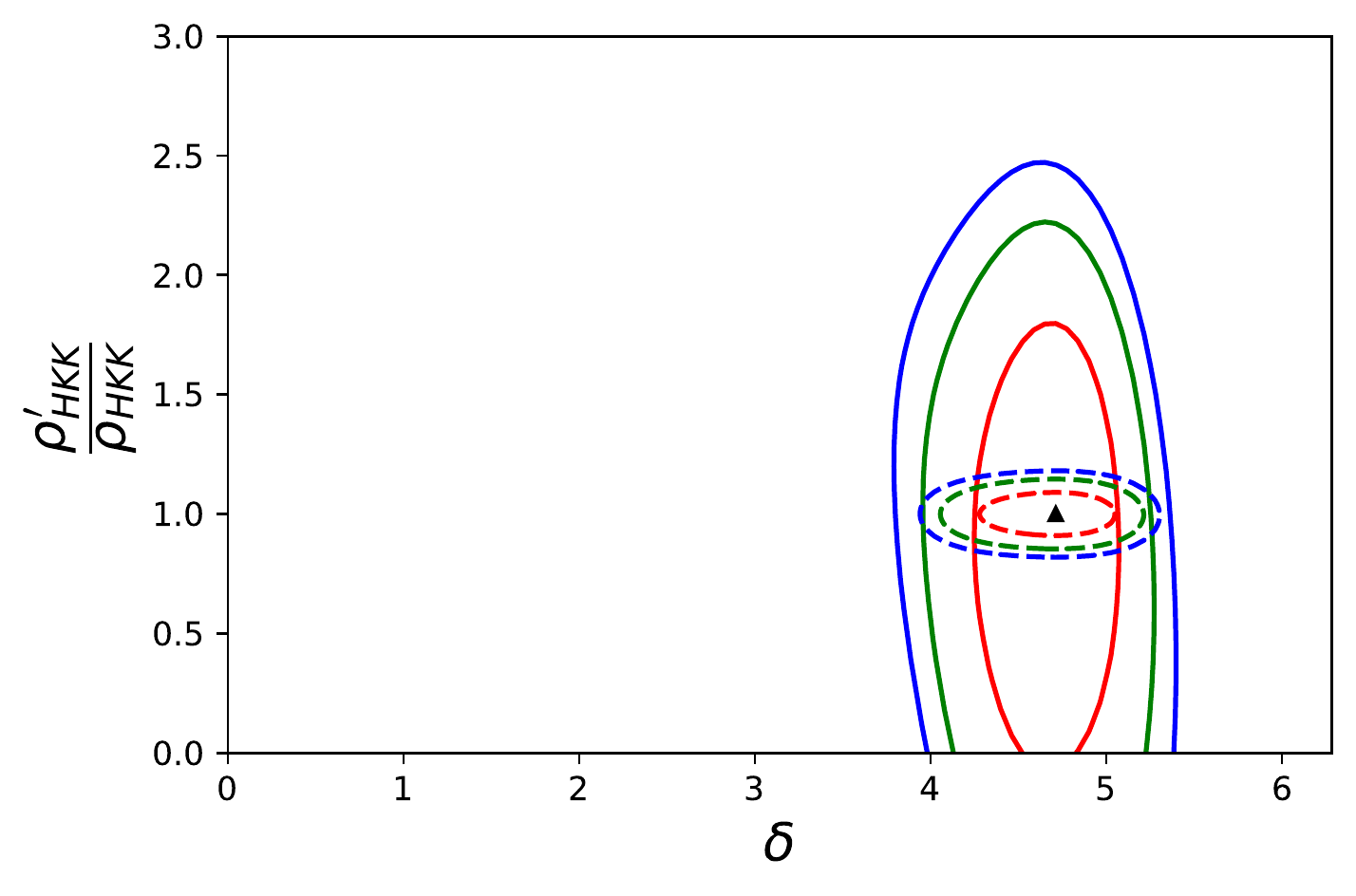}
	}
    \caption{
        \label{fig:Contour_Delta}
        Sensitivity of T2HK (left) and T2HKK (right) to the matter density scale $\rho'/\rho$ with $\delta_{CP}$, for true values of $\delta=0$, $\pi/2$, $\pi$ and $3\pi/2$ (from top to bottom). Contours correspond to $68.3\%$ (red), $95\%$ (green) and $99\%$ (blue) confidence regions with no prior constraint (solid) and a $6\%$ Gaussian prior constraint (dashed) on $\rho'/\rho$.}
\end{figure*}

\begin{figure*}[!htbp]
	\subfloat[$\theta_{23}$]{
        \includegraphics[width=0.4\textwidth]{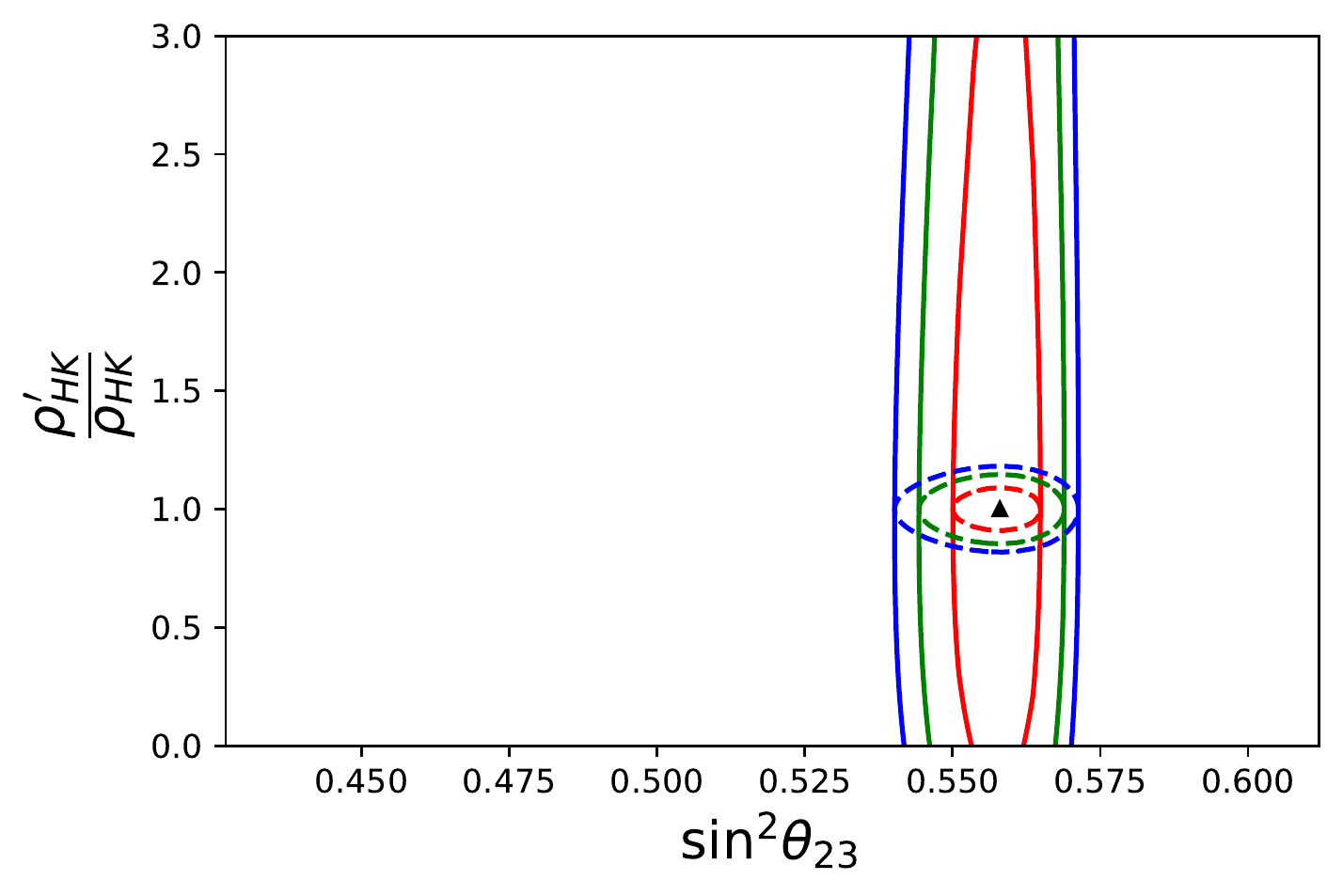}\qquad
    	\includegraphics[width=0.4\textwidth]{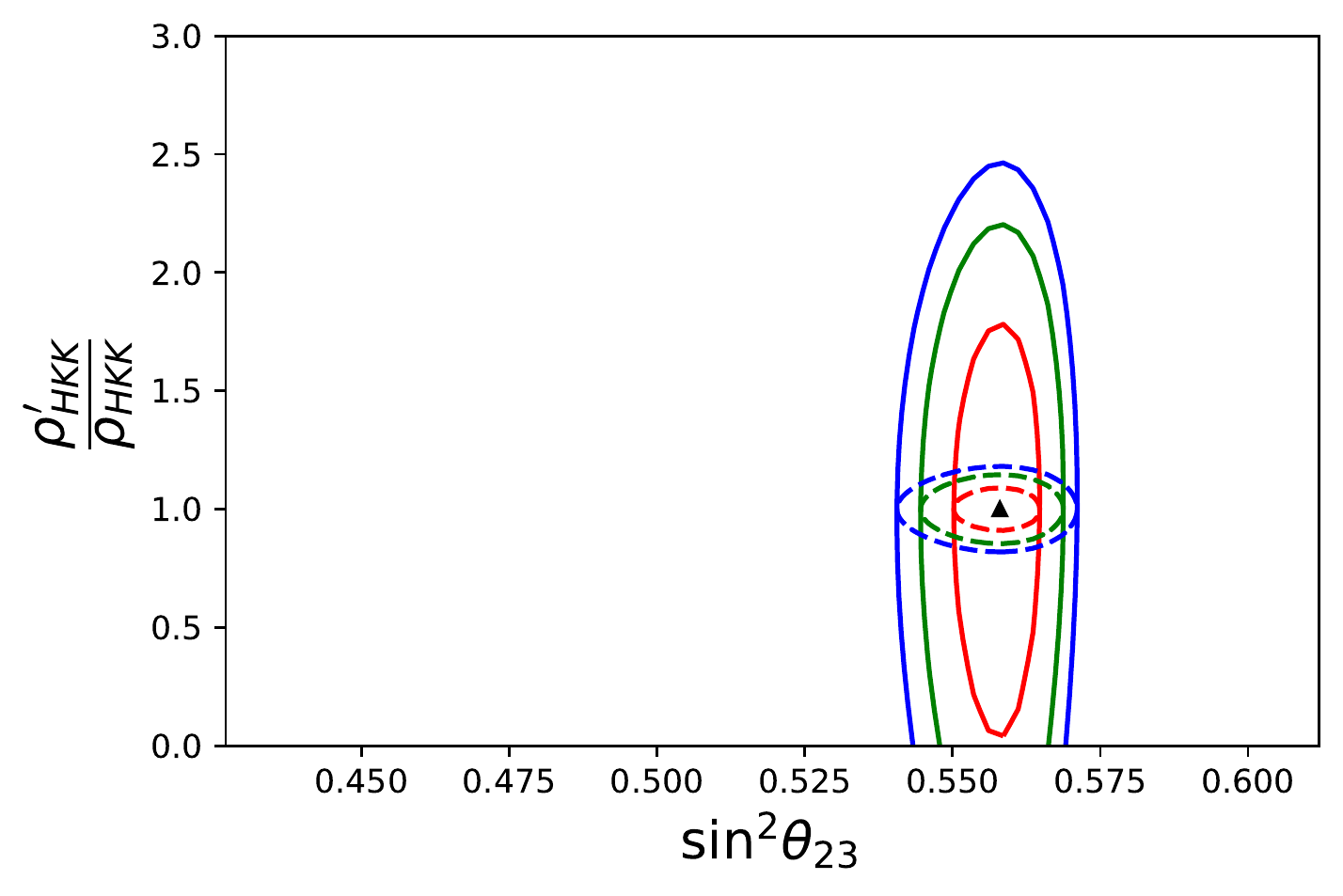}
	}\\
	\subfloat[$\theta_{13}$]{
        \includegraphics[width=0.4\textwidth]{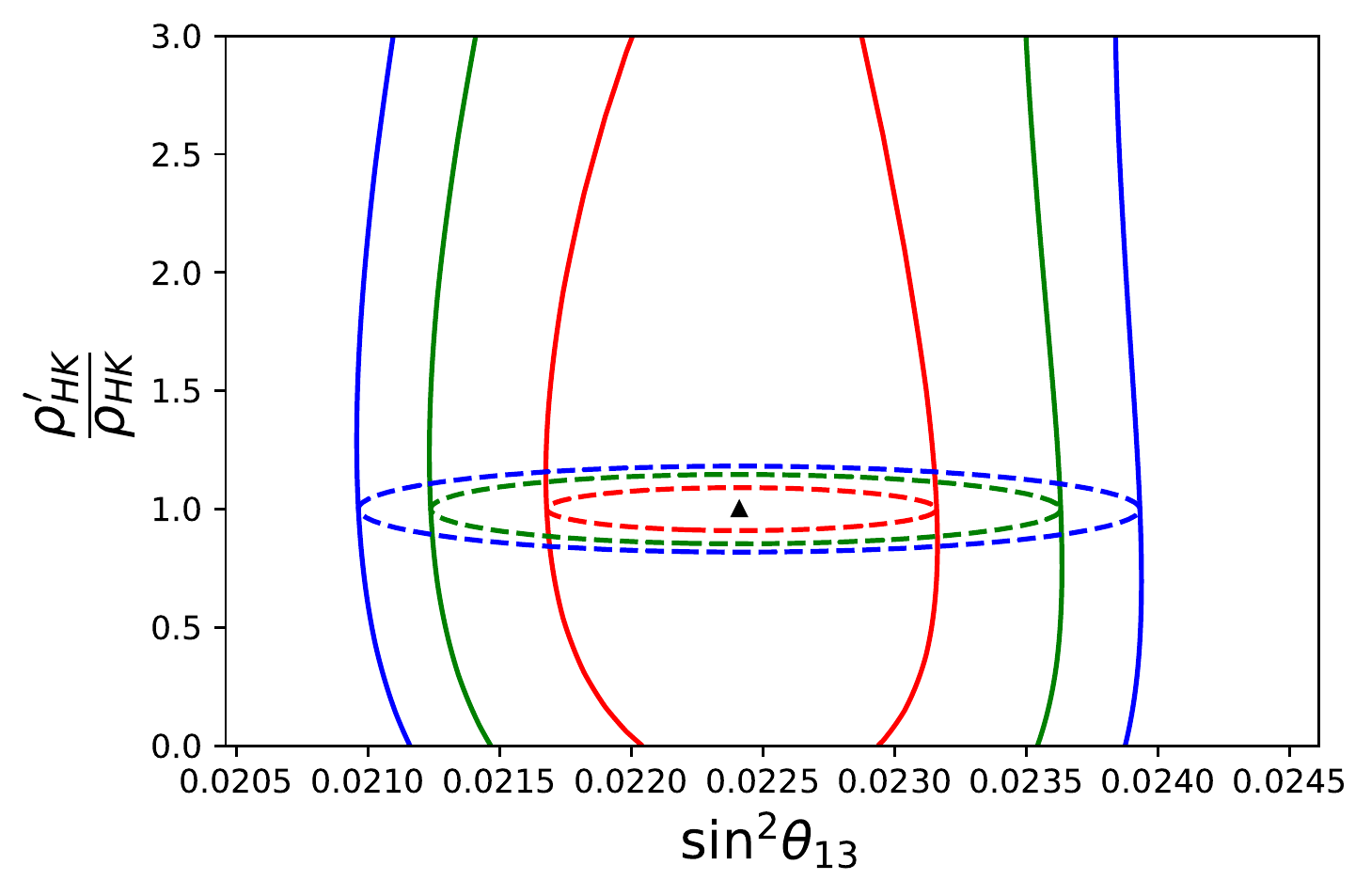}\qquad
    	\includegraphics[width=0.4\textwidth]{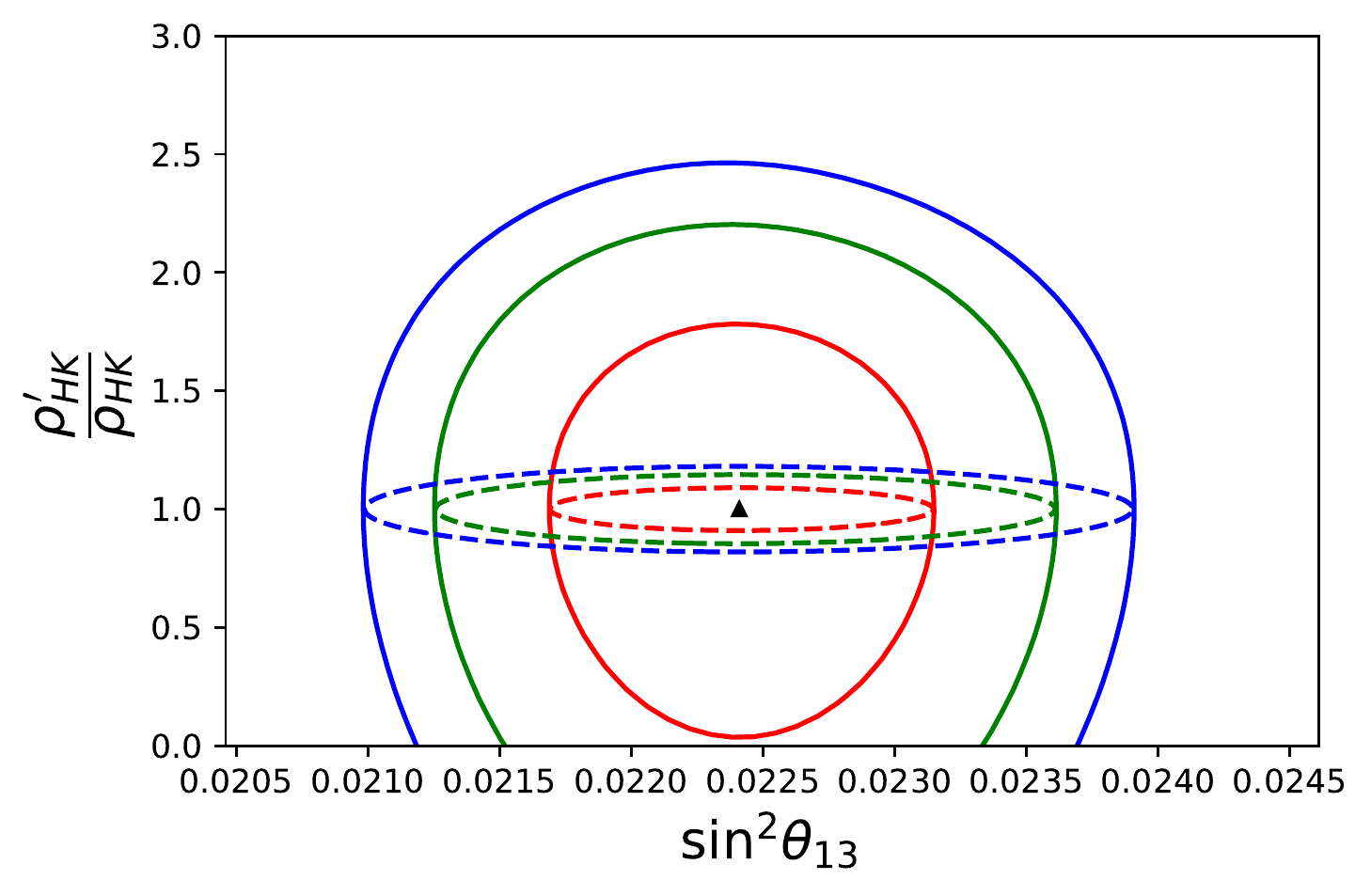}
	}\\
	\subfloat[$\Delta m^2_{31}$]{
    	\includegraphics[width=0.4\textwidth]{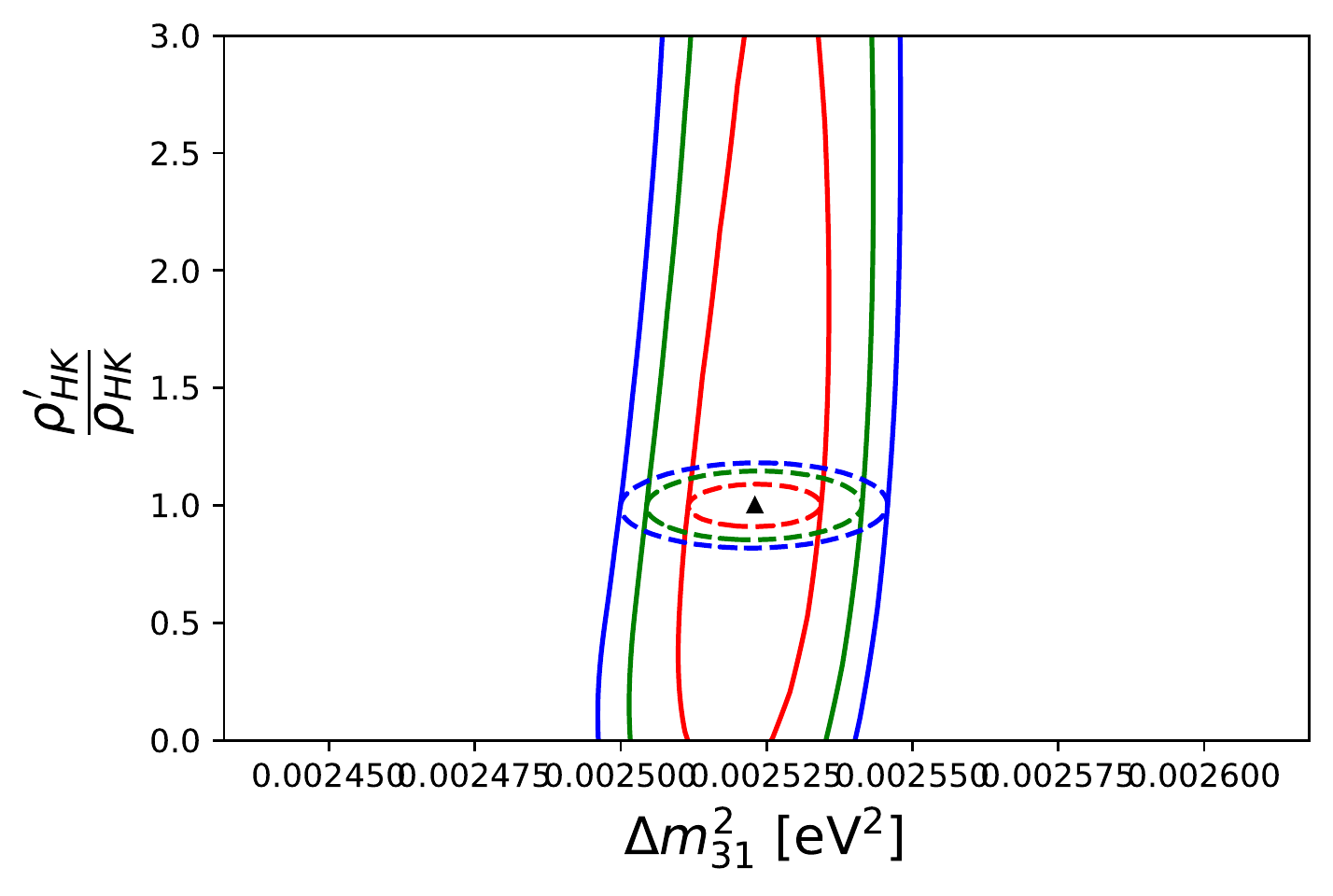}\qquad
    	\includegraphics[width=0.4\textwidth]{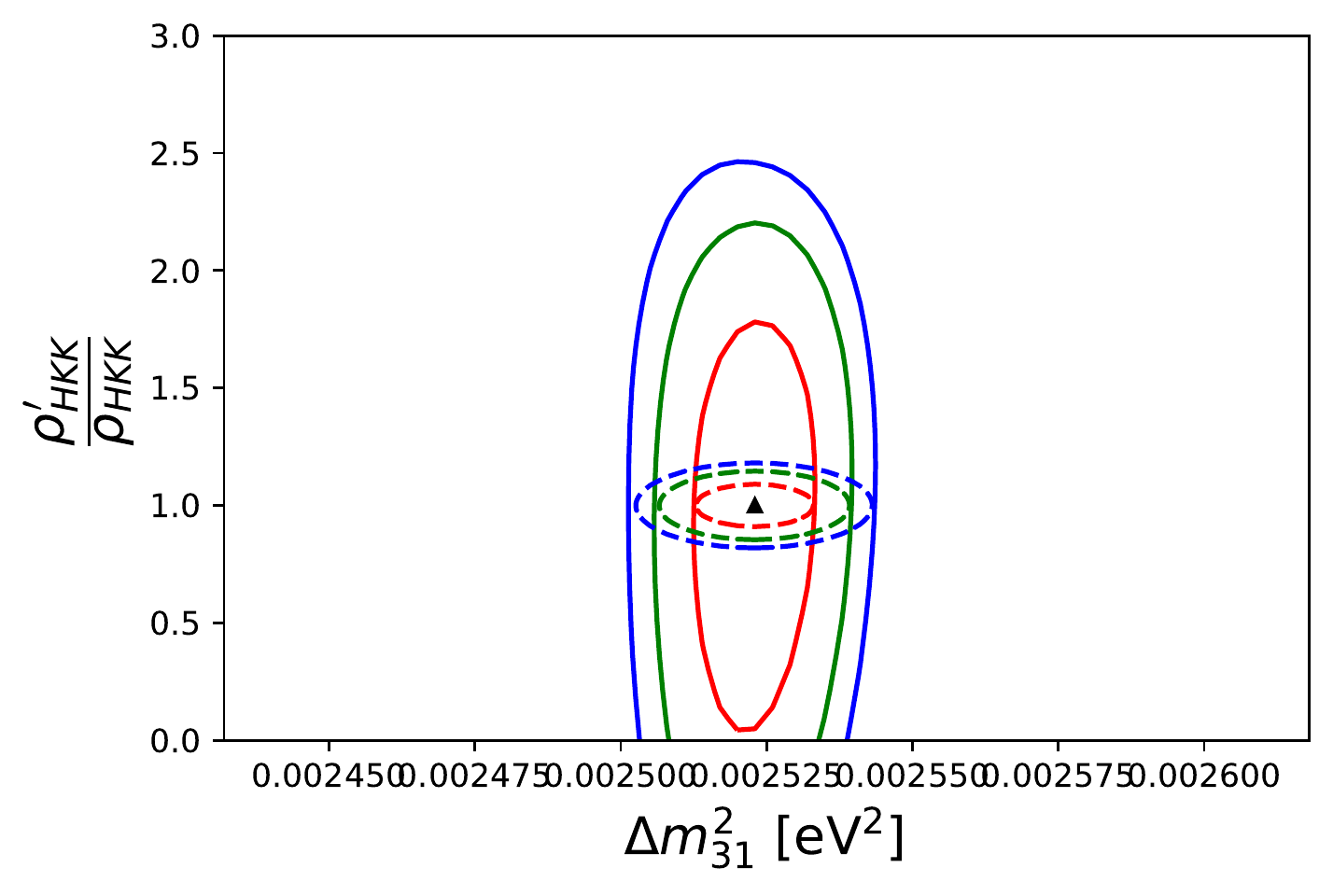}
	}
    \caption{
        \label{fig:Contour_Params}
        Sensitivity of T2HK (left) and T2HKK (right) to the matter density scale $\rho'/\rho$ with $\theta_{23}$ (top), $\theta_{13}$ (middle) and $\Delta m^2_{31}$ (bottom). Contours correspond to $68.3\%$ (red), $95\%$ (green) and $99\%$ (blue) confidence regions with no prior constraint (solid) and a $6\%$ Gaussian prior constraint (dashed) on $\rho'/\rho$.
        }
\end{figure*}

\section{Sensitivity to matter density and oscillation parameters}
\label{ap:MatterSensitivity}

Using the same GLoBES setup as in Section~\ref{Sens}, this study was able to produce estimates of the sensitivity of T2HK (assuming no Korean detector is in operation) and T2HKK (assuming both the Japanese and Korean detectors are in operation) to the scale of the matter density. For each experiment, the sensitivity to a scaling parameter $\rho'/\rho$, which scales up or down the matter density $\rho$ at every point along the baseline, was determined in combination with the sensitivity to the oscillation parameters. Unlike previous sections, the left and right hand panels correspond here to T2HK and T2HKK, respectively.

Fig.~\ref{fig:Contour_Delta} shows contour plots depicting the sensitivity to the mass density scales $\rho'/\rho$ for multiple potential true values of $\delta_{CP}$ ($\delta=0$, $\pi/2$, $\pi$ and $3\pi/2$), since its value is the least well-determined of the oscillation parameters.

In addition to the $\delta-\rho'/\rho$ plane, Fig.~\ref{fig:Contour_Params} includes contour plots for $\rho'/\rho$ vs. three other oscillation parameters: $\theta_{23}$, $\theta_{13}$ and $\Delta m^2_{31}$. As expected, the reactor angle $\theta_{13}$ proves to have the widest contours of all parameters studied, as the experiments will be least sensitive to this mixing angle.

\begin{figure*}[!htbp]
    \subfloat[$\delta=0$]{%
    	\includegraphics[width=0.4\textwidth]{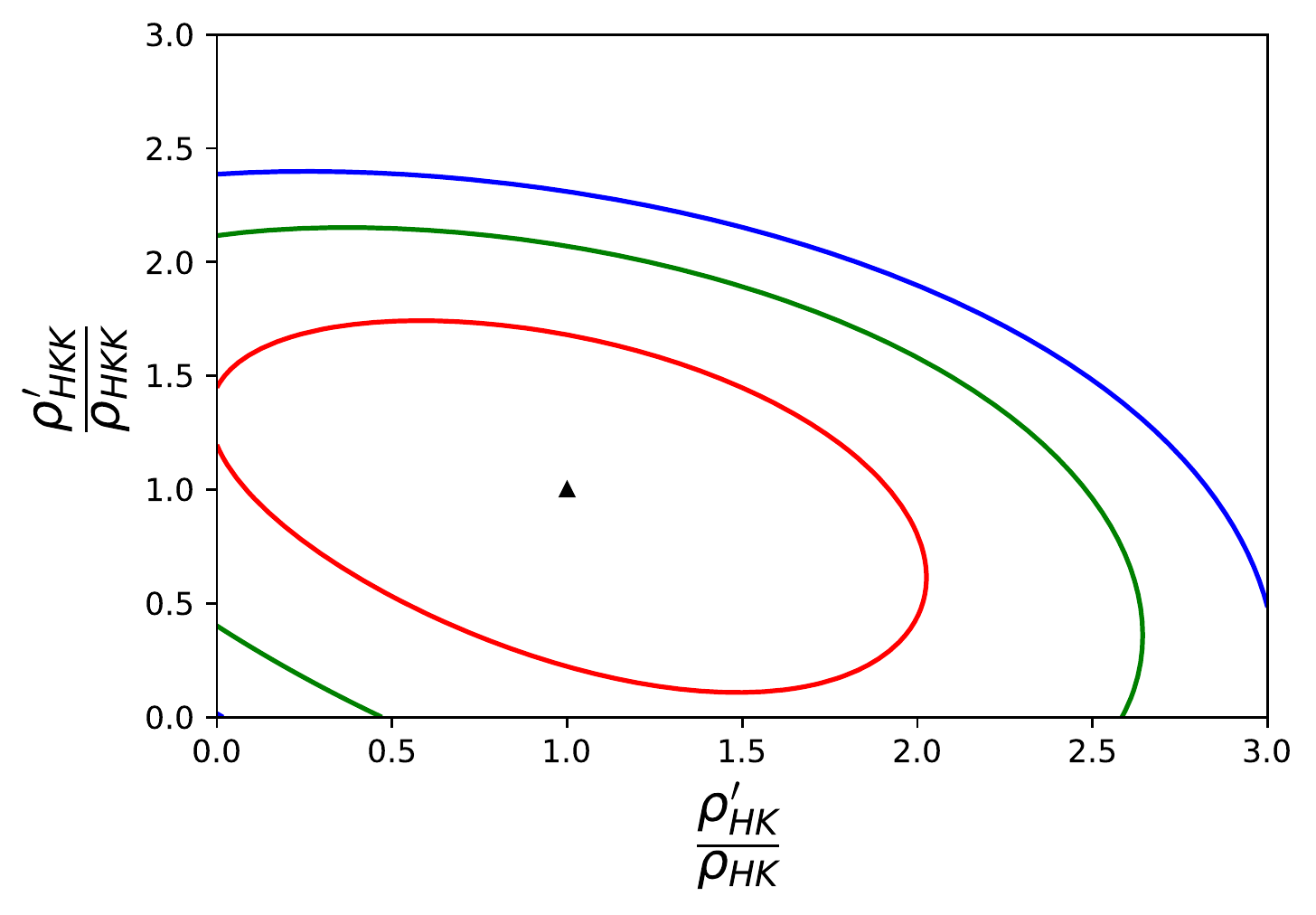}
    }\qquad
    \subfloat[$\delta=\pi/2$]{%
    	\includegraphics[width=0.4\textwidth]{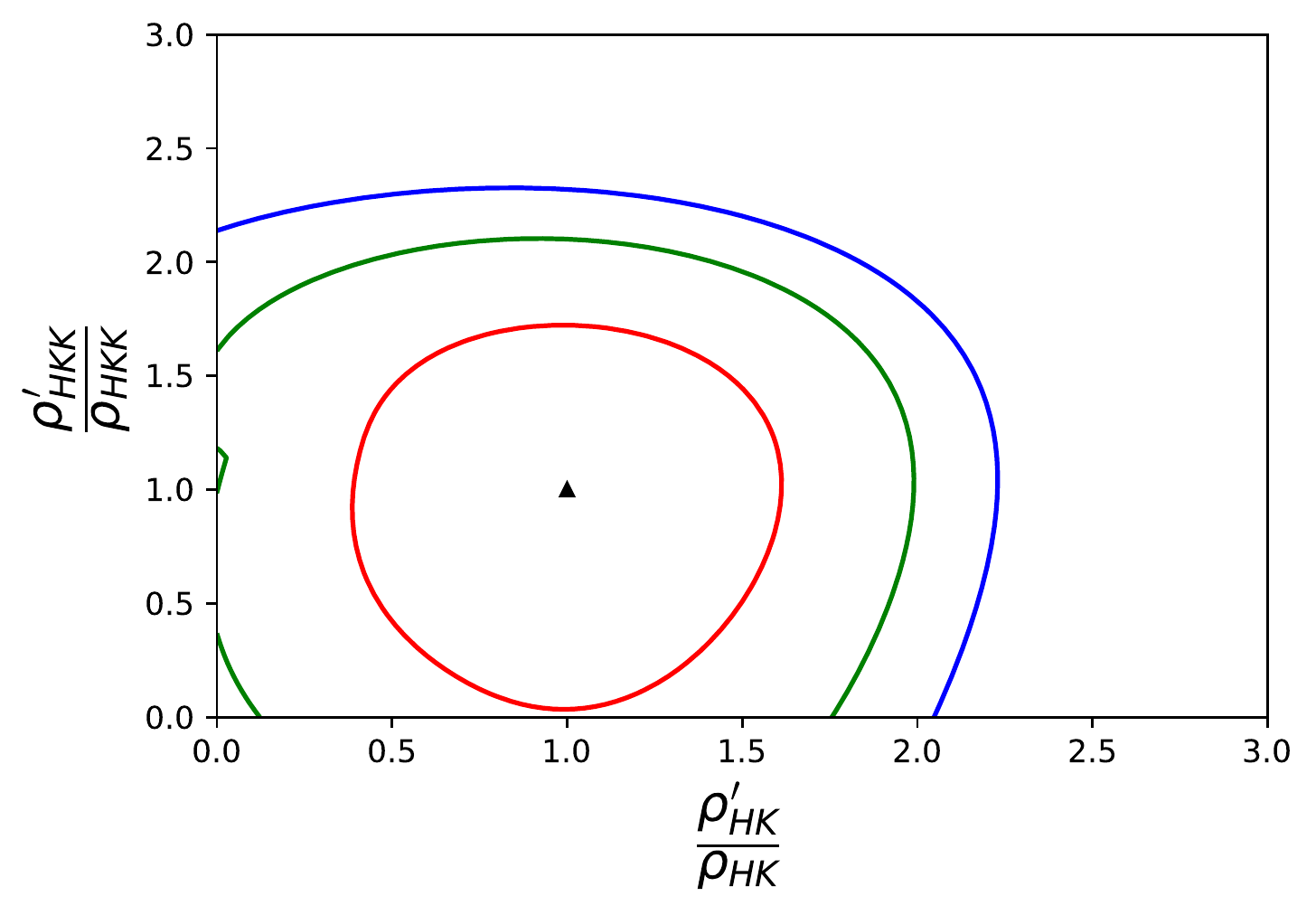}
    }\\
    \subfloat[$\delta=\pi$]{%
    	\includegraphics[width=0.4\textwidth]{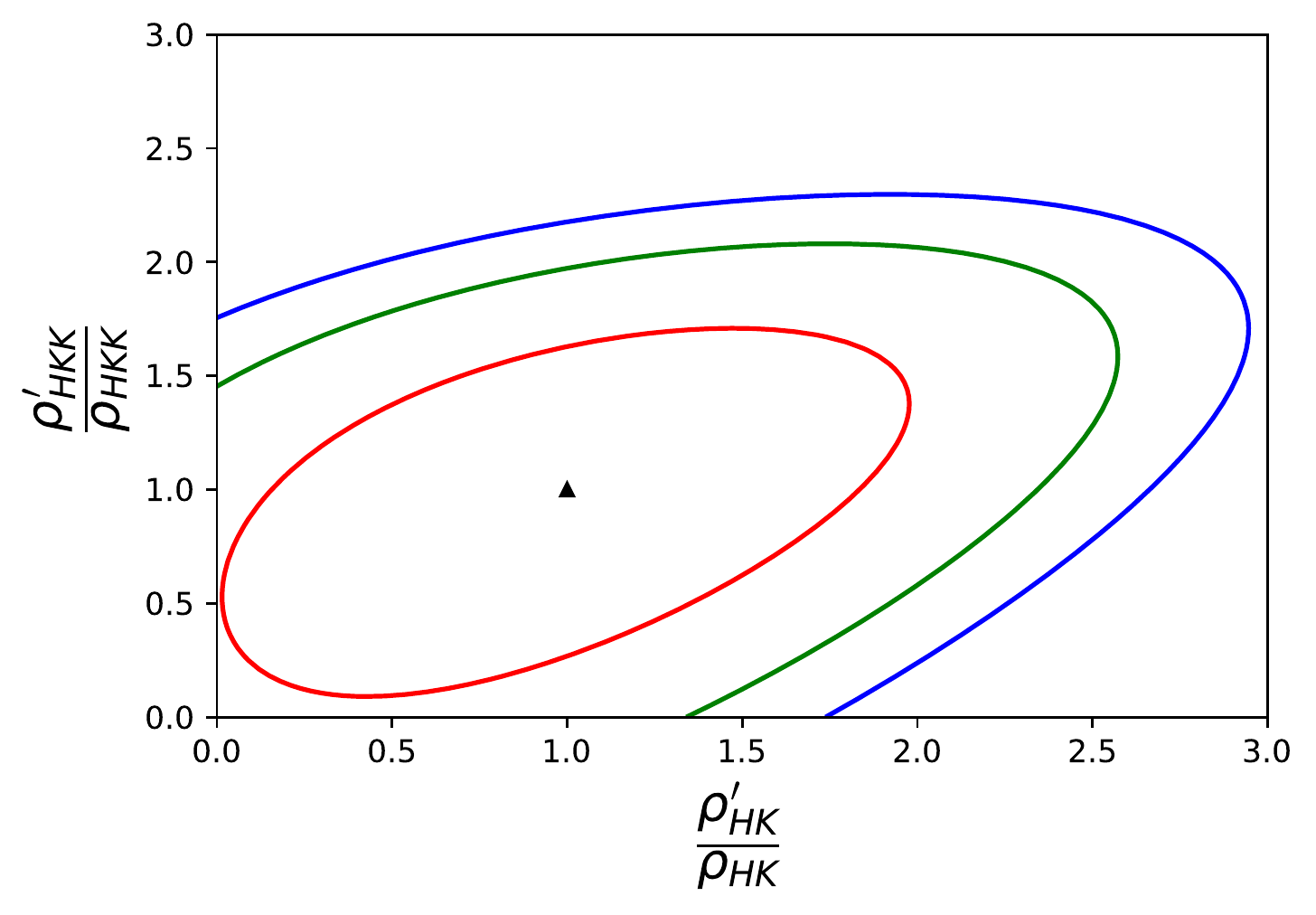}
    }\qquad
    \subfloat[$\delta=3\pi/2$]{%
        \includegraphics[width=0.4\textwidth]{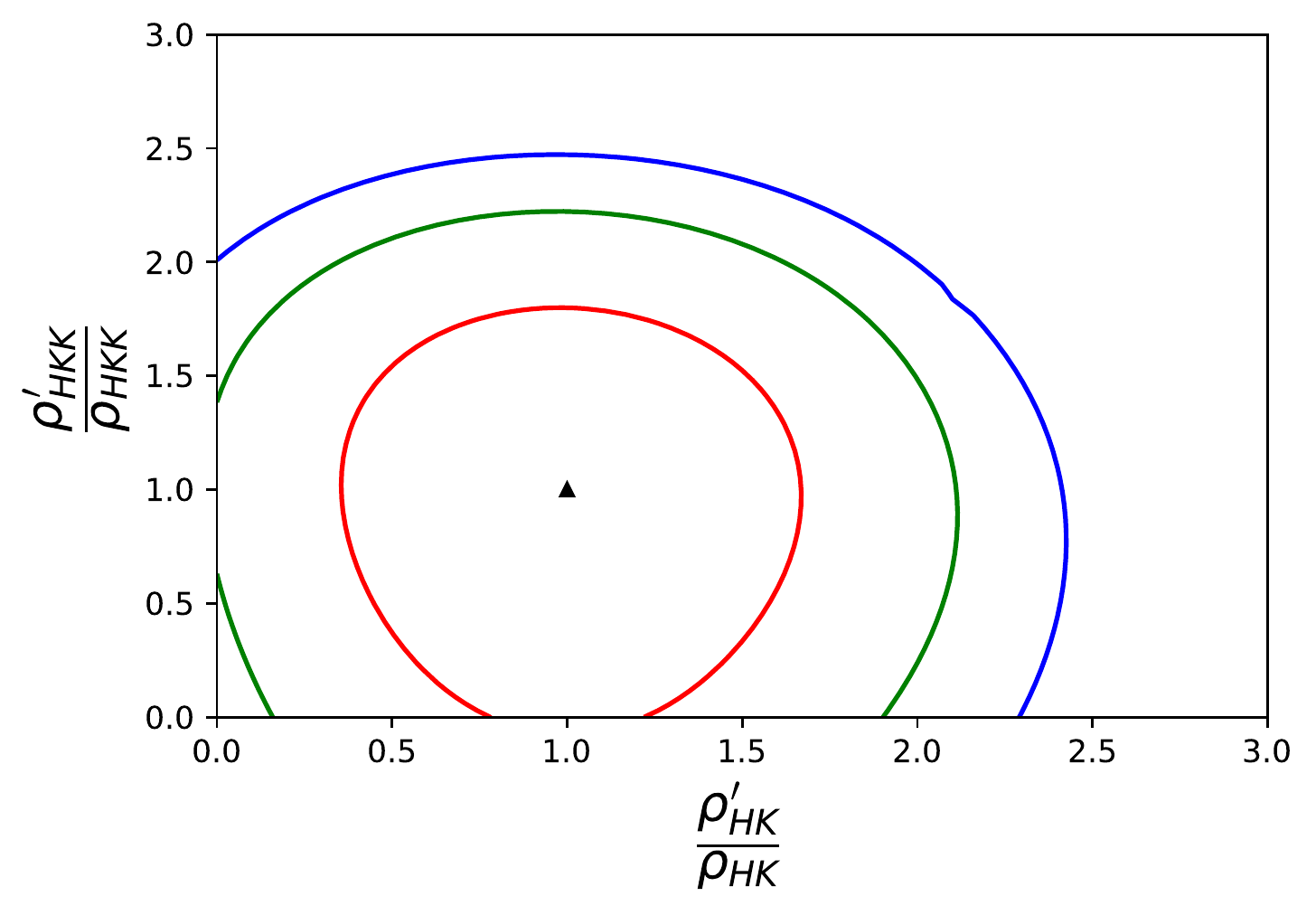}
    }
    \caption{
	    \label{fig:Contour_Rho}
	    Sensitivity to the matter density scale for T2HK ($\rho'_\textrm{HK}/\rho_\textrm{HK}$) vs. T2HKK ($\rho'_\textrm{HKK}/\rho_\textrm{HKK}$). The sensitivity is shown for different true values of $\delta$: 0, $\pi/2$, $\pi$ and $3\pi/2$. Contours correspond to $68.3\%$ (red), $95\%$ (green) and $99\%$ (blue) confidence regions (no prior constraint).
	}
\end{figure*}

The oscillation parameters to which T2HK and T2HKK are less sensitive -- namely $\theta_{12}$, $\theta_{13}$ and $\Delta m^2_{21}$ -- are constrained by Gaussian priors with central values and $1\sigma$ errors corresponding to the values given by the current NuFIT4.1 global fit~\cite{Esteban:2018azc}. In each case, the sensitivity was determined for $\rho'/\rho$ both with no constraint other than the measurements of T2HK(K) and also with a $6\%$ Gaussian prior constraint corresponding to the uncertainty given in Ref.~\cite{Hagiwara:2011kw}.

It is interesting to observe how the addition of a $6\%$ Gaussian prior constraint on $\rho$ has a very minimal effect on the measurement sensitivity of the different oscillation parameters; only the contours for the measurement of $\delta_{CP}$ at T2HKK become slightly narrower in comparison with those obtained when $\rho$ is left as a free parameter.

This suggests that the uncertainty on a measurement of $\delta$ at T2HK or T2HKK is in danger of being underestimated if the uncertainty on our prior knowledge of the matter density is also underestimated. For the other oscillation parameters, however, either the effect of the matter density is small enough or the ability of T2HK and T2HKK to directly constrain the matter density is great enough that measurements of those parameters are not affected.

Finally, the combined sensitivity to both the scale of the matter density for T2HK ($\rho'_\textrm{HK}/\rho_\textrm{HK}$) and the scale of the matter density for T2HKK ($\rho'_\textrm{HKK}/\rho_\textrm{HKK}$), assuming that both the Japanese and Korean detectors are operational, is shown in Fig.~\ref{fig:Contour_Rho}.

\section{Discussion and Conclusions}
\label{Conc}

\begin{figure*}[!htbp]
	\includegraphics[width=0.9\textwidth]{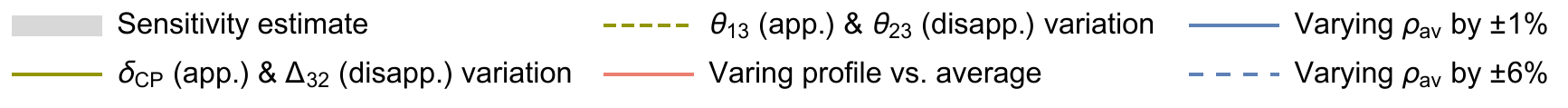} \\
	\includegraphics[width=0.485\textwidth]{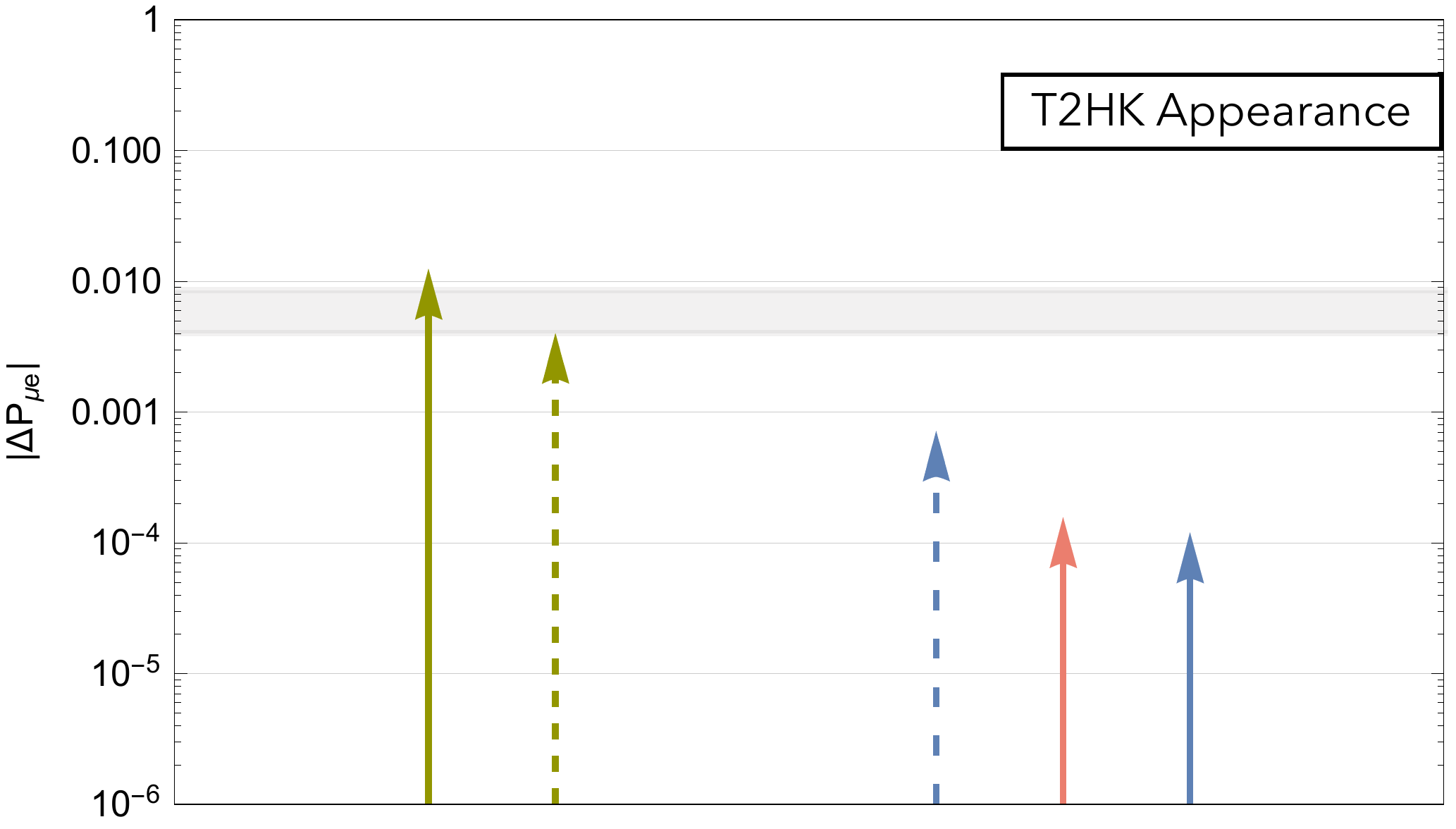}
	\includegraphics[width=0.485\textwidth]{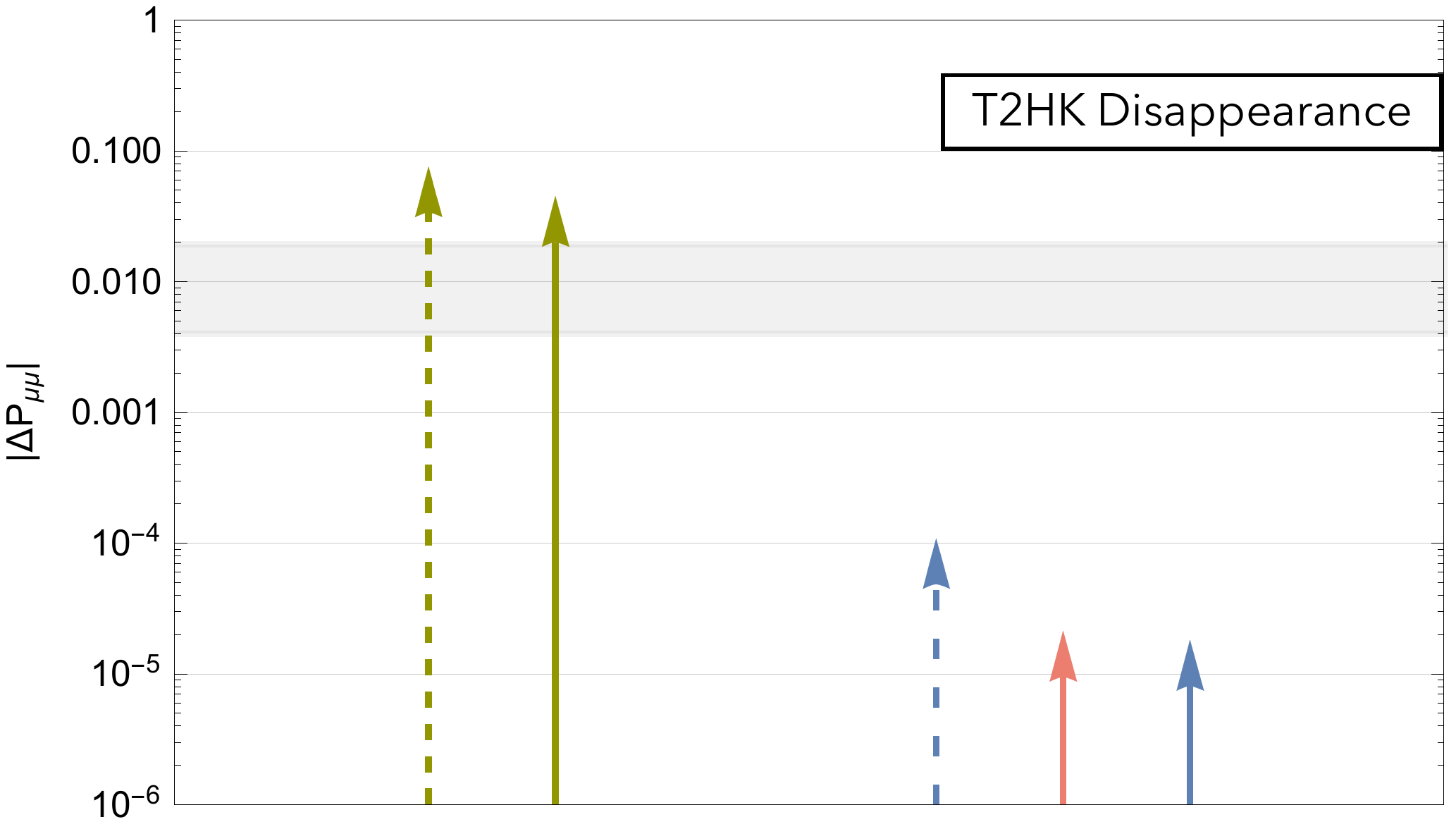}
    \caption{
        \label{fig:summaryHK}
        Summary of the main results obtained for T2HK. Results are grouped into sensitivity estimate, oscillation parameter variation and matter density profile studies. Left and right panels show appearance and disappearance channels, respectively.
    }
\end{figure*}

Fig.~\ref{fig:summaryHK} shows a summary plot for T2HK, where the sensitivity estimates obtained in Section~\ref{Sens} are depicted as a horizontal grey band to which the order of magnitude of the pertinent results can be compared. Returning to the previous convention, left and right panels show appearance and disappearance channels, respectively.

It is clear from these plots that changes in oscillation probability arising from variations in the matter density profile will not be detectable at T2HK, as the effects from the three scenarios studied in Section~\ref{Dens} are far below our estimates for the sensitivity of the experiment.

\begin{figure*}[!htbp]
    \includegraphics[width=0.8\textwidth]{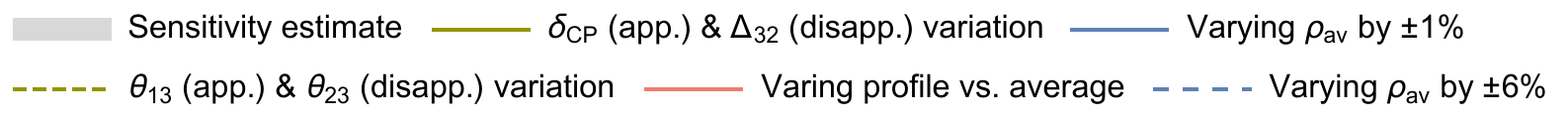}\\
    \vspace{-0.1cm}
    \includegraphics[width=0.75\textwidth]{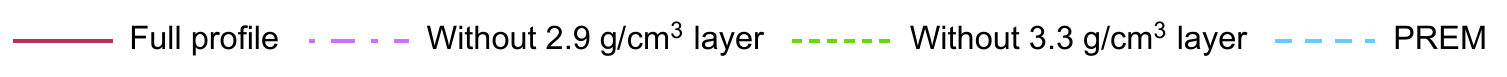}\\
	\subfloat[Mount Bisul]{
    	\includegraphics[width=0.485\textwidth]{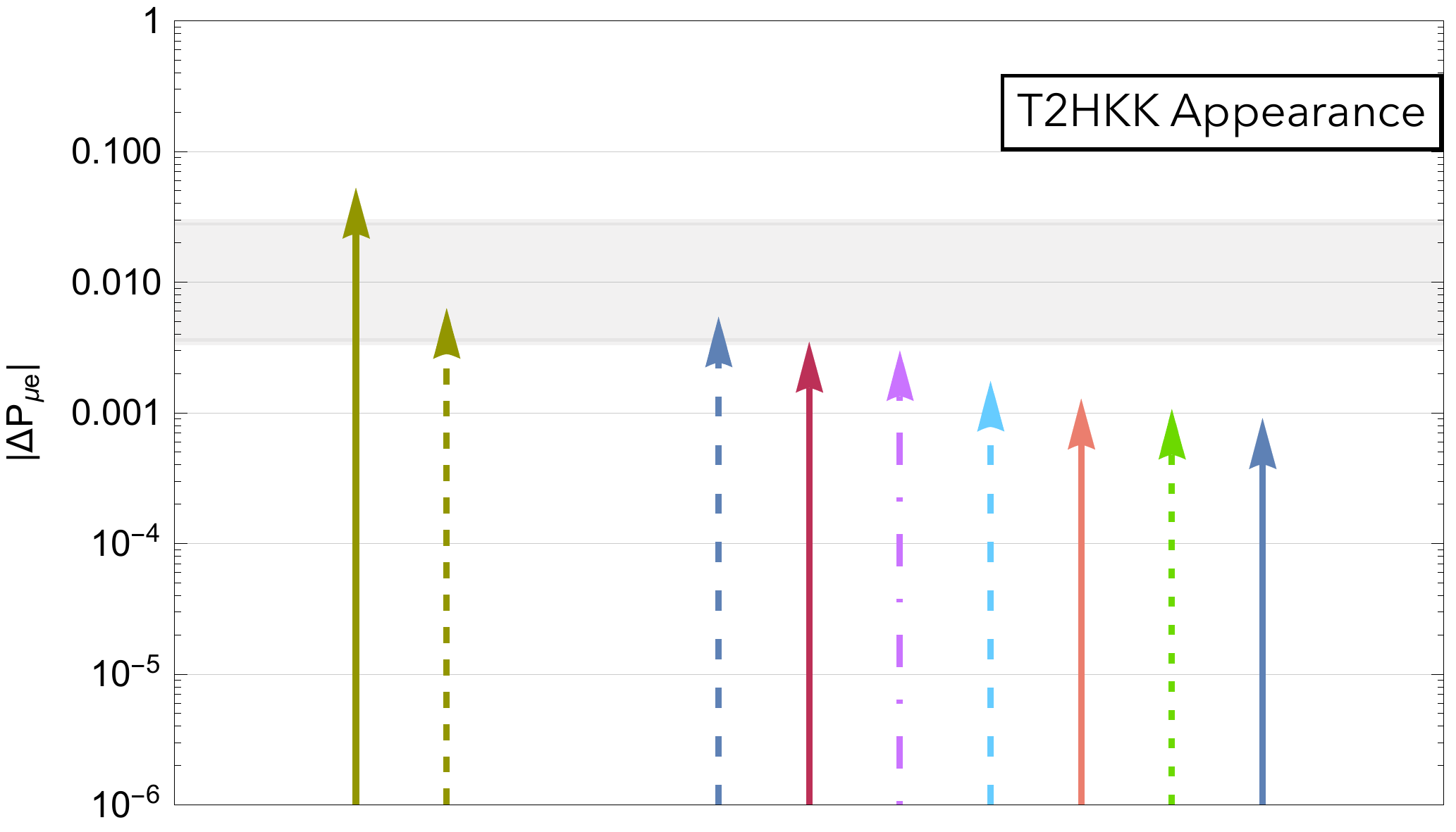}
    	\includegraphics[width=0.485\textwidth]{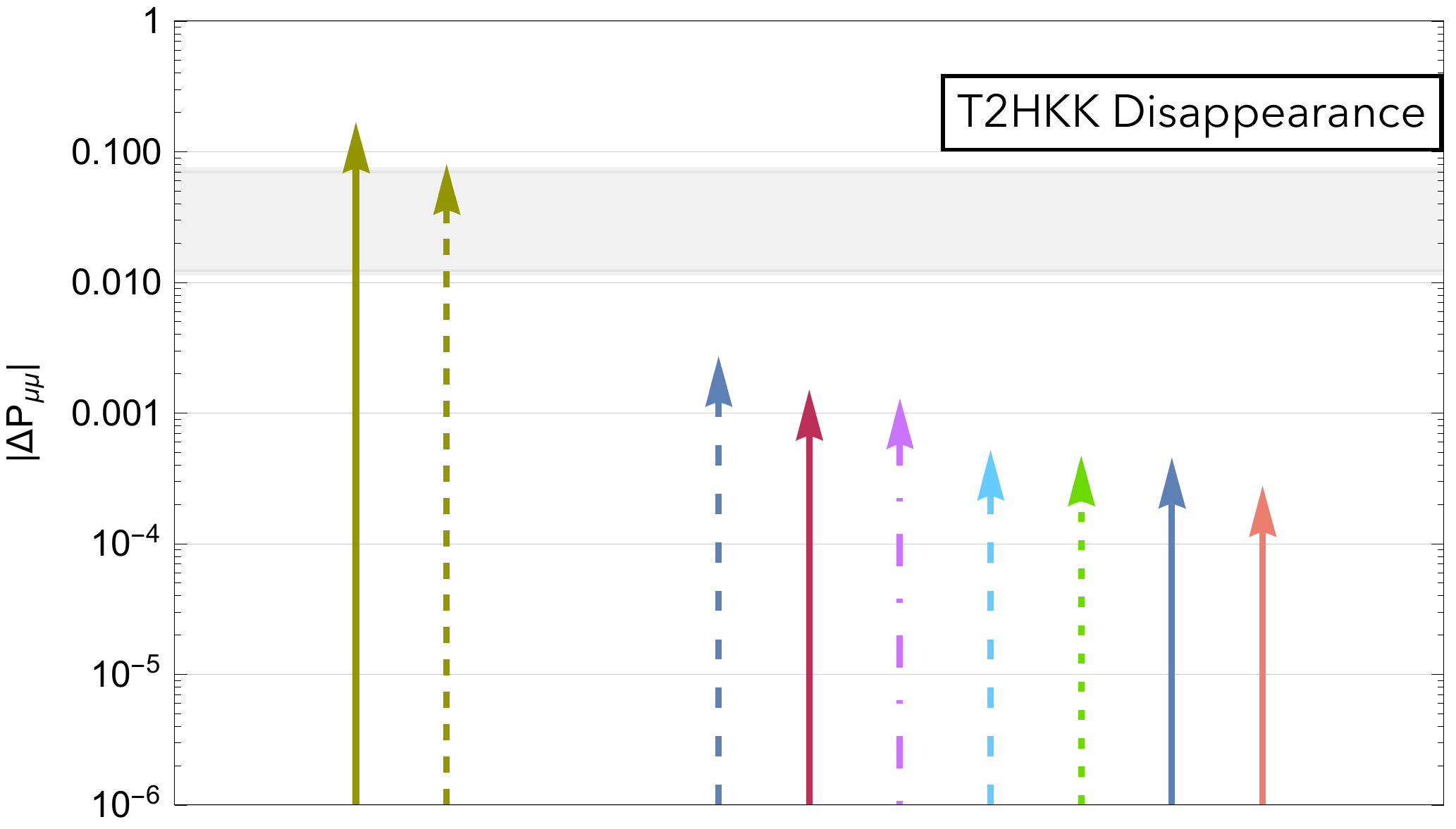}
	}\\
	\subfloat[Mount Bohyun]{
        \includegraphics[width=0.485\textwidth]{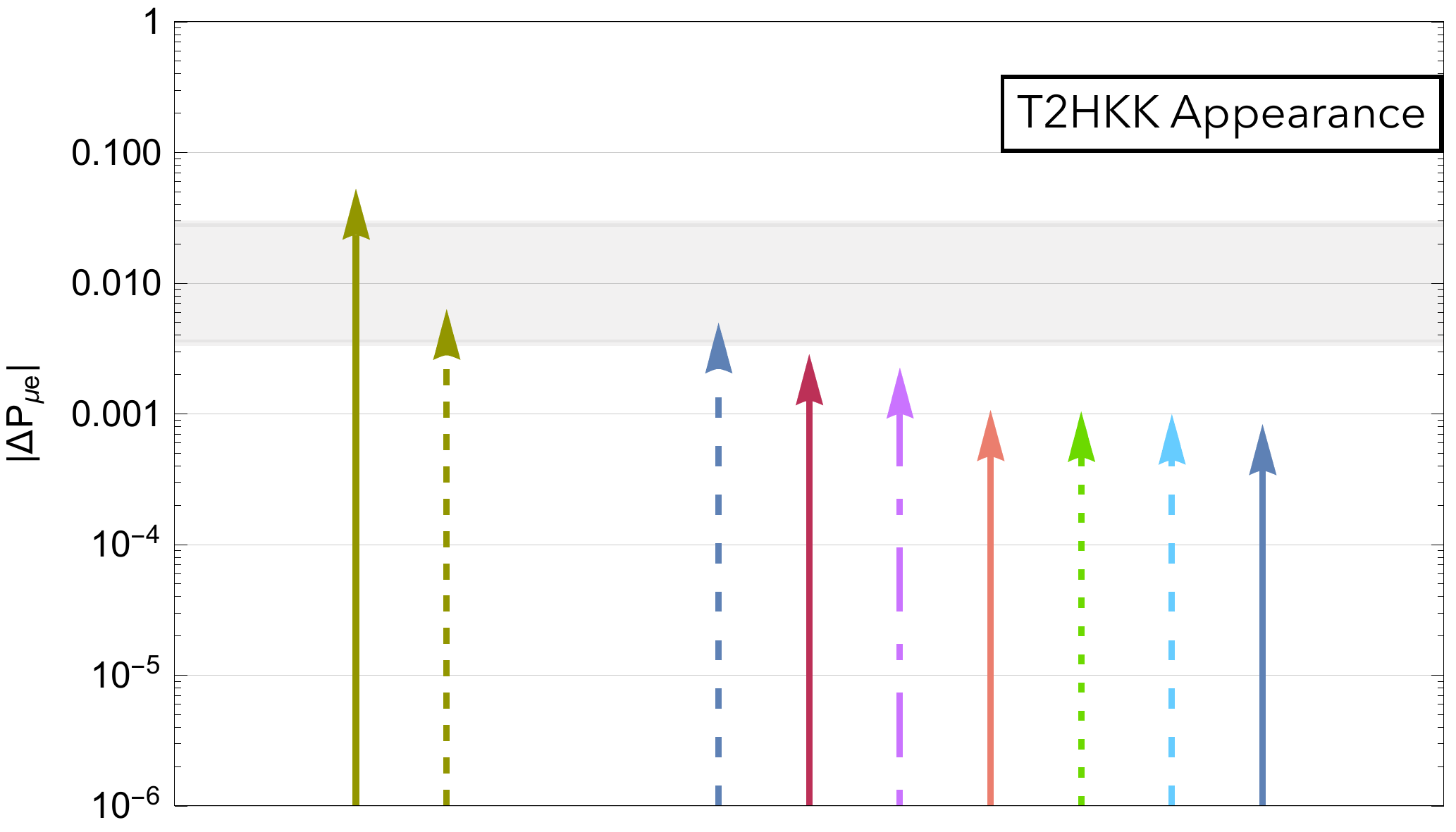}
    	\includegraphics[width=0.485\textwidth]{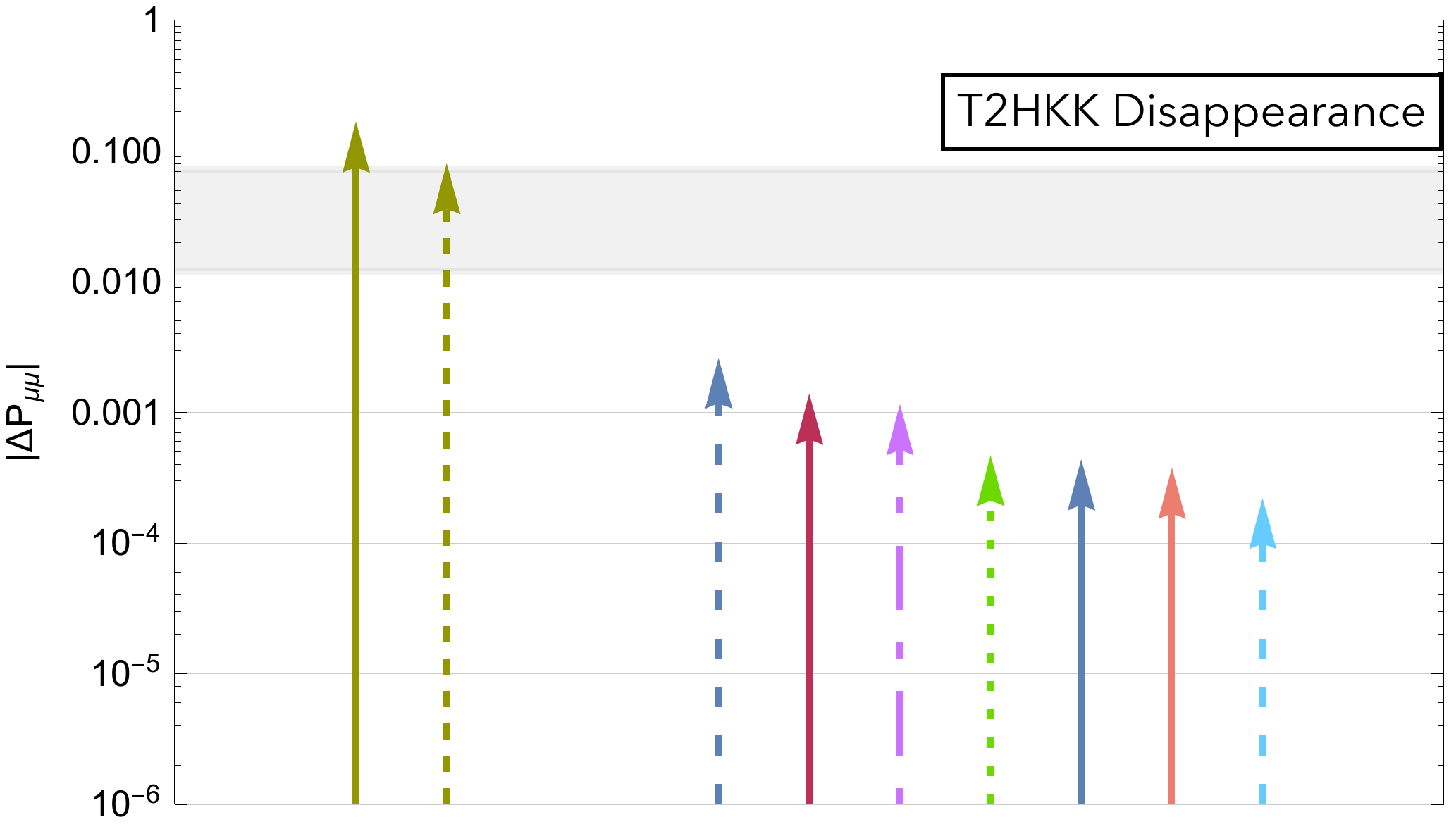}
	}
	\caption{
        \label{fig:summaryHKK}
        Summary of the main results obtained for Mount Bisul (above) and Mount Bohyun (below). Results shown are for all energies, and are grouped into sensitivity estimate, oscillation parameter variation and matter density profile studies. Left and right panels show appearance and disappearance channels, respectively.
    }
\end{figure*}

Similarly, Fig.~\ref{fig:summaryHKK} shows a summary plot for T2HKK, this time including the results from Section~\ref{Low-dens}, which are unique to the T2HKK experiment.

As expected, it can be seen from these plots that there are significant differences between matter effects on neutrino oscillation probabilities at T2HK and T2HKK. While the effects of changes in the matter density profile of T2HKK are in some cases over an order of magnitude larger than those same effects in T2HK, we can conclude that they will nevertheless be immeasurable at both experiments, as they remain smaller than the estimated measurement uncertainty.

Conversely, the results arising from a variation of the oscillation parameters within their $\pm1\sigma$ ranges do, in some cases, exceed this threshold. This is in line with the Hyper-Kamiokande collaboration's goal of determining the value of $\delta_{CP}$ with a sensitivity of $<23\,^\circ$ -- as well as to precisely measure $\sin^2\theta_{23}$ and $\Delta m_{32}^2$ -- within its first 10 years of running \cite{Abe:2018uyc}.

Aside from the parameter variation studies, it is interesting to note that the largest change in oscillation probability always arises from varying the average matter density of a particular profile by $\pm6\%$; the uncertainty on the average given by \cite{Hagiwara:2011kw}. At T2HKK, this is followed in every case by the effect of comparing a full profile from Fig.~\ref{fig:Baselines} to a basic low-density profile. On the other hand, the smaller effects vary in prominence depending on the candidate site and the channel in question.

In summary, we conclude that using a constant density profile in calculations of oscillation probabilities for both appearance and disappearance channels is sufficient for both the T2HK and T2HKK experiments, as is clear from Figs.~\ref{fig:summaryHK} and~\ref{fig:summaryHKK}, respectively.

\section*{Acknowledgments}

The authors would like to thank Alexei\,Yu.\,Smirnov for his positive comments and helpful suggestions.

This manuscript has been authored by Fermi Research Alliance, LLC under Contract No. DE-AC02-07CH11359 with the U.S. Department of Energy, Office of Science, Office of High Energy Physics.

S.\,F.\,K. acknowledges the STFC Consolidated Grant ST/L000296/1 and the European Union's Horizon 2020 Research and Innovation programme under Marie Sk\l {}odowska-Curie grant agreements Elusives ITN No.\ 674896 and InvisiblesPlus RISE No.\ 690575.

S.\,M.\,S. wishes to acknowledge the STFC Doctoral Training Grant held by Queen Mary University of London and additional support from the University of Southampton, the Valerie Myerscough Science and Mathematics Trust Fund at the University of London, and the Spanish grants FPA2017-85216-P (AEI/FEDER, UE) and PROMETEO/2018/165 (Generalitat Valenciana).\newline

\appendix

\section{Matter Density Profile Variations for the other Individual Baselines}
\label{ap:Baselines}

The results for each of the potential baselines studied for T2HKK are shown below in Fig.~\ref{fig:Dens_HKK}. These baselines range from 1000 km to 1200 km, and the results corresponding to 1100 km and 1050 km are used in Section~\ref{Dens} to represent Mount Bisul and Mount Bohyun, respectively.

\begin{figure*}[!htbp]
    \subfloat[Baseline: 1000 km]{%
        \includegraphics[width=0.45\textwidth]{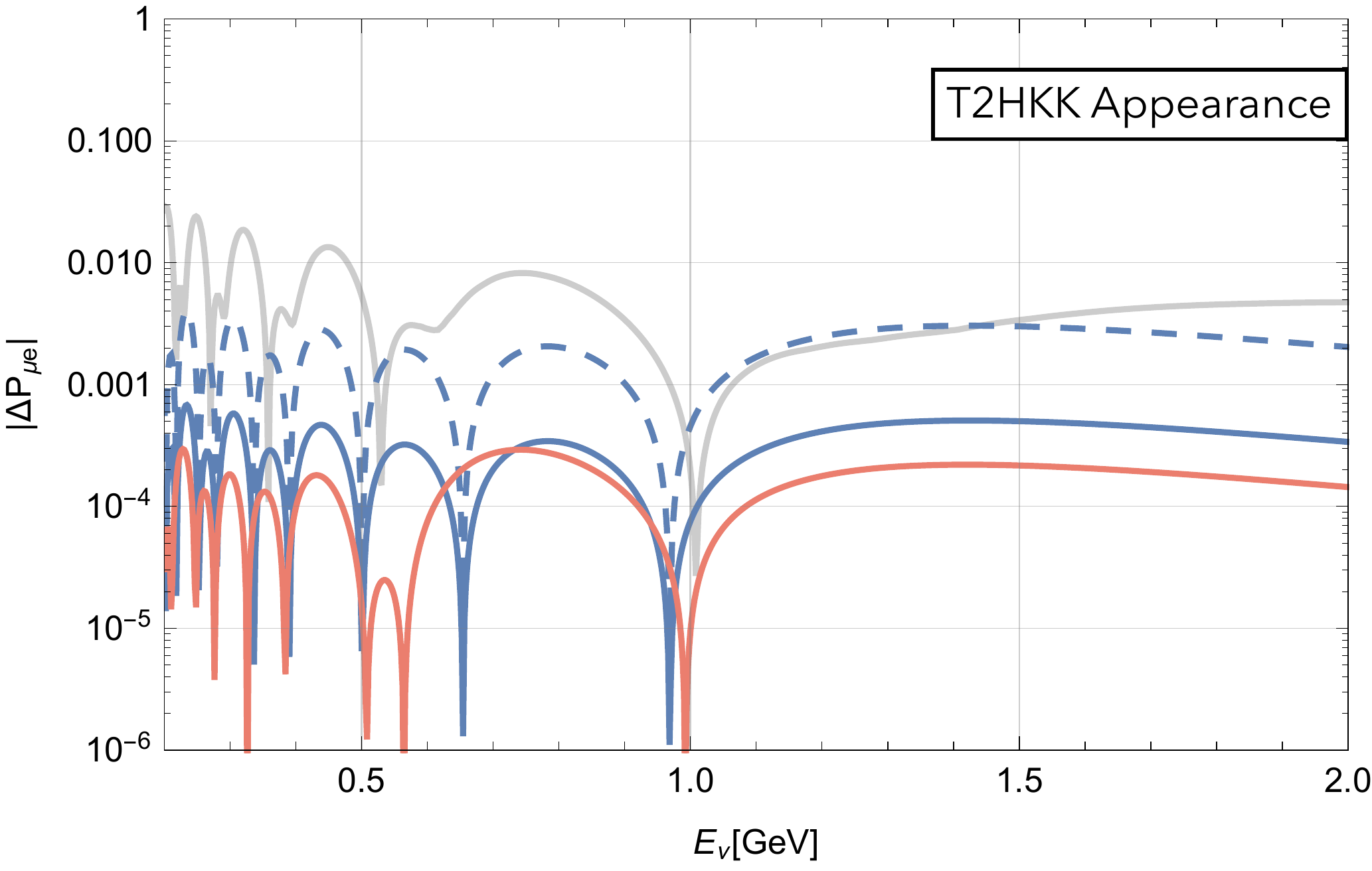}\qquad
	    \includegraphics[width=0.45\textwidth]{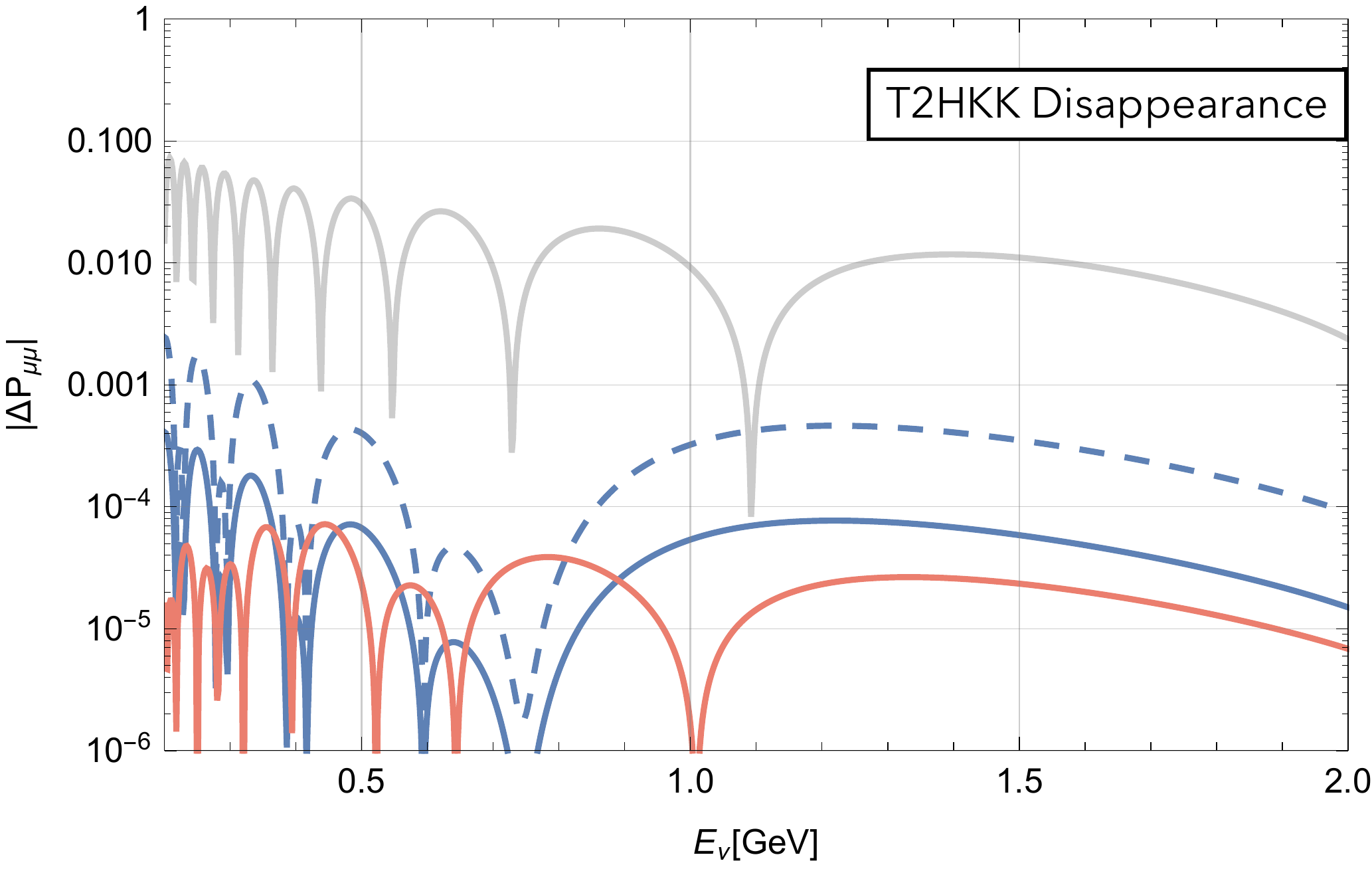}
    }\\
	\subfloat[Baseline: 1050 km]{%
	    \includegraphics[width=0.45\textwidth]{App_HKK_DeltaPvsE_1050_label.pdf}\qquad
	    \includegraphics[width=0.45\textwidth]{Dis_HKK_DeltaPvsE_1050_label.pdf}
    }\\
    \subfloat[Baseline: 1100 km]{%
    	\includegraphics[width=0.45\textwidth]{App_HKK_DeltaPvsE_1100_label.pdf}\qquad
    	\includegraphics[width=0.45\textwidth]{Dis_HKK_DeltaPvsE_1100_label.pdf}
	}\\
	\subfloat[Baseline: 1150 km]{%
	    \includegraphics[width=0.45\textwidth]{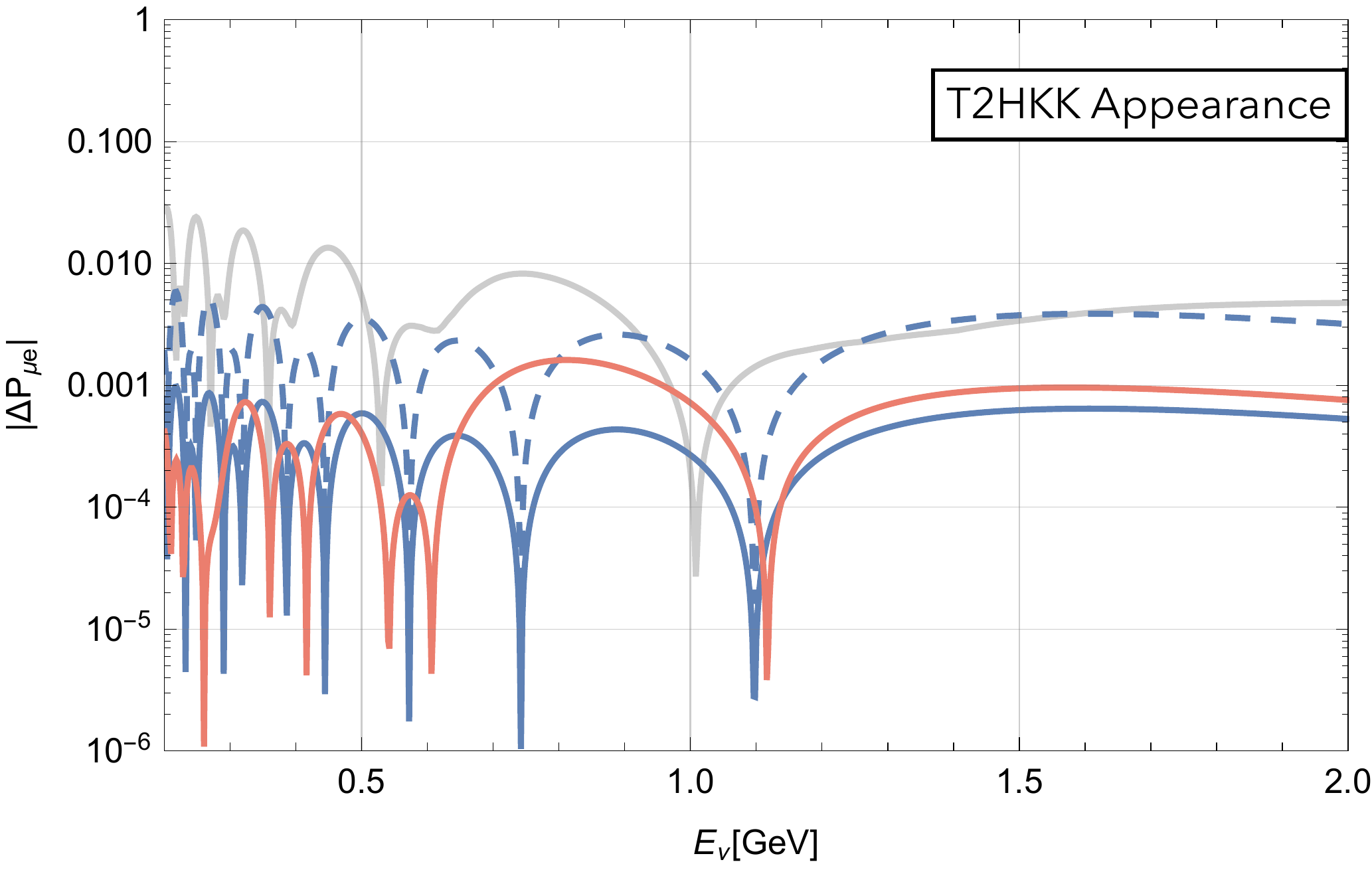}\qquad
	    \includegraphics[width=0.45\textwidth]{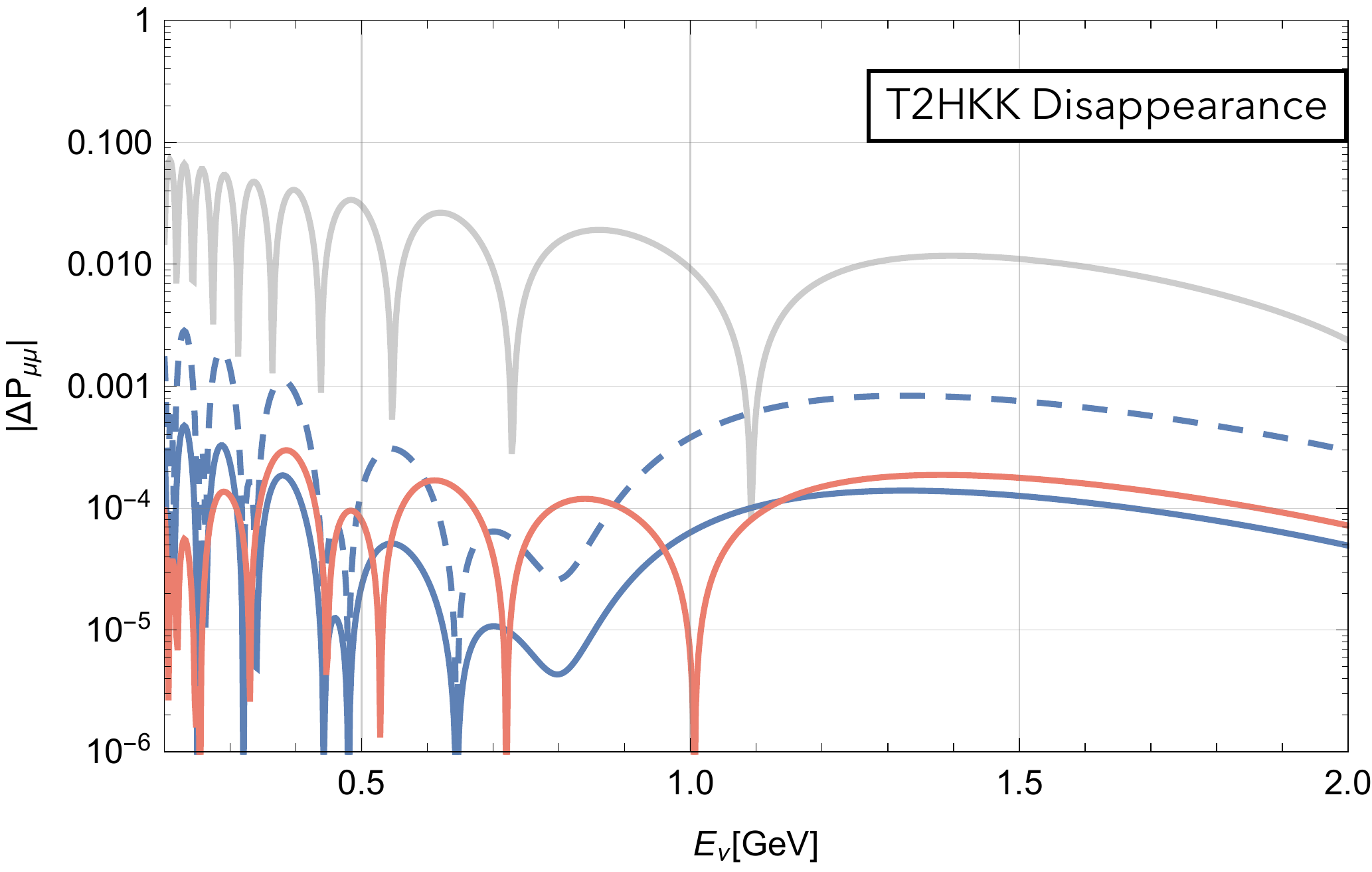}
	}\stepcounter{figure}
\end{figure*}
\begin{figure*}[!htbp]\ContinuedFloat
	\subfloat[Baseline: 1200 km]{%
    	\includegraphics[width=0.45\textwidth]{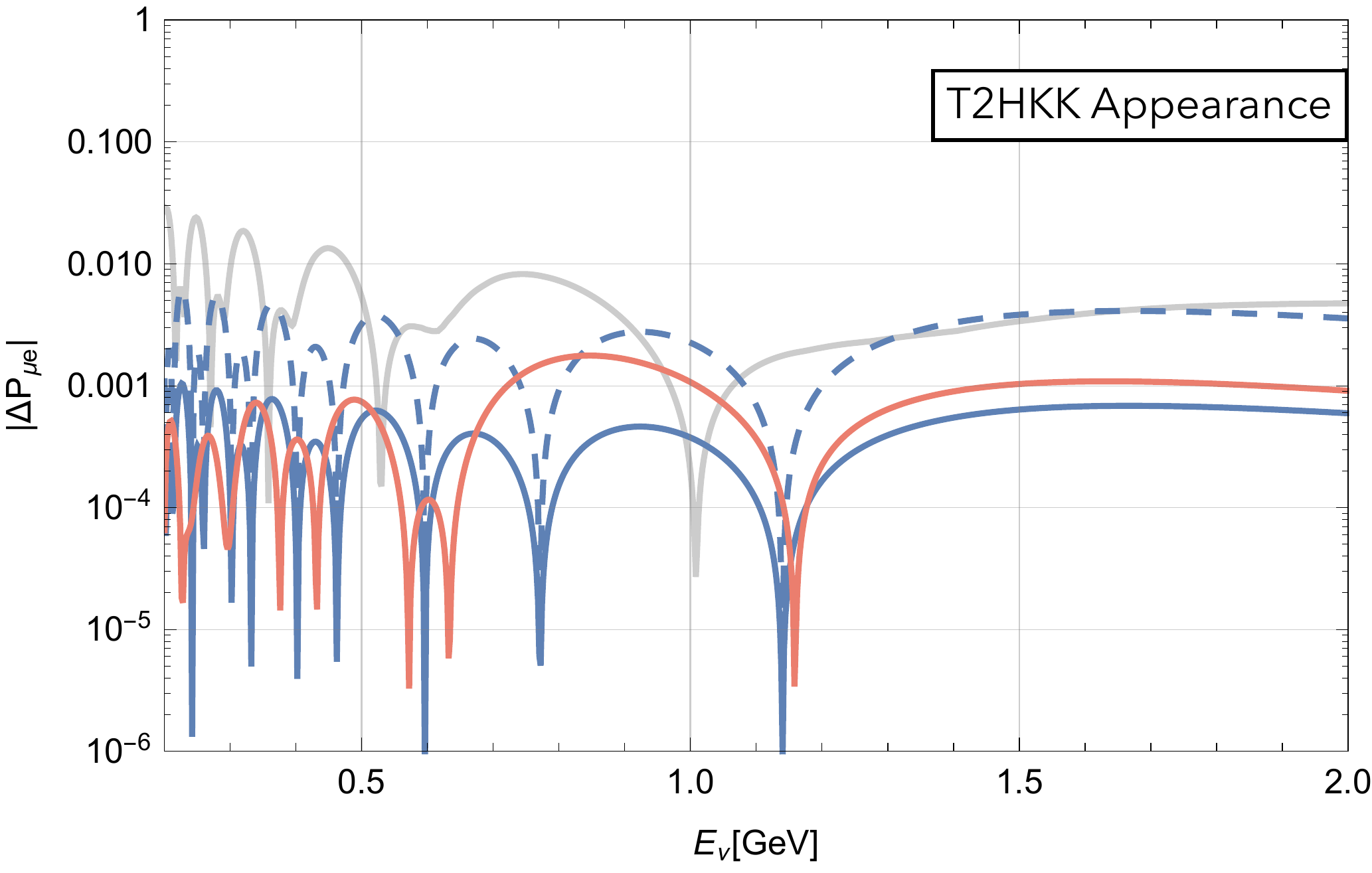}\qquad
    	\includegraphics[width=0.45\textwidth]{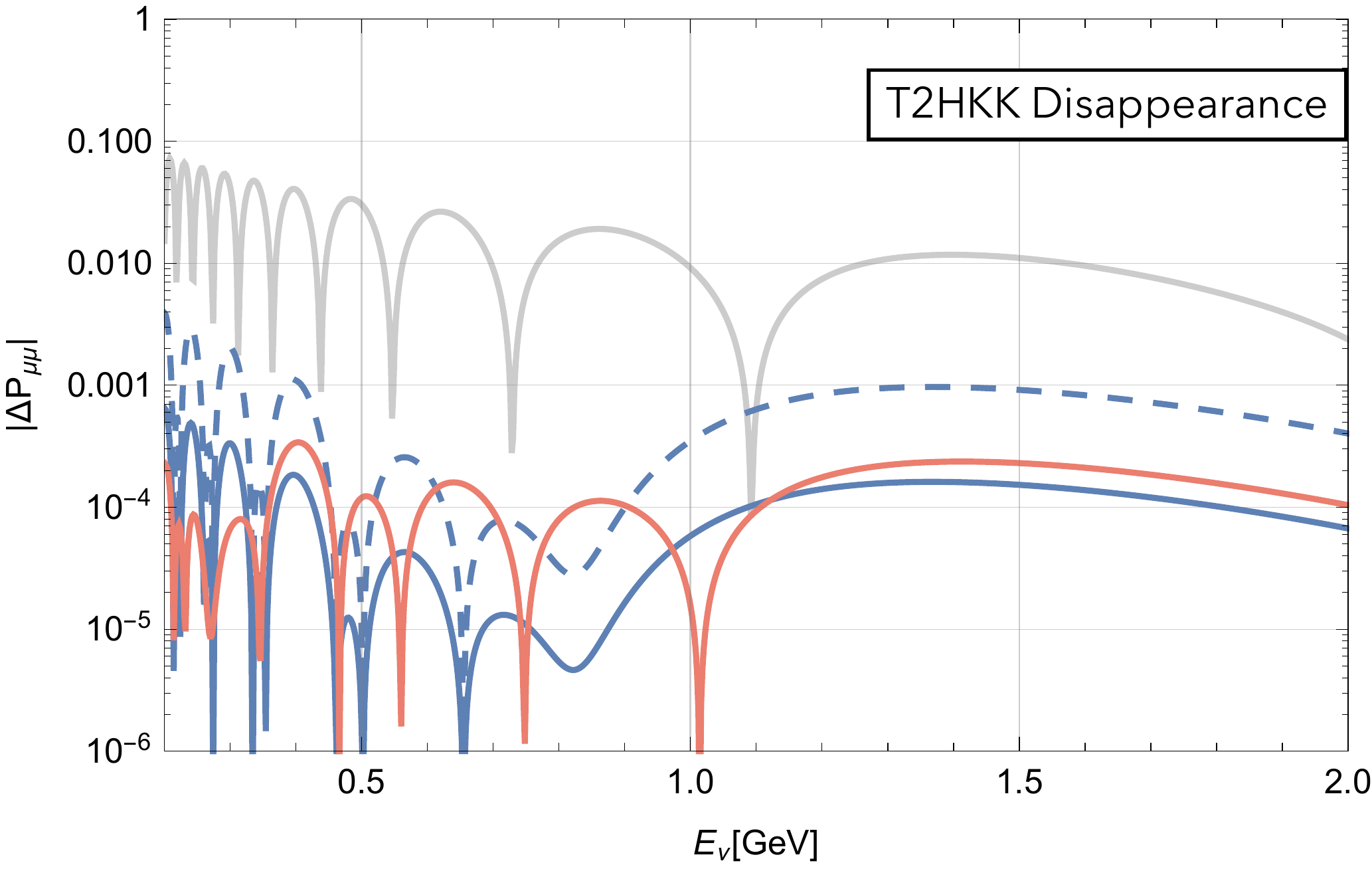}
    }
    \vspace*{5mm}
    \caption{
        Density profile changes with respect to average density corresponding to the five potential candidate sites for the Korean detector, with baselines ranging from 1000 km to 1200 km. The solid (dashed) blue curve is obtained by varying the average density by $\pm 1\%$ ($\pm 6\%$) and taking the difference between the two extremes at each point. The red curve is produced by taking the difference at each point between the probability calculated using the changing (real) matter density profile and the probability computed using its average density value as a constant. Left and right panels correspond to appearance and disappearance channels, respectively.
    }
    \label{fig:Dens_HKK}
\end{figure*}

\section{Combined 5-baseline studies}
\label{ap:Combined}

\begin{figure*}[!htbp]
\vspace*{5mm}
	\includegraphics[width=0.45\textwidth]{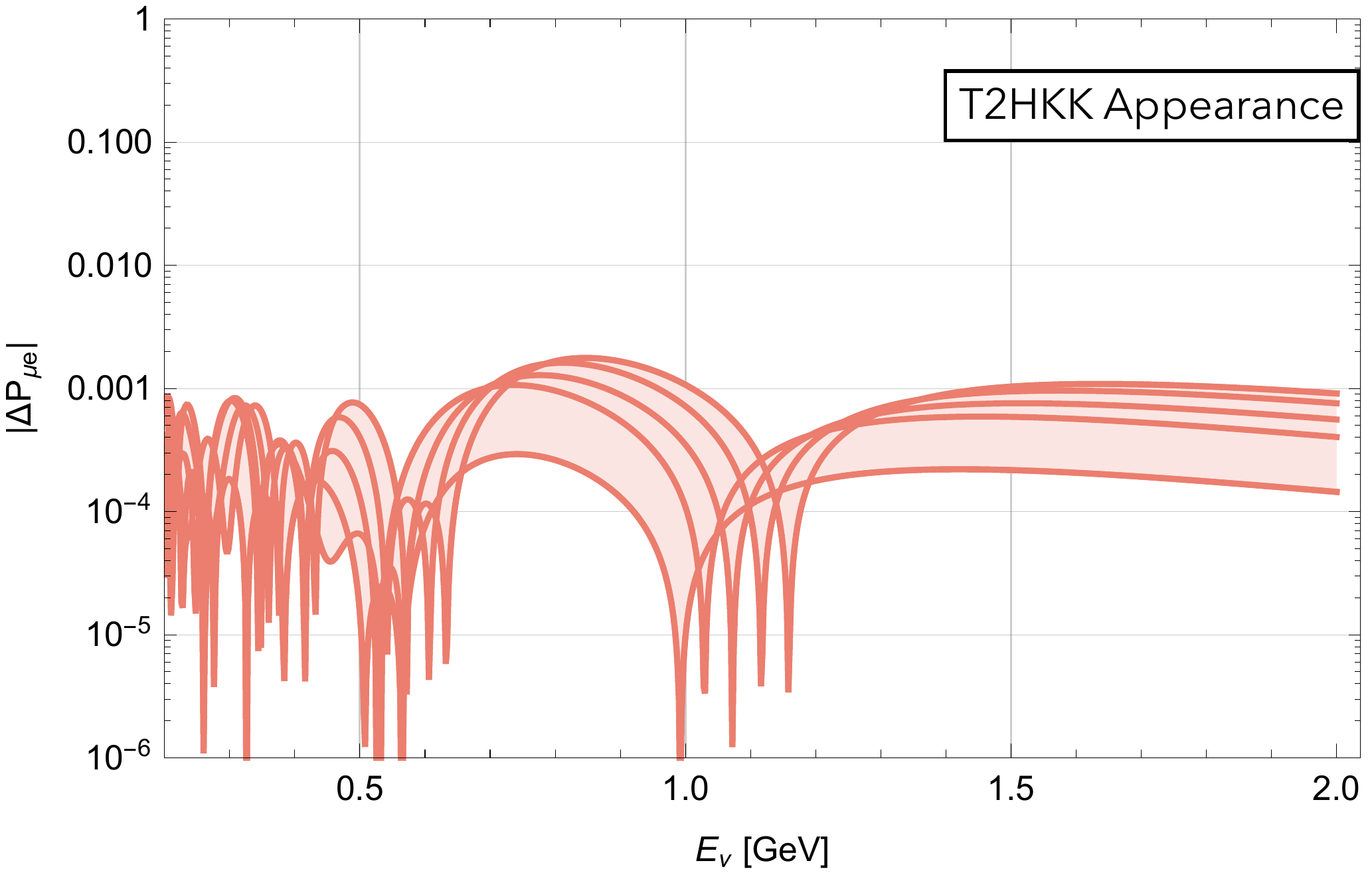}\qquad
	\includegraphics[width=0.45\textwidth]{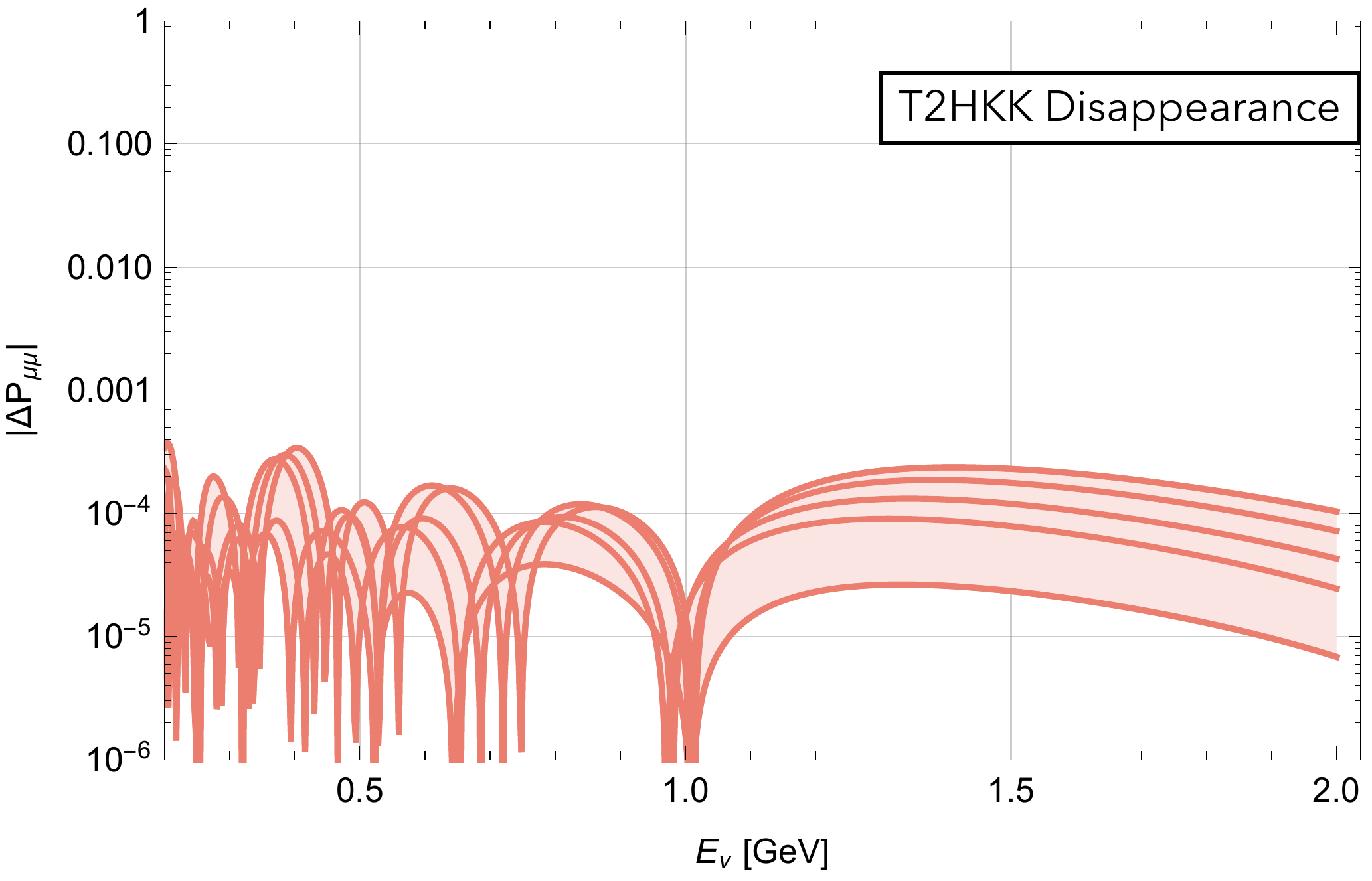}\\
	\includegraphics[width=0.45\textwidth]{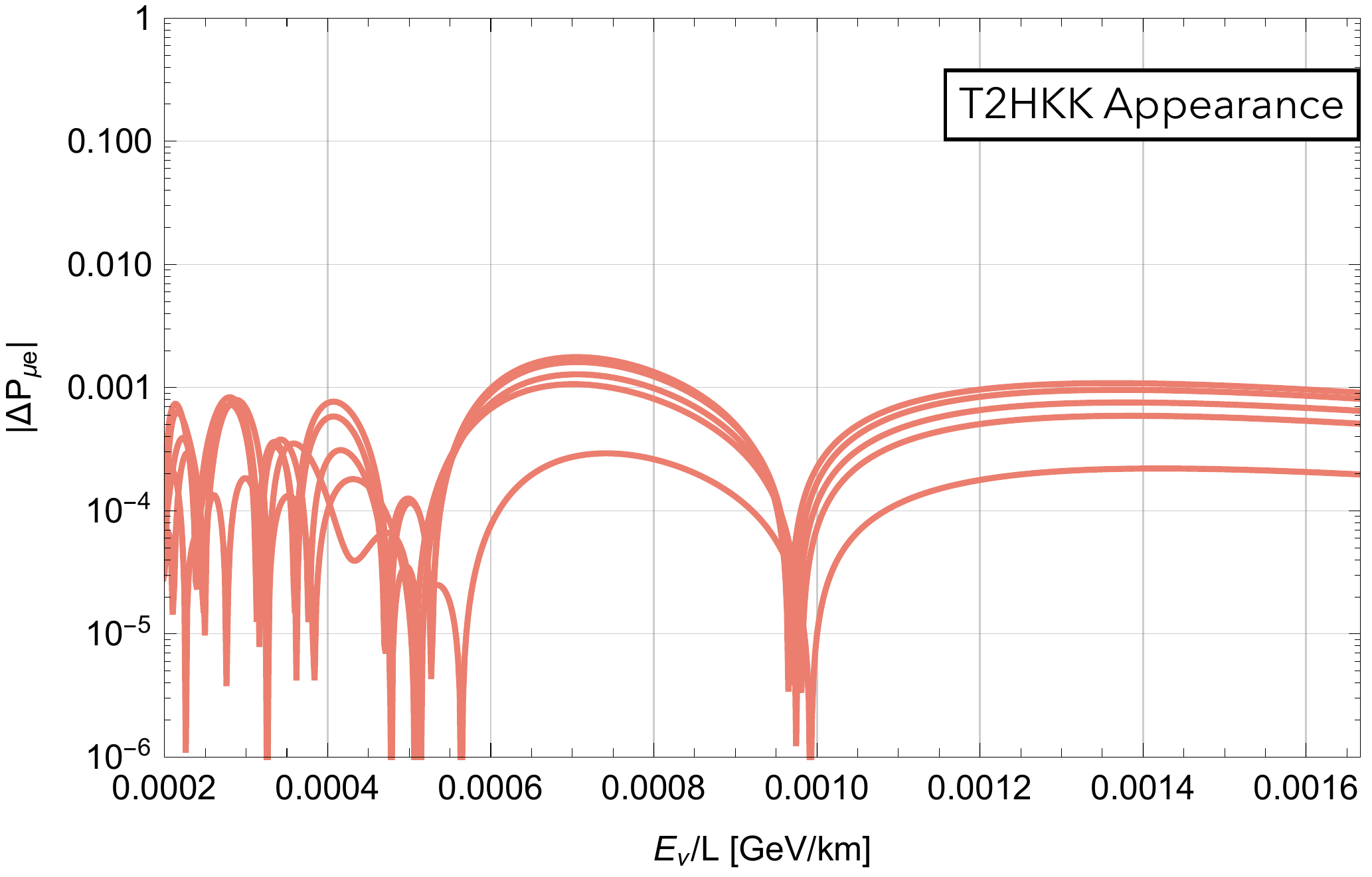}\qquad
	\includegraphics[width=0.45\textwidth]{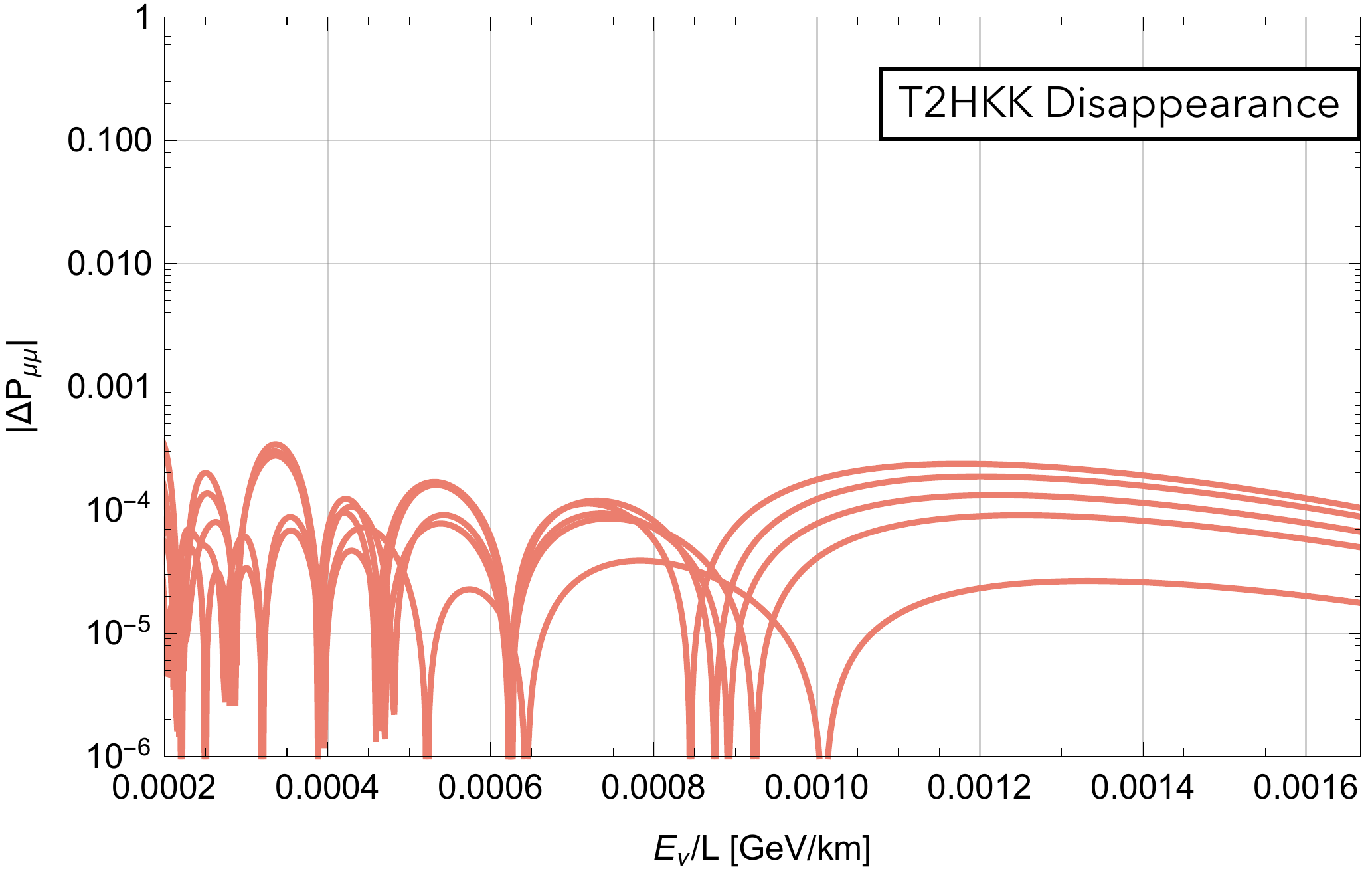}
\vspace*{5mm}
\caption{Changes in oscillation probability as a function of energy (top) and $E/L$ (bottom) for the five different potential baselines with respect to average density of T2HKK.}
\label{fig:DeltaProb_All_HKK}
\end{figure*}

Fig.~\ref{fig:DeltaProb_All_HKK} shows the individual changes in oscillation probability of each baseline profile with respect to their average density value, as listed in Appendix~\ref{ap:Baselines}, combined into one plot, as a function of energy and of $E/L$.

\begin{figure*}[!htbp]
	\includegraphics[width=0.45\textwidth]{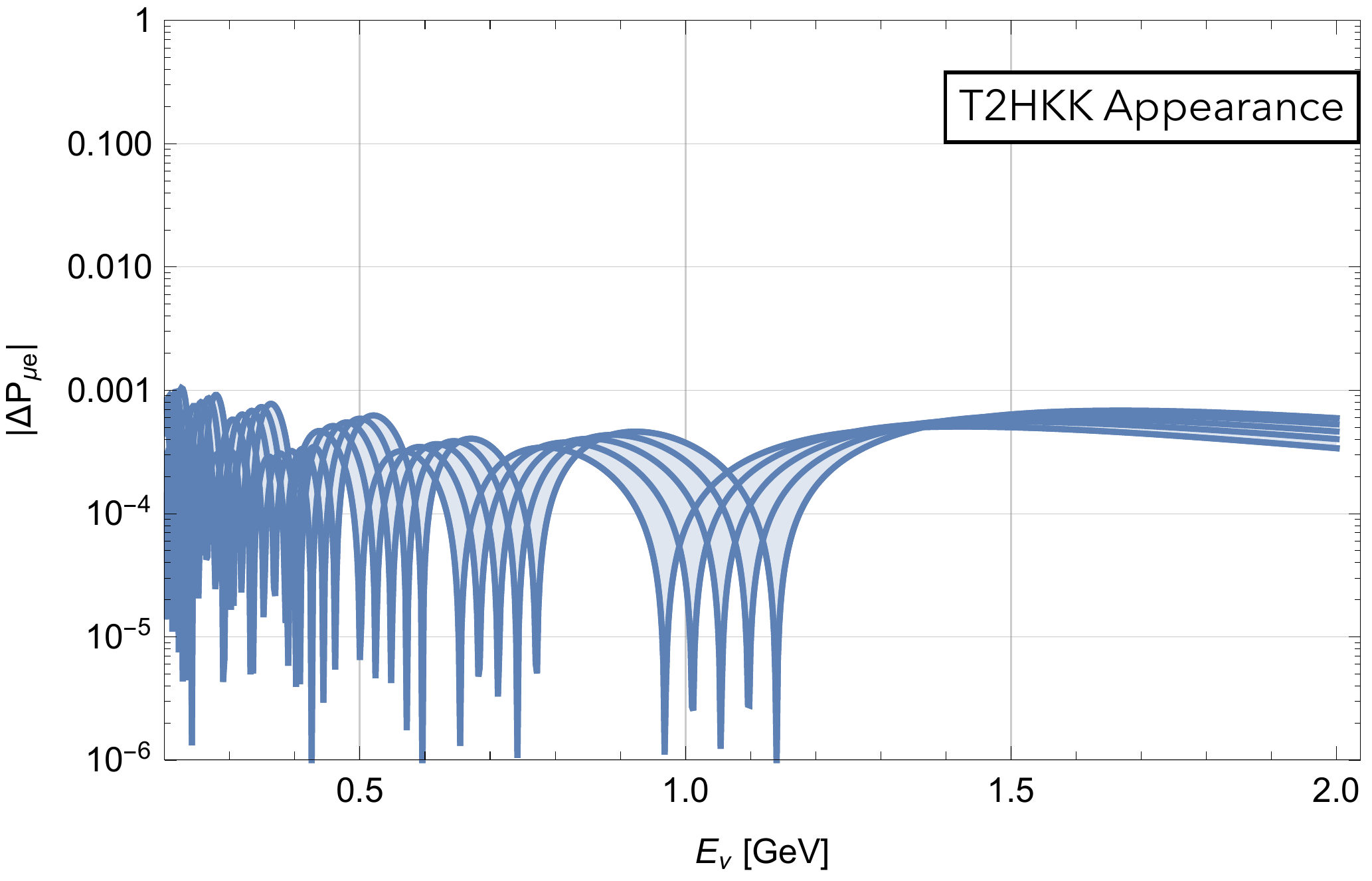}\qquad
	\includegraphics[width=0.45\textwidth]{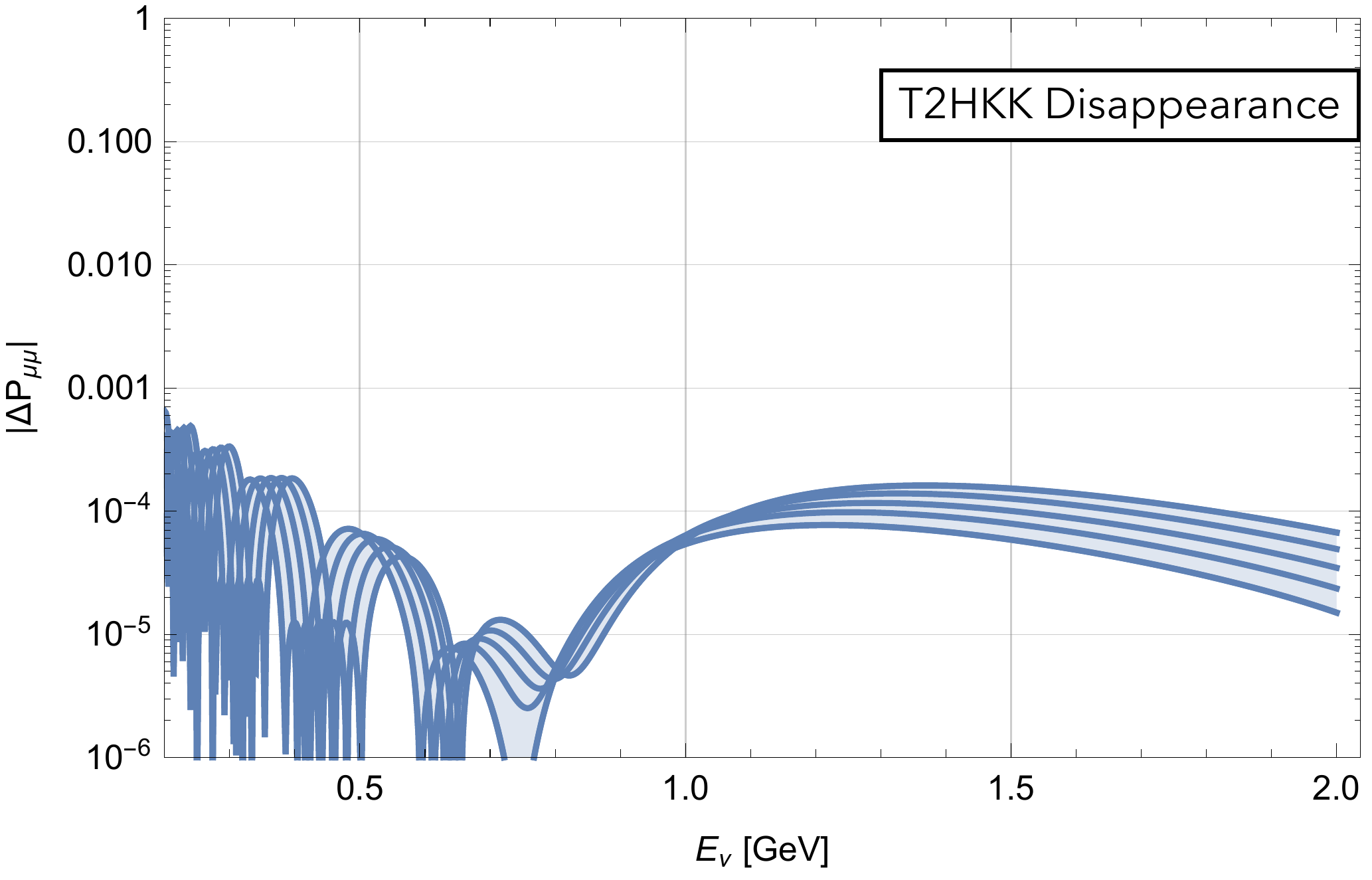}\\
	\includegraphics[width=0.45\textwidth]{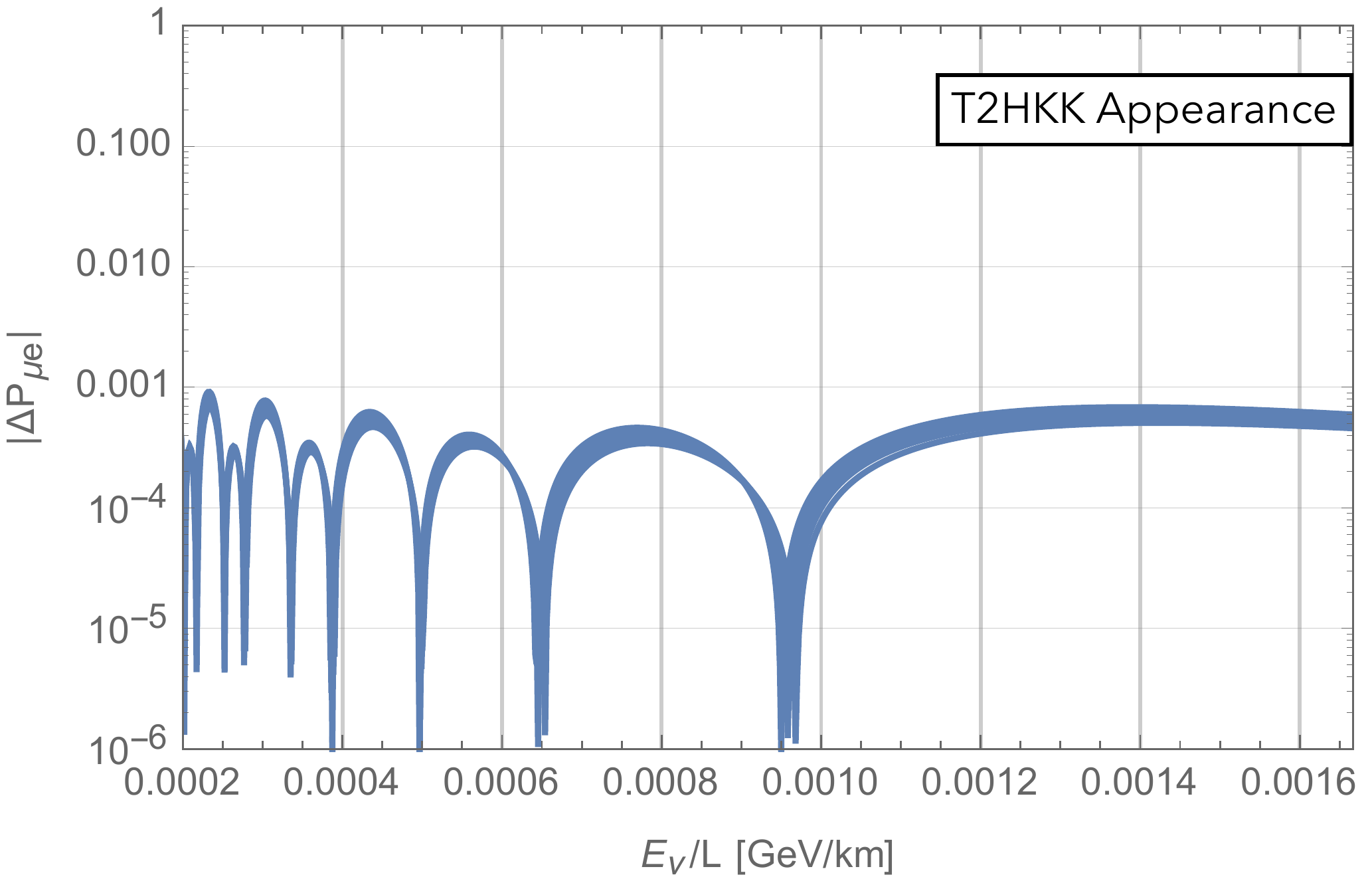}\qquad
	\includegraphics[width=0.45\textwidth]{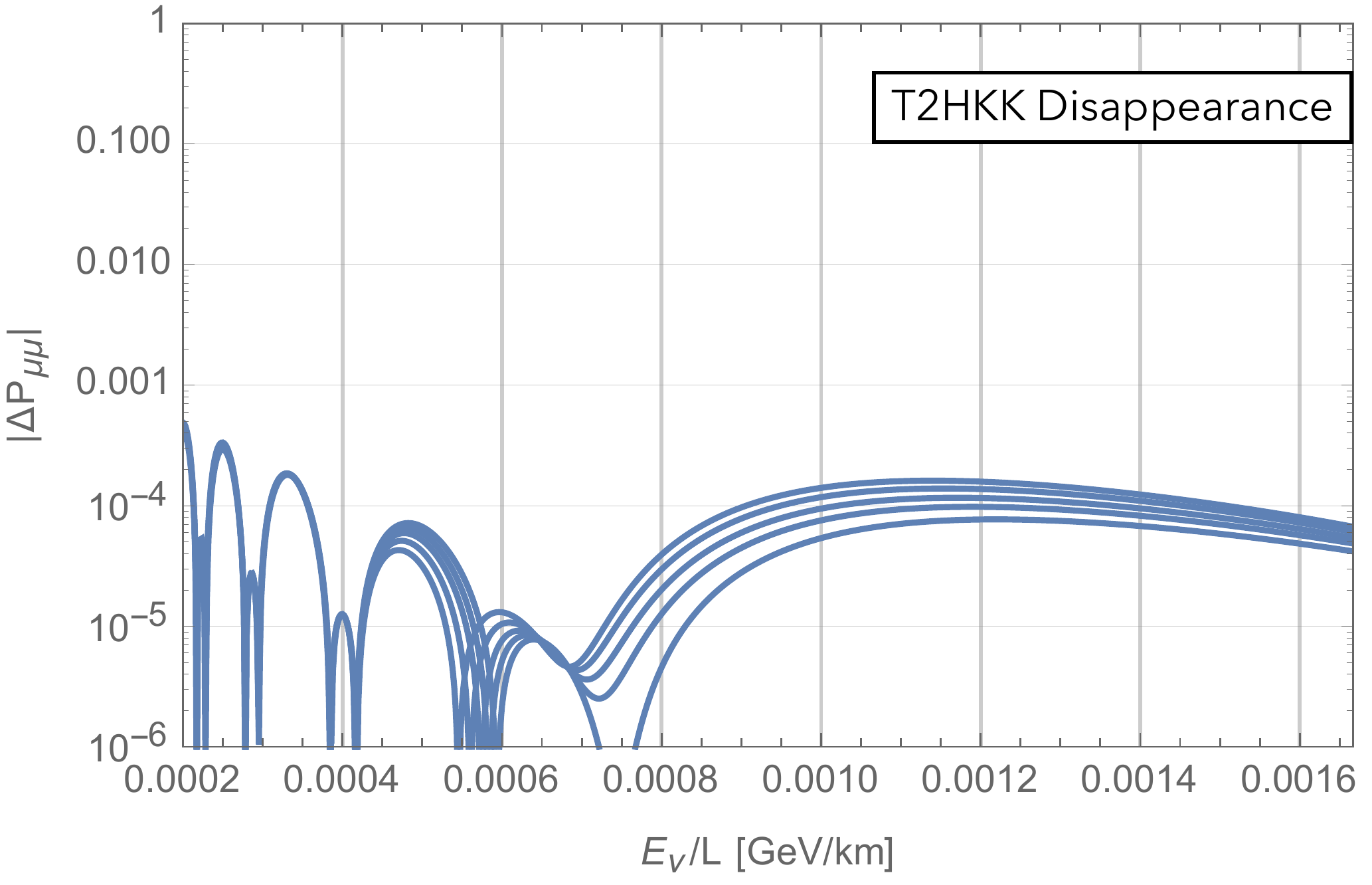}
\caption{Changes in oscillation probability as a function of energy (top) and $E/L$ (botton) arising from varying the average density of the five different potential baselines of T2HKK by $\pm 1\%$.}
\label{fig:1PercentProb_All_HKK}
\end{figure*}

Then, Fig.~\ref{fig:1PercentProb_All_HKK} shows the changes in oscillation probability given by varying the average density of each baseline profile by $\pm 1\%$, as listed in Appendix~\ref{ap:Baselines}, combined into one plot, as a function of energy and of $E/L$.

\begin{figure*}[!htbp]
	\includegraphics[width=0.45\textwidth]{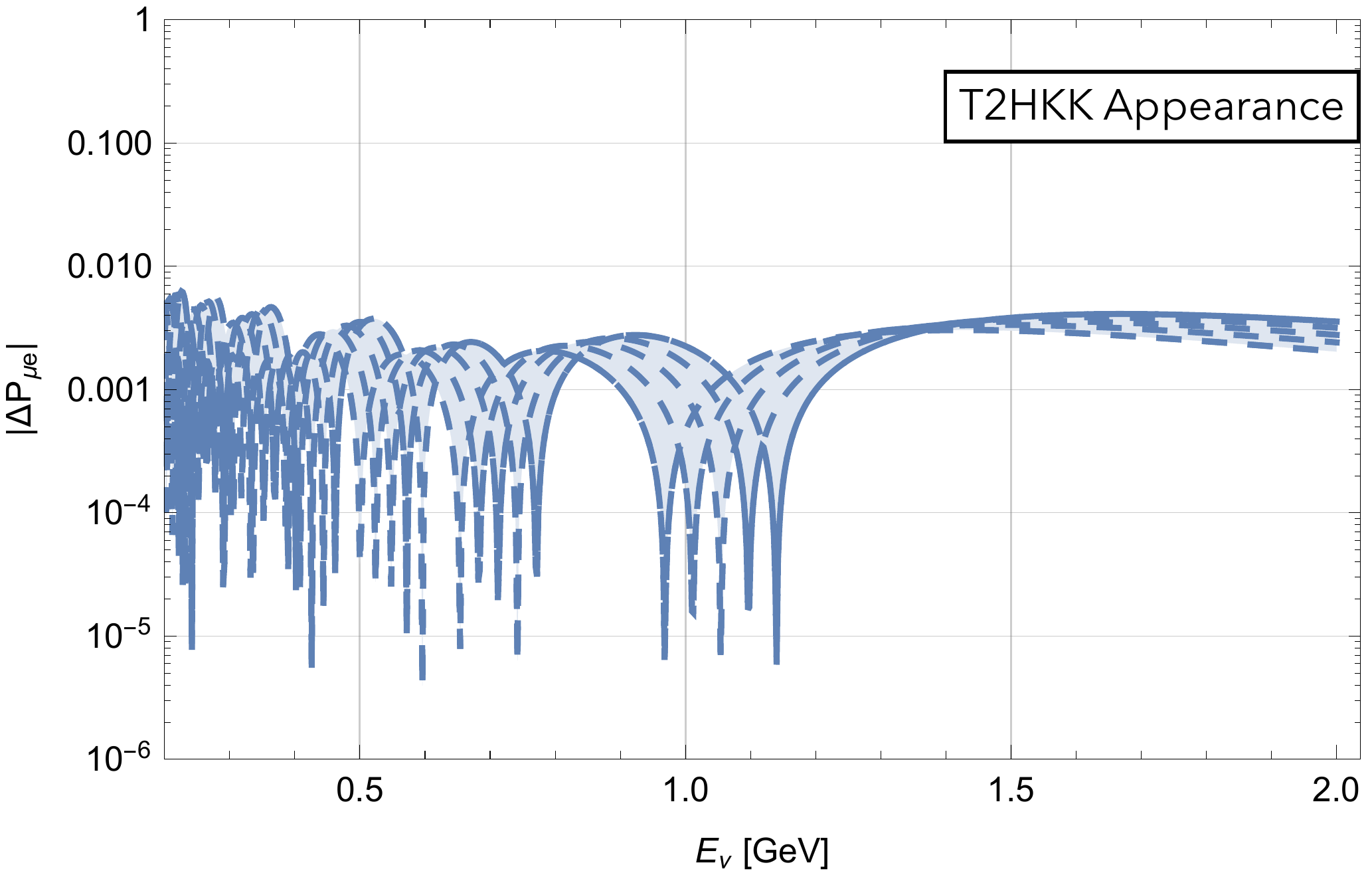}\qquad
	\includegraphics[width=0.45\textwidth]{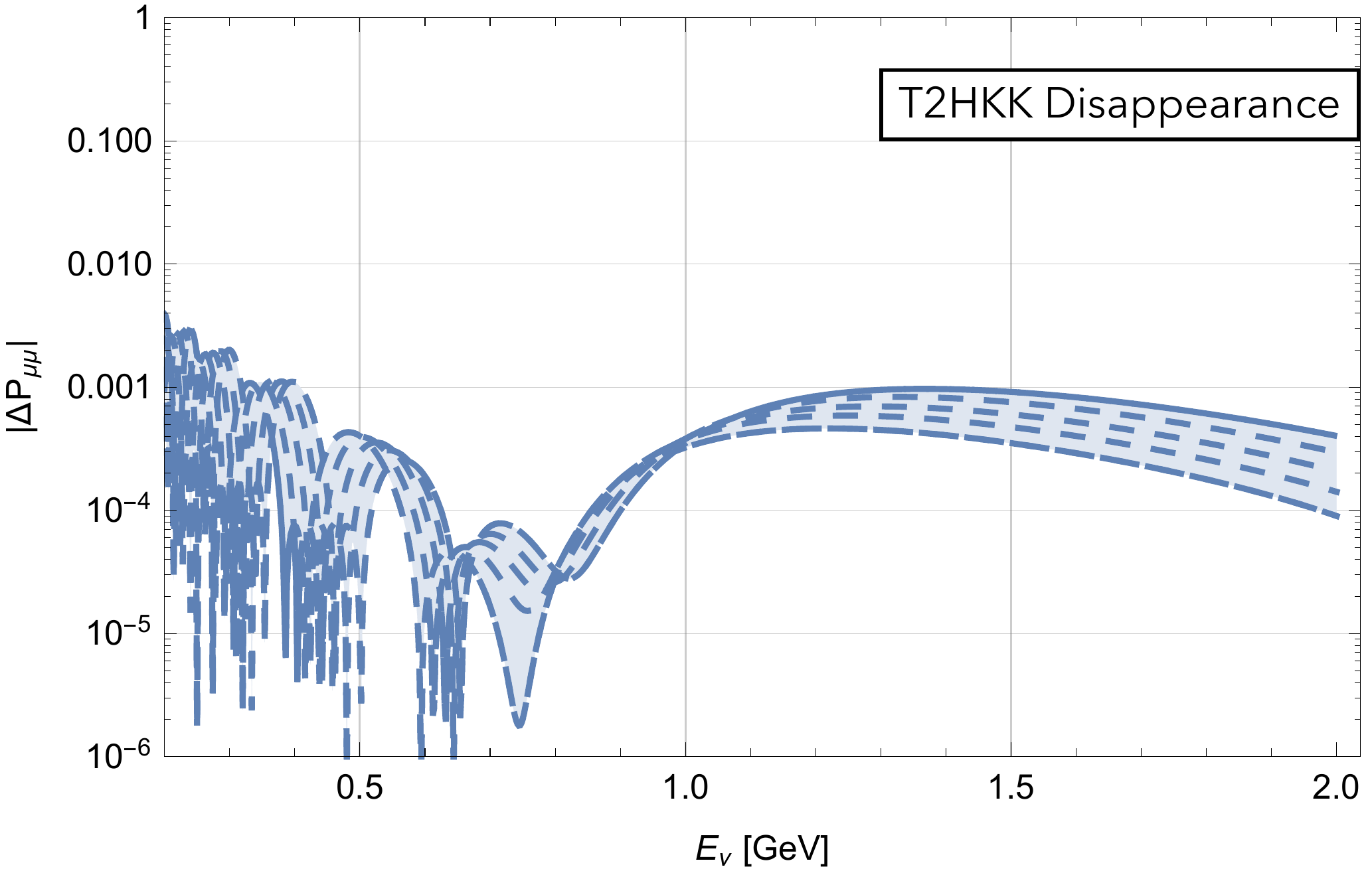}\\
	\includegraphics[width=0.45\textwidth]{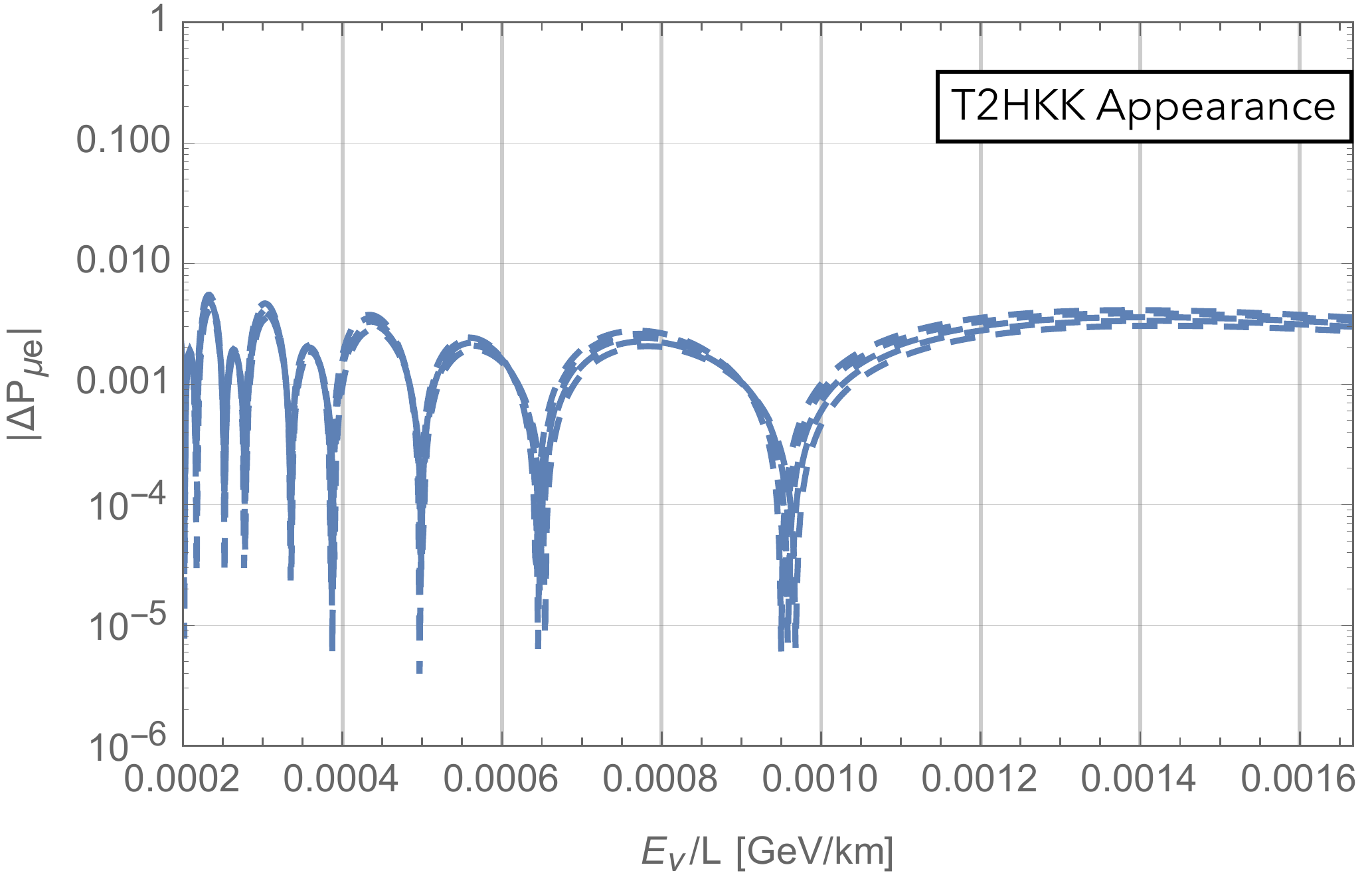}\qquad
	\includegraphics[width=0.45\textwidth]{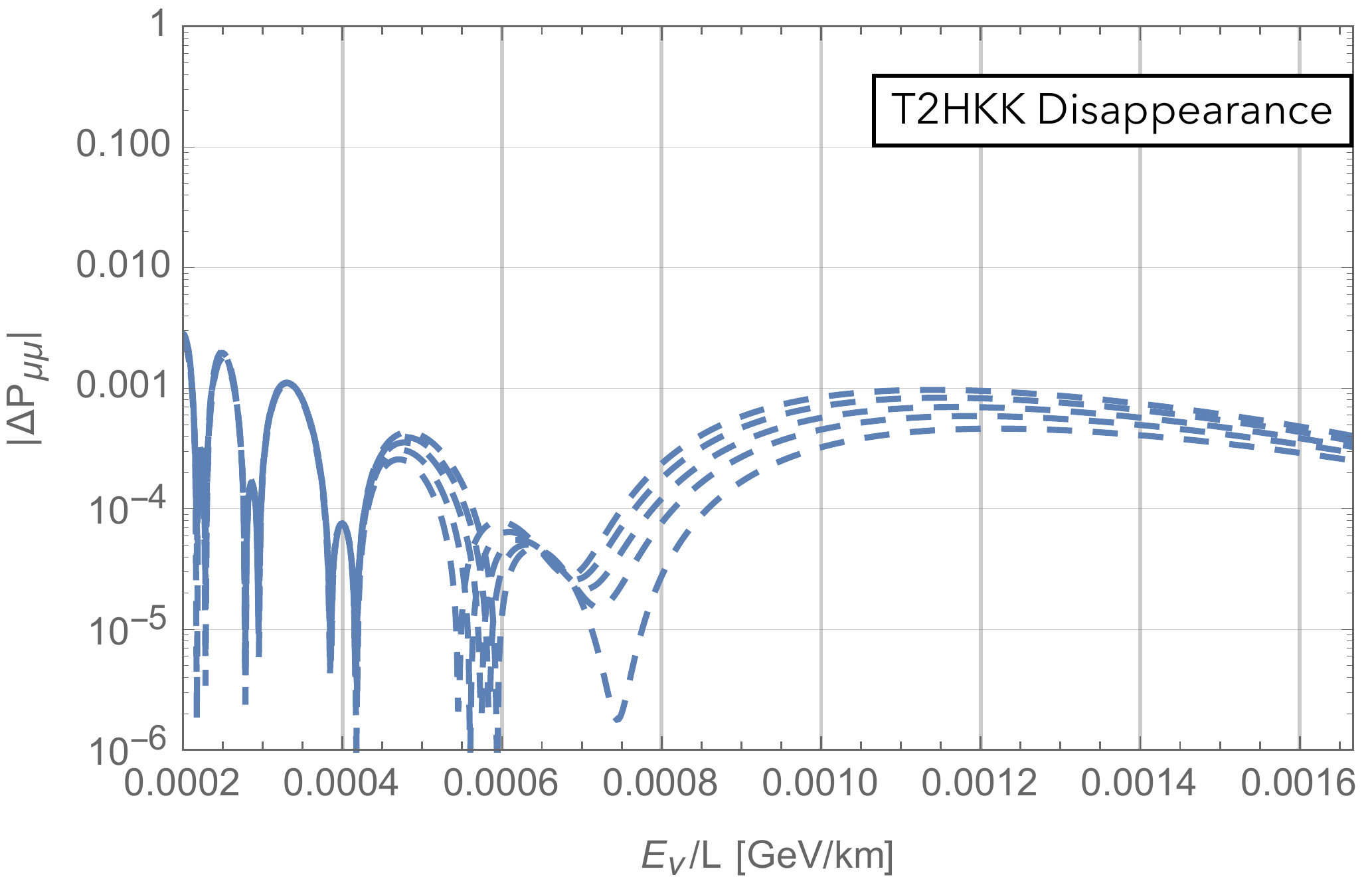}
    \caption{Changes in oscillation probability as a function of energy (top) and $E/L$ (bottom) arising from varying the average density of the five different potential baselines of T2HKK by $\pm 6\%$.}
\label{fig:6PercentProb_All_HKK}
\end{figure*}

Finally, Fig.~\ref{fig:6PercentProb_All_HKK} shows the changes in oscillation probability given by varying the average density of each baseline profile by $\pm 6\%$, as listed in Appendix~\ref{ap:Baselines}, combined into one plot, as a function of energy and of $E/L$.

\end{document}